\documentclass[aps,epsfigure,twocolumn,superscriptaddress,showkeys,nofootinbib]{revtex4-1}

\usepackage[colorlinks=true,linkcolor=blue,urlcolor=blue,citecolor=blue,pdfusetitle]{hyperref}
\usepackage[utf8]{inputenc}
\usepackage[english]{babel}
\usepackage{amsmath}
\usepackage[caption = false]{subfig}
\usepackage{graphicx,epstopdf}
\usepackage{blindtext}
\usepackage{float}
\usepackage{ragged2e}
\usepackage[table,xcdraw]{xcolor}
\usepackage{lipsum}
\usepackage{amsfonts}
\usepackage{bbm}
\usepackage{amssymb}
\usepackage{enumerate}
\usepackage{color}
\usepackage{latexsym}
\usepackage{times,txfonts}
\usepackage[normalem]{ulem}
\usepackage{amsmath}
\usepackage{tikz}
\usetikzlibrary{quantikz}
\usepackage{algpseudocode}
\usepackage{algorithm}
\usepackage{tikz}
\usepackage{threeparttable}
\usepackage{enumitem}
\usepackage{xcolor}

\usepackage[most]{tcolorbox}
\tcbuselibrary{theorems,skins}

\newtcbtheorem[auto counter]{examplebox}{Example}{%
  enhanced,
  colback=blue!4!white,
  colframe=blue!75!black,
  fonttitle=\bfseries,
  coltitle=white,
  colbacktitle=blue!75!black,
  sharp corners,
  boxrule=0.8pt,
  title style={left color=blue!100!black,right color=blue!80!black},
  separator sign=:\ %
}{ex}

\begin{document}

\title{Variational quantum {computing} for quantum simulation: principles, implementations, and {challenges}}

\author{Lucas Q. Galvão}
\email{lqgalvao3@gmail.com}
\affiliation{QuIIN - Quantum Industrial Innovation, Centro de Competência Embrapii Cimatec. SENAI CIMATEC, Av. Orlando Gomes, 1845, Salvador, BA, Brazil CEP 41850-010}
\affiliation{Universidade SENAI CIMATEC, Salvador, BA, Brazil}

\author{Anna Beatriz M. de Souza}
\affiliation{QuIIN - Quantum Industrial Innovation, Centro de Competência Embrapii Cimatec. SENAI CIMATEC, Av. Orlando Gomes, 1845, Salvador, BA, Brazil CEP 41850-010}
\affiliation{Universidade SENAI CIMATEC, Salvador, BA, Brazil}

\author{Marcelo A. Moret}
\affiliation{Universidade SENAI CIMATEC, Salvador, BA, Brazil}

\author{Clebson Cruz}
\email{Corresponding author: Clebson Cruz \newline Email: clebson.cruz@ufob.edu.br}
\affiliation{Centro de Ciências Exatas e das Tecnologias, Universidade Federal do Oeste da Bahia - Campus Reitor Edgard Santos. Rua Bertioga, 892, Morada Nobre I, 47810-059 Barreiras, Bahia, Brasil.}



\begin{abstract}
This work presents a comprehensive overview of variational quantum {computing} and their key role in advancing quantum simulation. 
This work {explores} the simulation of quantum systems and sets itself apart from approaches centered on classical data processing, {by focusing on the critical role of quantum data in Variational Quantum Algorithms (VQA) and Quantum Machine Learning (QML)}. We systematically delineate the foundational principles of variational quantum computing, establish their motivational {and challenges} context within the noisy intermediate-scale quantum (NISQ) era, and critically examine their application across a range of prototypical quantum simulation {problems}. Operating within a hybrid quantum-classical framework, these algorithms represent a  promising 
{yet problem-dependent pathway whose practicality remains contingent on trainability and scalability under noise and barren-plateau constraints.}
This review serves to complement and extend existing literature by synthesizing the most recent advancements in the field and providing a focused perspective on the persistent challenges and emerging opportunities that define the current landscape of variational quantum {computing for quantum simulation}.

\end{abstract}
\maketitle

\section{Introduction}

The simulation of physical systems constitutes a promising field of research, since it enables the identification of fundamental properties of natural phenomena \cite{Rohrlich_1990}. {The advent of digital computing significantly expanded this capability, allowing researchers to numerically investigate scientific problems that resist analytical solution \cite{brookshear2009computer, feynman2018feynman}. However, the inherent computational complexity of simulating quantum mechanical systems presents a formidable challenge, as their probabilistic nature necessitates resources that grow exponentially with system size \cite{feynman2018simulating}. This fundamental limitation delineates the computational boundary where classical resources encounter their most profound constraint: the inability to efficiently solve problems in a polynomial time  \cite{nielsen2010quantum, PhysRevLett.94.170201, Qin2022}.}

{This challenge establishes the central motivation for the emergence of quantum computing: } the potential to circumvent these fundamental limitations by employing inherently quantum architectures to emulate quantum systems~\cite{doi:10.1126/science.273.5278.1073, RevModPhys.86.153}. {Indeed, Feynman's seminal insight established the visionary framework of a universal quantum simulator \cite{feynman2018simulating}, which catalyzed foundational research in quantum computing \cite{doi:10.1126/science.273.5278.1073, RevModPhys.86.153, PhysRevLett.82.5381}. This paradigm subsequently expanded beyond its original scope to encompass diverse computational domains \cite{doi:10.1137/S0036144598347011, 10.1145/237814.237866, PhysRevLett.103.150502, Biamonte2017, nielsen2010quantum}. } While classical simulations operate through the manipulation of discrete binary digits ({bits}), quantum simulations utilize quantum bits, or qubits, as their fundamental unit of information~\cite{deutsch1985quantum}. This unique capability has enabled significant advances across diverse applications, including molecular energy calculations~\cite{doi:10.1073/pnas.0808245105, cao2019quantum}, quantum dynamics~\cite{PhysRevLett.82.5381, doi:10.1126/science.273.5278.1073}, and the study of strongly correlated many-body systems~\cite{PhysRevLett.101.070503, PhysRevLett.118.140403}, establishing quantum simulation as a cornerstone of modern quantum information science~\cite{RevModPhys.86.153}.

However, a primary impediment to the widespread application of quantum simulation is its reliance on fault-tolerant quantum computation, a technological milestone that has not yet been realized \cite{preskill2025battling}. This requirement severely constrains the scope of current implementations, as existing quantum processors are limited to the so-called Noisy Intermediate-Scale Quantum (NISQ) {hardware}, characterized by limited
numbers of qubits and noise processes that limit circuit depth ~\cite{preskill2018quantum}. In this context, variational quantum computing emerges as a promising paradigm designed to leverage these imperfect devices with a hybrid quantum-classical architecture: a parameterized quantum circuit (ansatz) is executed on the quantum processor to prepare a trial state and estimate a cost function, while a classical optimizer iteratively adjusts the parameters to minimize this cost ~\cite{cerezo2021variational, RevModPhys.94.015004}. {In principle,} this approach mitigates the impact of noise and depth constraints by distributing the computational workload, positioning variational quantum computing as a leading strategy for achieving practical quantum utility without full error correction \cite{preskill2018quantum}.


In the literature, the use of quantum variational algorithms for quantum simulations has demonstrated applications in areas such as quantum chemistry \cite{peruzzo2014variational, TILLY20221}, dynamics of conservative \cite{li2017efficient}, and open systems \cite{PhysRevLett.125.010501, doi:10.1021/acs.jpclett.4c00576}, quantum thermodynamics \cite{PhysRevLett.123.220502, verdon2019quantumhamiltonianbasedmodelsvariational}, many-body simulation \cite{PhysRevResearch.3.023095}, fundamentals of quantum physics \cite{arrasmith2019variational}, among others \cite{cerezo2021variational, RevModPhys.94.015004}. {However, recent studies have revealed fundamental challenges related to barren plateaus (BP) and classical simulability that may limit its scalability \cite{larocca2025barren}.} Thus, evaluating the applicability of variational quantum computing for quantum simulation remains a central challenge in current quantum computing research.

This work provides a comprehensive overview of variational quantum {computing} and their applications to quantum simulation. Our focus is restricted to the domain of simulating quantum systems, explicitly excluding its applications to classical data processing. We delineate the foundational principles, challenges, and illustrative applications. {While previous reviews have separately covered quantum simulation \cite{RevModPhys.86.153, Zhang2025, Bauer2023} and variational quantum computing \cite{cerezo2021variational, RevModPhys.94.015004}, this work specifically addresses their intersection by focusing on the critical role of \emph{quantum data}. We define quantum data as initial states prepared by a quantum mechanical process of interest, typically generated within a quantum device. By integrating advances in variational methods with quantum simulation tasks reliant on quantum data, this work provides a unified perspective that highlights practical strategies, challenges, and future directions for achieving the ultimate goal of practical ``quantum advantage".}

{
The structure of this paper is organized as follows. Section \ref{secII} surveys quantum-simulation paradigms emphasizing strategies on NISQ devices. Section \ref{secIII} develops the variational toolkit: \ref{secIII}\textcolor{blue}{.A} frames the motivation and provides evidence-based contrasts with classical baselines; \ref{secIII}\textcolor{blue}{.B} discusses the physics-motivated ansätze and expressibility–trainability trade-offs; \ref{secIII}\textcolor{blue}{.C} presents the fundamentals of cost functions and gradients, while \ref{secIII}\textcolor{blue}{.D} reviews classical optimization under noise;\ref{secIII}\textcolor{blue}{.E} provides a critical view of BP, their origins, when they can be avoided, and the attendant trade-offs with classical simulability. Section \ref{secIV} presents implementations relevant to quantum simulation: \ref{secIV}\textcolor{blue}{.A} shows the ground/excited states (VQE/VQD) simulations, \ref{secIV}\textcolor{blue}{.B} highlights quantum dynamics (including open-system settings), \ref{secIV}\textcolor{blue}{.C} presents the finite-temperature preparation (free-energy–based methods), and, finally \ref{secIV}\textcolor{blue}{.D} describes the quantum machine-learning models for the simulation of quantum systems. Section \ref{secV} concludes the review by synthesizing the state of the field with a balanced outlook, outlining the research directions that align algorithm design with hardware realities.}

\section{Quantum Simulation}\label{secII}

Recent technological advances have significantly impacted the scientific scenario, enabling the development of new theories and approaches to various natural phenomena \cite{bachelard1984new, wolfram2002new}. Particularly, the development of increasingly powerful computers has enabled the approach to problems in physics that were considered to have no analytical solution, which has contributed to the advancement of numerical simulations \cite{simulations}. In fact, the local, reversible, and causal aspects of classical physics do not, in principle, present difficulties in terms of adaptation to classical computers \cite{feynman2018feynman}, allowing the application of temporal discretization techniques, as long as in extremely small intervals \cite{press2007numerical}.

However, the probabilistic nature of quantum mechanics imposes a barrier to conventional simulations, since a large number of computational resources are required to achieve a certain accuracy \cite{nielsen2010quantum}. For example, a quantum system composed of $N$ particles would require knowledge of the probabilities of these particles at distinct points $x_1, x_2, \ldots, x_N$ for a time $t$ in order to achieve its complete description. If the representation of these particles requires $S$ points in space, the number of possible configurations would be of the order of $S^N$ \cite{RevModPhys.86.153}. Thus, a time evolution operator for this system will require around $S^{N} \times S^N$ inputs. This exponential dependence on the number of particles makes the problem intractable for conventional computers when $N \gg 1$ (see Ex. \ref{ex:spin-1/2}).

\begin{examplebox}{Simulating spin-1/2 systems on classical computers}{spin-1/2} 

Spin-1/2 systems require $S = 2$ points in space to be represented, so the computer needs to store $2^N$ configurations. In the literature, $N = 40$ is considered an intractable problem for the vast majority of current computers, since it results in $2^{40} \approx 10^{12}$ configurations, while the time evolution operator (unitary matrix of size $2^N \times 2^N$) would require $2^{40} \times 2^{40} \approx 10^{24}$ elements to be represented \cite{feynman2018simulating, RevModPhys.86.153, doi:10.1126/science.273.5278.1073}. Converting to bits, this would result in about $\backsim 3.2 \times 10^{13}$ bits. The effect of exponential dependence is visible when doubling the number of particles, now resulting in $\backsim 3.8 \times 10^{25}$ bits. \newline

For comparison purposes, \citeauthor{doi:10.1126/science.1200970} (2011) \cite{doi:10.1126/science.1200970} estimated the amount of information stored by all of humanity in 2007, resulting in $\backsim 2.4 \times 10^{21}$ bits --- ten thousand times smaller than the simulation of a system with $N = 80$ particles! In 2024, this number was updated to $\backsim 1.9 \times 10^{24}$ bits \cite{bartley2023big}, still smaller than the amount of bits required for the simulation. 
\end{examplebox}

In this scenario, quantum computing emerges as a powerful tool to simulate physical systems using genuinely quantum architectures. In fact, the genesis of the proposition of quantum computers is marked by Richard Feynman's series of lectures on computer simulations in 1981 \cite{feynman2018simulating}, although other works have independently pointed out ways for this insertion \cite{PhysRevLett.48.1581}. This is because, being quantum systems themselves, the storage of large amounts of information could be encapsulated in a relatively small amount of physical space. This led Feynman to a foundational question that would fuel decades of research in quantum computing:

\begin{quote}
\textit{What is the universal quantum simulator? [...] If you had discrete quantum systems, what other quantum systems are exact imitators of it, and is there a class against which everything can be matched?} (\citeauthor{feynman2018simulating}, 2018) \cite{feynman2018simulating}.
\end{quote}

In 1985, ~\citeauthor{deutsch1985quantum} \cite{deutsch1985quantum} established a crucial theoretical foundation by proposing a quantum generalization of the Turing machine, thereby demonstrating the principle of universal quantum computation. Over a decade later, ~\citeauthor{doi:10.1126/science.273.5278.1073} (1996) \cite{doi:10.1126/science.273.5278.1073} provided a direct response to Feynman's conjecture by formulating a framework for a universal quantum simulator capable of emulating the dynamics of a quantum system with local interactions. For systems whose Hamiltonian terms commute, the time-evolution operator can be exactly decomposed into a sequence of unitary quantum operations. For the general case involving non-commuting observables, ~\citeauthor{doi:10.1126/science.273.5278.1073} (1996) \cite{doi:10.1126/science.273.5278.1073}discussed an algorithm leveraging the \textit{Trotter-Suzuki decomposition} to approximate the time-evolution operator as a product of exponentials of the individual Hamiltonian terms.

Within this framework, the first-order approximation is given by
\begin{equation} \label{eq:trotter}
U(t) = e^{-i H t / \hbar} \approx \left( \prod_{j=1}^{n} e^{-i H_j \Delta t} \right)^{t / \Delta t},
\end{equation}
where the total Hamiltonian $H$ is decomposed into $n$ local terms, $H = \sum_{j=1}^n H_j$, each acting on a limited subset of particles. The accuracy of this approximation is controlled by the number of steps $t / \Delta t$; a finer discretization (larger $n$) yields a more precise simulation at the expense of increased computational resources~\cite{kluber2023trotterization, kalos2012monte, PhysRevLett.106.170501}. Each exponential $e^{-i H_j \Delta t}$ must be further decomposed into a sequence of fundamental quantum gates. Consequently, the algorithmic precision is intrinsically linked to the circuit's gate complexity and depth~\cite{doi:10.1126/sciadv.aau8342}. This formalism establishes that any physical system governed by a well-defined Hamiltonian can, in principle, be simulated on a quantum computer, a property that encompasses the vast majority of quantum mechanical problems ~\cite{sakurai2020modern}.

These propositions played a fundamental role in the consolidation of the theory of quantum simulations and have been implemented in several contexts, such as quantum chemistry \cite{cao2019quantum}, quantum thermodynamics \cite{somhorst2023quantum}, and many-body systems \cite{cornish2024quantum,PhysRevLett.101.070503}. The practical application of these simulations promises to transform materials science, offering solutions to problems considered intractable by classical computing \cite{RevModPhys.86.153}. Furthermore, the field of quantum simulations was responsible for some of the first experimental demonstrations of quantum advantage, as evidenced in recent works \cite{doi:10.1126/science.abn7293, daley2022practical}.

\begin{figure}
    \centering
    \includegraphics[width=1\linewidth]{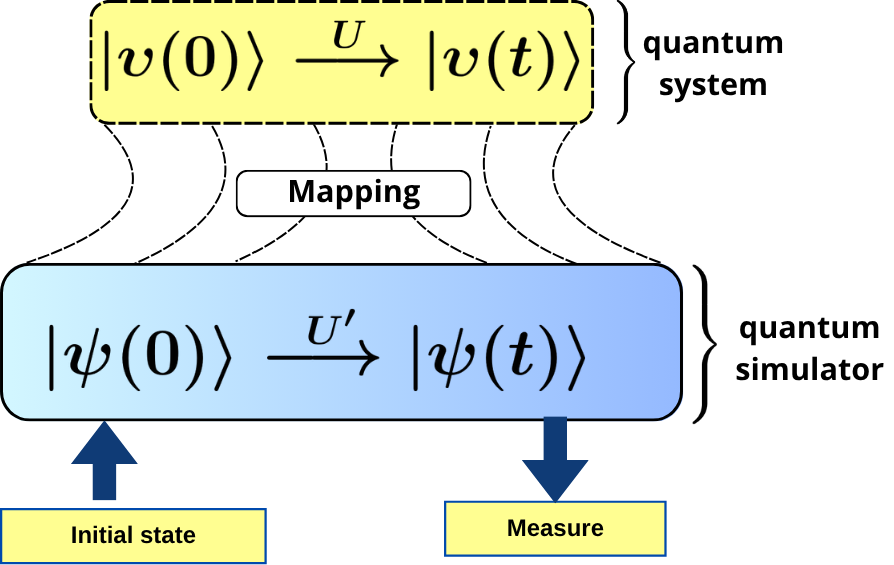}
    \caption{Schematic of digital and analog quantum simulations. First, an initial state \(\ket{\psi}\) is prepared, and then the desired evolution \(U\) is mapped onto an effective operator \(U'\) implementable in the simulator. After applying \(U'\), measurements of the system are performed to extract the physical quantities of interest associated with the simulated state \(\ket{\upsilon(t)}\).}
    \label{fig:model_sim}
\end{figure}

Formally, quantum simulations are divided into three main categories: inspired, analog, and digital. Inspired simulations use classical algorithms motivated by quantum structures to achieve greater computational efficiency, being especially promising for many-body problems \cite{PhysRevLett.106.170501,PhysRevA.79.032316}. In analog and digital simulations, the purpose is to reproduce the dynamics of a quantum system $\ket{\upsilon}$ in a dedicated quantum device. In this case, an initial state \(\ket{\psi}\) is prepared, a mapping of the desired time evolution \(U\) to an effective operator \(U'\) is implemented in the simulator, and, finally, the system is measured to extract the physical quantities of interest (see Fig.~\ref{fig:model_sim}).

In analog quantum simulations, a known quantum system is controlled to mimic the behavior of another system of interest, as shown in the diagram below:
\[
\begin{array}{ccc}
\text{Simulated (Sim)} & & \text{Physical (Phy)} \\
\ket{s} & \xrightarrow{\ \ \phi\ \ } & \ket{p} \\
\Big\downarrow \scriptstyle{U} & & \Big\downarrow \scriptstyle{V_T} \\
\ket{s(T)} & \xleftarrow{\ \ \phi^{-1}\ \ } & \ket{p_T}
\end{array}
\]
where the evolution from $\ket{s}$ to $\ket{S(T)}$ occurs through the time evolution operator $U$. However, note that this evolution occurs indirectly: the state $\ket{s}$ is taken to the state of the physical system $\ket{p}$ from the map $\phi$, which has its evolution mapped from the operator $V_T = \phi^{-1} U\phi$ \cite{PhysRevLett.82.5381}. At the end of the process, the mapped state $\ket{p_T}$ is then taken to the corresponding state of the simulator $\ket{S(T)}$ through the map $\phi ^{-1}$. This mechanism is useful when the goal is to perform specific quantum simulations on simple manipulation systems (see Ex. \ref{ex:AQS}).

\begin{examplebox}{Analog quantum simulation of the Bose–Hubbard model}{AQS}[htpb]
A controllable quantum system can be represented by a Hamiltonian describing a gas of bosonic atoms interacting in a periodic potential.
\begin{equation*}
H_{sim} = -J \sum_{i, j} \hat a_i^\dagger \hat a_j + \sum_i \varepsilon_i \hat n_i + \frac{1}{2} U \sum_i \hat n_i (\hat n_i - 1) \ ,
\end{equation*}
where \(\hat a_i^\dagger\) and \(\hat a_i\) are, respectively, the creation and annihilation operators for an atom at site \(i\), \(\hat n_i = \hat a_i^\dagger \hat a_i\) counts the number of atoms at that site, \(\varepsilon_i\) is the energy gap due to the external confinement potential, \(J\) quantifies the tunneling rate between neighboring sites, and \(U\) measures the interaction strength between atoms at the same site. This model formally coincides with the Bose–Hubbard Hamiltonian,
\begin{equation*}
    H = -J \sum_{\langle i,j\rangle} \hat b_i^\dagger \hat b_j
+ \tfrac{1}{2}U \sum_i \hat n_i(\hat n_i - 1)
- \mu \sum_i \hat n_i\,,
\end{equation*}
where \(\mu\) is the chemical potential and \(\hat b_i^{(\dagger)}\) obey the same site interpretation. The analog simulation with atoms in an optical lattice directly follows this Hamiltonian. In contrast, in systems such as Josephson junction arrays, the quantum phase model is used, in which the operators \(\hat a_i\) are rewritten in terms of the amplitude and phase of the superconducting order parameter in each circuit element, thus establishing the mapping \(H_{\mathrm{sys}}\leftrightarrow H_{\mathrm{sim}}\). For a more detailed description, see ref. \cite{PhysRevLett.76.4947}. 
\end{examplebox}

\begin{figure}
    \centering
    \includegraphics[width=0.8\linewidth]{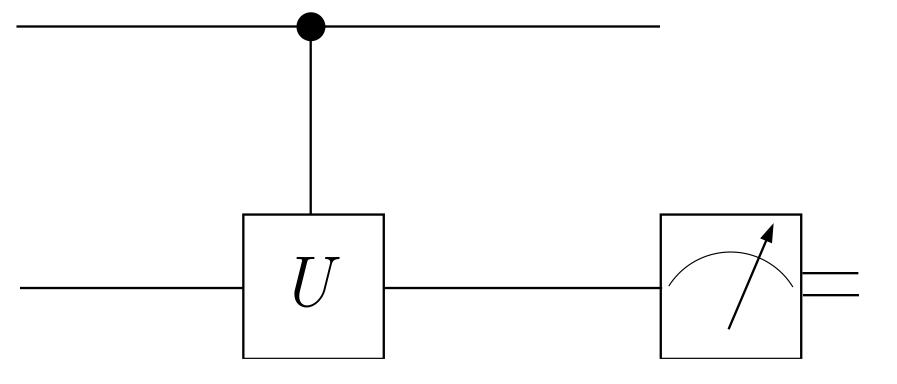}
    \caption{General representation of a quantum circuit. The circuit is read from left to right and represents a controlled operation: the top line is the control qubit and the bottom is the target qubit. When the control qubit is in the $\ket{1}$ state, the unitary operation $U$ is applied to the target qubit. Finally, the target qubit is measured in an appropriate basis (represented by the semicircular gauge).}
    \label{fig:circ}
\end{figure}

Digital quantum simulations employ the quantum circuit model (see Fig.~\ref{fig:circ}) to reproduce the evolution of physical systems \cite{nielsen2010quantum}. In their most basic form, an initial state \(\ket{\psi}\) is prepared, a sequence of carefully designed unitary gates that approximate the dynamics generated by a Hamiltonian of interest is applied, and finally, measurements are performed to extract the desired observables \cite{benenti2008quantum}, as demonstrated in Ex. \ref{ex:DQS}. Although they share the same goal of imitating a physical system in quantum hardware, digital simulations explicitly assume that such a system can be described by universal digital logic gates \cite{doi:10.1126/science.1121541}. This paradigm enables the construction of general simulation algorithms, in which local interactions in Hamiltonians can be implemented in a controlled and scalable manner \cite{doi:10.1126/science.273.5278.1073}.

\begin{examplebox}{Digital quantum simulation of the Lamor precession}{DQS}
The precession of a state $\ket{\psi}$ can be understood by applying a transverse magnetic field, represented by the Hamiltonian
\begin{equation}
H = \frac{\omega_0}{2} \sigma_Z
\end{equation}
where $\omega_0$ is the frequency of the field. The time evolution operator is then given by

\begin{align}
U = \exp{\left (- \frac{i}{\hbar}H_0 t \right )} = &\ \exp{\left (- \frac{i}{2\hbar}\omega_0 t \sigma_Z \right )} \\ = &\ \text{diag} \ [e^{- \frac{i}{2\hbar}\omega_0 \ t} \quad e^{\frac{i}{2\hbar}\omega_0 t}] \notag
\end{align}
which is implemented by applying the gate
\begin{equation}
Rz(\phi) = \text{diag} \ [e^{- \frac{i}{2}\phi} \quad e^{ \frac{i}{2}\phi}]
    \end{equation}
    decoding $\phi = \omega t$. For a more detailed description, see ref. \cite{alves2020simulating}. 
   
\end{examplebox}

Considering the many possibilities offered by quantum simulations, especially for problems intractable for classical computers, it is natural that much of the scientific community's attention is focused on the development of these approaches \cite{RevModPhys.86.153, https://doi.org/10.1002/wcms.70020}. However, although digital quantum simulations (DQS) present significant theoretical advantages, their practical application still faces significant limitations. These restrictions are largely related to the current state of quantum hardware technology, whose construction and control remain considerable technical challenges, as will be discussed in more detail in the following section.

\section{Variational Quantum Computing}\label{secIII}
\begin{figure*}
    \centering
    \includegraphics[width=0.9 \linewidth]{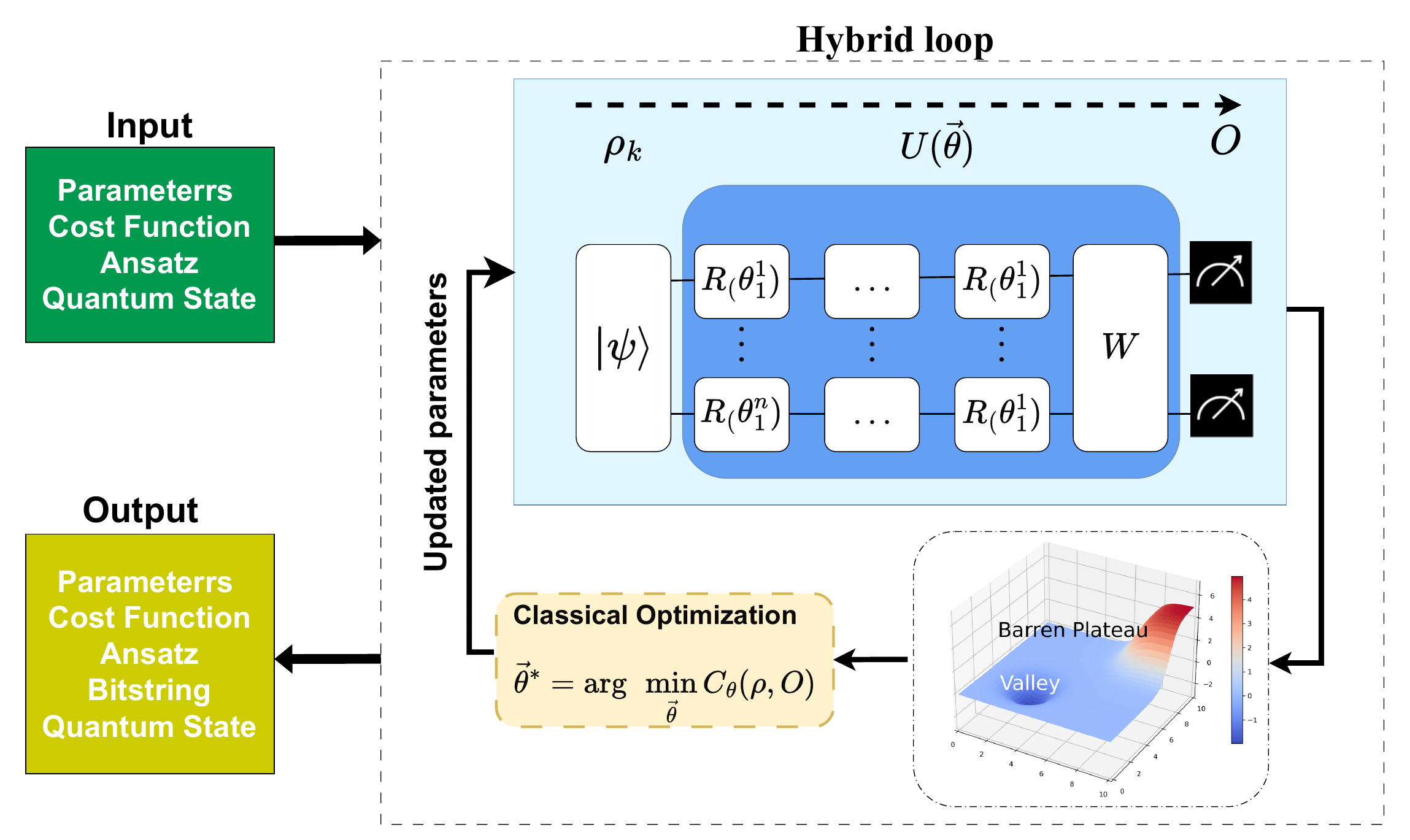}
    \caption{General representation of a variational quantum algorithm. Inputs: initial parameters ${\theta}_0$, cost function $C({\theta})$, parametrized ansatz $U({\theta})$, and the input state $\lvert\psi\rangle$. The ansatz is applied on an input state $\lvert\psi\rangle$ or training instances $\rho_k$, followed by an optional basis–change or post–processing block $W$ and measurements of observables $\{O_k\}$. The measured expectations define the cost function $C({\theta})=\sum_k f_k({\theta},\rho_k)$, which a classical optimizer minimizes by updating the initial parameters ${\theta}\!\to\!{\theta}^\ast$. The procedure outputs the optimized parameters ${\theta}^\ast$, minimized cost $C({\theta}^\ast)$, the optimized ansatz $U({\theta}^\ast)$, and the prepared state $\lvert\Psi({\theta}^\ast)\rangle$. The inset cost landscape (bottom right) illustrates the “hypersurface”, where variational training corresponds to searching for the global minimum under NISQ constraints.
}
    \label{fig:VQA}
\end{figure*}

{A standard classification in variational quantum computing delineates two primary paradigms: \textit{Variational Quantum Algorithms} (VQAs), which are problem-driven and typically target specific objectives such as ground-state preparation of physical systems \cite{cerezo2021variational, RevModPhys.94.015004}, and \textit{Quantum Machine Learning} (QML) models, which are data-driven and encompass diverse tasks including supervised learning and generative modeling \cite{Biamonte2017, cerezo2022challenges}. Despite these distinct operational contexts, where VQAs focus on solving well-defined physical problems and QML models on learning patterns from data, both paradigms share a fundamental variational structure comprising parameterized quantum circuits and classical optimization \cite{cerezo2022challenges}. This work consequently adopts a unified analytical framework, examining VQAs and QML models jointly to extract insights that transcend their particular applications.}

\subsection{Motivation}

Advances in quantum computing have allowed the first quantum simulation experiments to be carried out on real devices, reinforcing their potential \cite{PhysRevLett.82.5381, doi:10.1073/pnas.1801723115, Daley2022}. In 2016, IBM made the first quantum computer available in the cloud, marking the beginning of public access to quantum processors based on superconducting circuits \cite{IBM2016QuantumCloud}. Years later, other companies began offering similar services, exploring different physical platforms, such as architectures based on trapped ions, neutral atoms, and photonics, contributing to the expansion of the range of available technologies \cite{gebheim2021state, spinquanta2025top}.

However, current quantum computers still face significant challenges, such as high noise levels and qubit limitations, either due to restricted connectivity or decoherence. These characteristics define Noisy Intermediate-Scale Quantum (NISQ) devices, which operate with a limited number of qubits and without real-time error correction \cite{preskill2018quantum}. In contrast, the era of fault-tolerant quantum computing (FTQC) envisions systems that can operate reliably even in the presence of errors, due to the full implementation of error correction protocols \cite{PRESKILL1998, 548464, aharonov1997fault, PhysRevA.57.127}. 

In this context, any proposal for leveraging quantum computing during the NISQ era must account for the stringent limitations of contemporary hardware to yield practical outcomes. These constraints include a limited qubit count, the necessity to minimize the number of logic gates, and the construction of quantum circuits with minimal depth to mitigate error accumulation~\cite{preskill2018quantum, PRXQuantum.2.040335}. A central challenge in quantum simulation thus emerges: while the number of qubits required to represent a system scales linearly with its size, the number of logical operations needed to simulate its evolution scales super-linearly or exponentially. This disparity creates a fundamental bottleneck, as the computational overhead grows prohibitively quickly compared to the system's physical scaling.

In an exploratory study, Kassal \textit{et al.} (2008) simulated a quantum system composed of $N$ particles subjected to a paired Coulomb potential, using a total of $n(3N - 6) + 4m$ qubits, where $n$ is the number of basis levels used in the representation and $m$ is the number of ancillas required to achieve a given accuracy \cite{doi:10.1073/pnas.0808245105}. Although this number scales linearly with $N$, the number of steps required to evaluate the potential is on the order of $\mathcal{O}(N^2m^2)$, which implies an equivalent number of logic gates. Note that this number still follows a polynomial scale, maintaining the quantum advantage but remaining incompatible with current quantum computers \cite{preskill2018quantum}. A summary of these results is shown in figure 2 of reference \cite{doi:10.1073/pnas.0808245105}
, where the authors shows the qubit resources and elementary gates for the quantum simulation of particles interacting via the Coulomb potential, indicating the need for at least $100$ qubits and more than $200\ 000$ gates to overcome the limit to classical simulations. 


As a potential alternative to these limitations, VQAs emerge as a class of digital quantum algorithms with the promise of a quantum advantage in the NISQ era, using computational resources in a hybrid way --- The quantum architecture represents the quantum state information through a parameterized circuit, while the classical part of the algorithm is responsible for adjusting these parameters based on the minimization of a cost function $C$ \cite{cerezo2021variational}, as shown in the diagram in Fig. \ref{fig:VQA}. {This hybrid approach} leverage a powerful synergy between quantum and classical processors. The quantum device is tasked with executing a parameterized quantum circuit (PQC) that prepares a trial state and computes a cost function, a task often intractable for classical computers. The result of this quantum computation is then fed to a classical optimizer, which calculates a new, improved set of parameters for the quantum circuit. This iterative feedback loop continues, dynamically refining the quantum operation until the cost function converges to an optimal value, effectively training the quantum system to solve a specific problem.

{Unlike classical variational methods, VQA and QML approaches natively exploit the exponentially large Hilbert space \cite{cerezo2021variational, Biamonte2017}. While classical approaches require explicit representation of quantum states, demanding memory that grows exponentially with system size or facing severe sampling barriers. These limitations of the classical approaches become particularly relevant in systems affected by the sign problem, where the simulations face exponential sampling overhead. 
This situation is exemplified in Example ~\ref{ex:SignProblem}, which discusses a typical model applied to quantum many-body systems where the signal-to-noise ratio decays exponentially with system size and inverse temperature. As a result, the classical sampling error diverges, making simulations infeasible in physically relevant regimes.}

{In contrast, VQAs leverage parameterized circuits whose structure can incorporate problem-specific knowledge, often avoiding the overhead associated with classical state descriptions.
This shifts the computational burden from storing exponential state representations to optimizing a compact parameter vector and estimating expectation values through quantum measurements. These results illustrate the type of problems where VQAs may offer tangible advantages. As we will see in Sec. \ref{sec:BP}, the resulting scaling advantage is conditional: it requires an expressive yet trainable ansatz and must overcome challenges like measurement shot noise, device decoherence, and BP. Nevertheless, emerging benchmarks demonstrate regimes where VQAs achieve superior accuracy and more favorable scaling than leading approximate methods based on tensor networks and neural networks \cite{doi:10.1126/science.abn7293, daley2022practical, doi:10.1126/science.ado6285}.}

\begin{examplebox}{The sign problem in Quantum Monte Carlo}{SignProblem}
{Consider a classical estimator for the average of an observable:
\begin{equation*}
\langle O\rangle = \frac{1}{Z} \mathrm{Tr}[O \exp{(-\beta H)}]= \frac{\sum_{C} p_C\, O_C}{\sum_{C} p_C},
\end{equation*}
where  $C$ enumerates configurations and the (classical) weights $p_C\in\mathbb{R}$ (or $\mathbb{C}$) can be signed or phased. 
The standard approach to handle negative weights in fermionic systems is to sample using the bosonic system with absolute weights $|p(c)|$, while assigning the sign $s(c) \equiv \operatorname{sign} p(c)$ to the measured quantity:
\begin{equation*}
\langle A\rangle = \frac{\sum_c A(c)p(c)}{\sum_c p(c)} = \frac{\langle A s\rangle'}{\langle s\rangle'},
\end{equation*}
where $\langle \cdot \rangle'$ denotes sampling with respect to $|p(c)|$.
Although this enables Monte Carlo simulations, the errors grow exponentially with system size $N$ and inverse temperature $\beta$. The average sign $\langle s\rangle = Z/Z'$ is the ratio of fermionic and bosonic partition functions. Since partition functions are exponentials of free energies, this ratio becomes:
\begin{equation*}
\langle s\rangle = e^{-\beta N \Delta f},
\end{equation*}
where $\Delta f$ is the free energy density difference. Consequently, the relative error scales as:
\begin{equation*}
\frac{\Delta s}{\langle s\rangle} \sim \frac{e^{\beta N \Delta f}}{\sqrt{M}} \,,
\end{equation*}
where $M$ is the number of samples. Similarly, the error in measured observables grows exponentially, and the computation time required for a fixed accuracy scales exponentially with $N$ and $\beta$ (see \cite{PhysRevLett.94.170201} for detailed complexity analysis).}

\end{examplebox}

\subsection{Ansatz}

Similar to conventional variational algorithms, the efficiency of VQAs depends explicitly on the definition of the cost function. This is because it represents a hypersurface in which the solution of the desired problem lies (see Fig. \ref{fig:VQA}), so that the optimization process must lead to this solution with a certain accuracy \cite{combarro2023practical}. To this end, we need to represent the state considering the application of a parametrized operator

\begin{equation}
\ket{\Psi(\vec \theta)} = U(\vec \theta) \ket{\psi} \ ,
\end{equation}
where $U(\vec \theta)$ represents the quantum circuit parameterized in $\vec \theta = \{\theta_1, \ldots, \theta_J \}$, which is applied to the input state of the algorithm $\ket{\psi}$. This circuit is called an \textit{ansatz}, and can be represented as
\begin{equation}
    U(\vec \theta) = U_J(\\\theta_J) \ldots U_j(\theta_j) \ldots U_2(\theta_2)U_1(\theta_1) ,
\end{equation}
where
\begin{equation} \label{eq:U_j}
    U_j(\theta_j) = \prod_m e^{-i\theta_m H_m}W_m \ .
\end{equation}

The representation provided by Eq. \eqref{eq:U_j} allows to write the \textit{ansatz} in terms of the optimization parameters $\{\vec \theta\}$, unparameterized unitary $\{W\}$ and Hermitian operators $\{H\}$ \cite{cerezo2021variational, TILLY20221, yuan2019theory}. Note that the parameters $\vec \theta$ can, in theory, be freely controlled in the circuit, by changing the probability amplitudes of the state $\ket{\Psi (\vec \theta)}$, as demonstrated in Ex. \ref{ex:ansatz}. In this sense, they are able to encode the desired information of the quantum system, such as its temporal evolution or, simply, a specific amplitude \cite{cerezo2021variational}. Therefore, the choice of \textit{ansatz} defines the expressibility and trainability of the algorithm, since they delimit the search space covered during the optimization \cite{PRXQuantum.3.010313, TILLY20221}.

\begin{examplebox}{{One-qubit ansatz}}{ansatz} 
The solution sought by the optimization process in the Hilbert space of a $1$ qubit basically consists of traversing the Bloch sphere in search of the state of interest. The representation of any qubit in the Bloch sphere is given by
\begin{equation*}
\ket{\psi( \theta, \phi)} = U(\theta, \phi) \ket{\psi} = \cos(\theta/2) \ket{0} + e^{-i \phi}\sin(\theta/2) \ket{1}
\end{equation*}
which can be obtained by applying the gate $Ry(\theta)$ and $Rz(\phi)$. Therefore, the task of VQA is to find the optimal parameters $\{ \theta ^*, \phi^*\}$ that lead to the state of interest $\ket{\psi( \theta ^*, \phi^*)}$.
    
\end{examplebox}

In general, \textit{ansätze} can be divided into two main classes: \textit{Physical Motivated Ansatz} (PMA) and \textit{Hardware Heuristic Ansatz} (HHA) \cite{cao2019quantum}. As the name suggests, PMAs are built inspired by the physical problem of interest, and are commonly used in quantum simulations \cite{PhysRevResearch.3.023092, lee2018generalized, PRXQuantum.1.020319, park2024hamiltonian}. On the other hand, HHAs are built considering the architecture of the quantum computer on which the algorithm will be executed, aiming to optimize the connectivity of the qubits \cite{leone2022practical, kandala2017hardware}. Thus, both meet specific requirements and must be chosen according to the objective of the VQA. For a detailed description of the most well-known types of \textit{ansatz} in the literature, see \cite{Qin_2023, TILLY20221, cerezo2021variational, cao2019quantum}.

\subsection{Cost Function}

After applying the \textit{ansatz}, the state $\ket{\Psi(\vec \theta)}$ is measured considering the set of observables $\{O_k\}$ \cite{nielsen2010quantum}. The value of these measurements, together with the parameters, is then forwarded to the optimization process considering the cost function $C(\vec \theta)$. This process is repeated until the value of the parameters that minimizes the function is found, represented as $\vec \theta^* = \{\theta_1^*, \ldots, \theta_J^*\}$ (see Fig. \ref{fig:VQA}). Thus, VQA problems are reduced to optimization problems of the type 
\begin{equation}
\vec \theta ^* = \arg \min_{\vec{\theta}} C(\vec \theta) \ .
\end{equation}

In general, the cost function must depend explicitly on the circuit parameters. However, it is constructed implicitly from the quantum state $\rho_k$, which results from the application of the \textit{ansatz} $U(\vec{\theta})$ on the initial state $\ket{\psi_k}$, together with the measurement operators ${O_k}$. Thus, its general form is written as
\begin{equation}
C(\vec \theta) = f(\{\rho_k\}, \{O_k\}, \{U(\vec \theta)\}) \ ,
\end{equation}
where $f$ is the chosen cost function. In the context of VQAs, it is useful to define it as a function that depends explicitly on the trace of $\rho_k$ written in the basis of $U(\vec \theta)$, i.e., 
\begin{equation}
C(\vec \theta) = \sum_k f_k\left ( Tr[O_kU(\vec \theta) \rho_k U^{\dagger}(\vec \theta)]\right ) \ ,
\end{equation}
with $\{f_k\}$ being the set of functions used to define the cost function. This representation enables generalizations of VQAs from their variations in terms of the parameters of interest \cite{cerezo2021cost}, using the parameter shift-rule:
\begin{align} \label{leq:psr}
\frac{\partial C}{\partial \theta_l} = \sum_k \frac{1}{2 sin \alpha} \big ( Tr[O_k U^\dagger (\theta_+) \rho_k U(\theta_+)] \\ - Tr[O_k U^\dagger (\theta_-{)} \rho_k U(\theta_-))] \big ), \notag
\end{align}
where $\theta_{\pm} = \theta \pm \alpha e_l$ for an arbitrary number $\alpha \in \mathbb{R}$, where $e_l$ is a vector containing the value 1 in its first component and 0 in the others \cite{guerreschi2017practical, mitarai2018quantum, schuld2019evaluating}. Note that the accuracy of eq. \ref{leq:psr} is directly influenced by the term $1/2\sin{\alpha}$, which is minimized for $\alpha = \pi/2$. This definition allows to analytically evaluate the effect of gradient-based optimizers, which defines an important step in the progress of VQAs \cite{cerezo2021variational}. This is because choosing the right optimizer is essential for greater accuracy in convergence, since it will define how the search will be performed in the solution space \cite{10646539}. 

{\subsection{Classical optimization}}

The efficacy of a Variational Quantum Algorithm is fundamentally governed by the optimization of its parametrized quantum circuit. This process constitutes a classical optimization loop nested within the hybrid quantum-classical architecture. The classical optimizer's objective is to navigate a high-dimensional, non-convex parameter landscape to minimize a cost function \( C(\vec{\theta}) \), which is evaluated on the quantum processor. This task is particularly challenging due to the noisy nature of the quantum hardware (which introduces stochasticity into cost function evaluations), the presence of BP that exponentially suppress gradients, and the inherent computational overhead of each quantum evaluation \cite{cerezo2021variational}.

{There are various classical optimization methods available for minimizing cost functions in variational algorithms, such as Constrained Optimization BY Linear Approximation (COBYLA), Nelder-Mead, BFGS, and L-BFGS-B \cite{bressert2012scipy}. COBYLA and Nelder-Mead are derivative-free approaches, making them robust and straightforward to apply without requiring gradient calculations, but they may converge more slowly on smoother problems \cite{powell1994advances,cheng2024quantum,singh2023benchmarking,pellow2021comparison}. In contrast, BFGS and L-BFGS-B are gradient-based quasi-Newton methods that utilize information about the function's slope to determine efficient search directions, often achieving faster convergence rates \cite{Xie2019Analysis, Berahas2017A}. The choice between these methods depends on factors like problem differentiability, computational resources, and the trade-off between robustness and speed.}


{\subsection{Barren plateau: origins, scope, and mitigation.} \label{sec:BP}}

\begin{figure}
    \centering
    \includegraphics[width=0.45\linewidth]{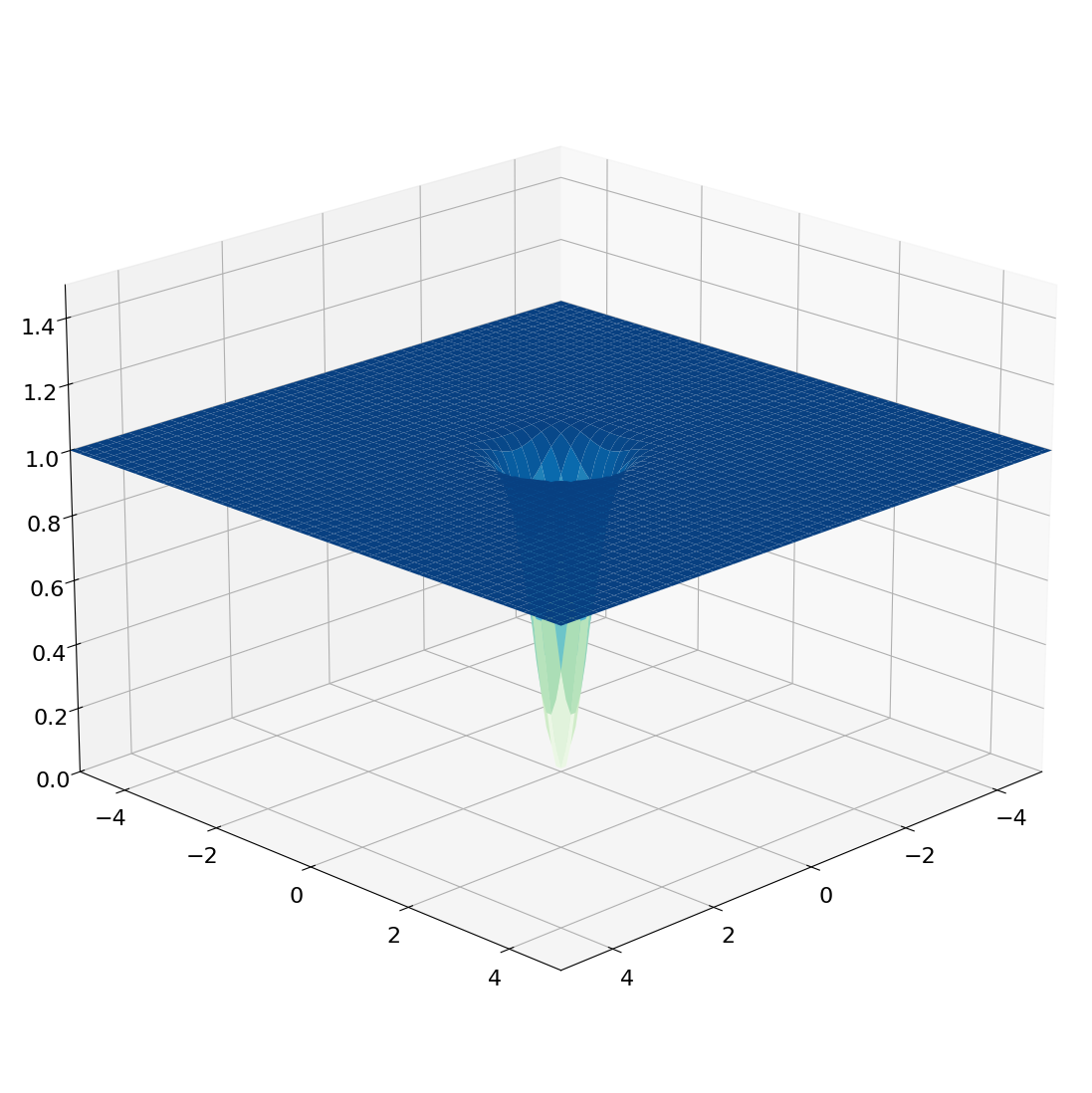}
    \includegraphics[width=0.45\linewidth]{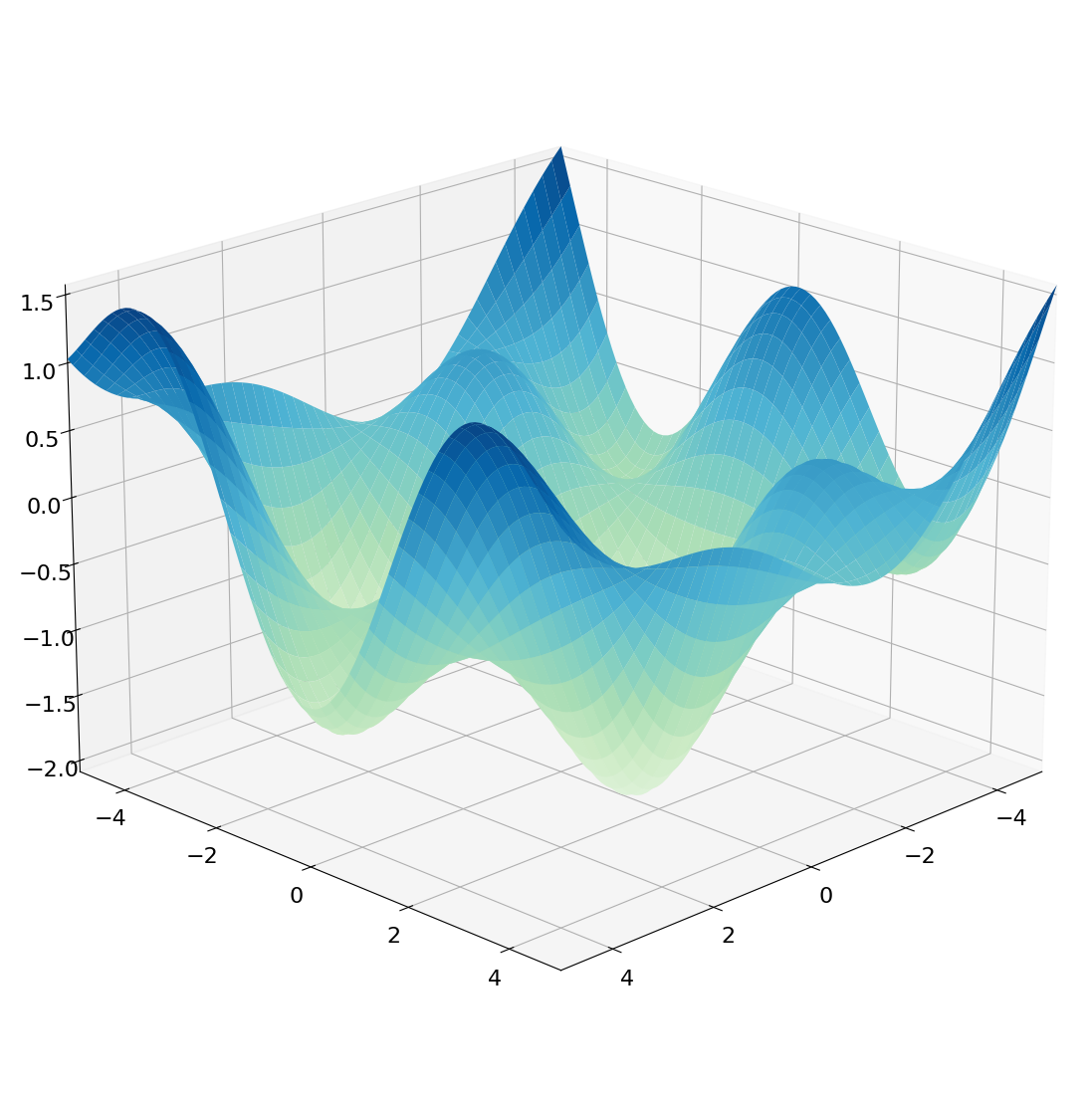}
    \caption{{The cost function landscapes with different behaviors. On the left, the mostly flat  cost function represents the vanishing of the gradient function, which makes the minimum inaccessible. On the right, the smooth loss function allows for the identification of the minimum point.}}
    \label{fig:BP_comparison}
\end{figure}

{The performance of variational quantum computing is constrained by fundamental challenges inherent to their optimization landscapes \cite{cerezo2021variational, RevModPhys.94.015004}. Principal among these are the limited expressibility of the parameterized ansatz, which may fail to capture the exact solution \cite{Sim2019, PhysRevLett.124.090504}; the proliferation of sub-optimal local minima that trap classical optimizers \cite{PhysRevLett.127.120502, Anschuetz2022, Fontana2022nontrivial}; and the emergence of BP, where the cost function's gradient vanishes exponentially with system size \cite{uvarov2021barren}. The BP problem is widely considered the most prohibitive obstacle to scaling these algorithms \cite{larocca2025barren}. This phenomenon is illustrated in Fig. \ref{fig:BP_comparison}: a landscape afflicted by a BP (left) exhibits a vanishing gradient, providing no directional preference for optimization. In contrast, a well-behaved, smooth landscape (right) features a well-defined gradient that guides the optimizer toward a minimum.}

{Intuitively, this implies a predominantly flat and featureless optimization landscape, where infinitesimal variations in the parameters $\bm{\theta}$ induce only exponentially small changes in either the loss $C_{\bm{\theta}}(\rho,O)$ or its partial derivatives $\partial_{\theta_i}C_{\bm{\theta}}(\rho,O)$ \cite{Fontana2024, Qi2023, mcclean2018barren}. This creates a fundamental challenge for gradient-based optimization, as identifying a descent direction requires resolving minute differences in the loss function, which is a task severely complicated by the statistical uncertainty inherent to quantum measurement. Since expectation values are estimated from a finite number of shots $N$, the resulting uncertainty scales as $\mathcal{O}(1/\sqrt{N})$, which can easily obscure exponentially small gradients and make determining a minimizing direction infeasible without an exponentially large number of measurements \cite{larocca2025barren}, as shown in Ex. \ref{ex:BP}.}

\begin{examplebox}{{Finite-size effects on sampling}}{BP} 
    {In practical implementations, the expectation value of the loss function, $C_{\bm{\theta}}(\rho,O) = \mathbb{E}_{\rho(\bm{\theta})}[O]$, is not computed exactly but is statistically estimated using a finite number of measurement shots $N$. We denote this finite-sample estimator as $\overline{C}_{\bm{\theta}}(\rho,O)$. When these $N$ measurement outcomes are independent, the sample variance (or standard error) of this estimator is given by:
    \begin{equation*}
    \mathrm{Var}_{s}[\overline{C}_{\bm{\theta}}(\rho,O)] = \frac{\mathrm{Var}_{\rho(\bm{\theta})}[O]}{N} = \frac{\mathbb{E}_{\rho(\bm{\theta})}[O^2] - (\mathbb{E}_{\rho(\bm{\theta})}[O])^2}{N}.
    \end{equation*}
    This variance quantifies the statistical uncertainty in the computed loss value, which arises from two factors: the intrinsic quantum mechanical variance of the observable $O$ in the state $\rho(\bm{\theta})$ (numerator), and the finite-sample error scaling as $1/N$ (denominator). Applying Chebyshev's inequality confirms that the statistical fluctuations of the estimator are of the order $\mathcal{O}(1/\sqrt{N})$ (see \cite{casella2024statistical}). A critical consequence of this relationship is that in regions of the parameter landscape where the true gradient is exponentially small, resolving a descent direction typically requires an exponentially large number of measurement shots $N$ to overcome this inherent statistical noise (see \cite{larocca2025barren} for a detailed analysis).}
\end{examplebox}

{The origin of BP can be attributed to several factors, such as circuit expressiveness \cite{PRXQuantum.3.010313, Larocca2022diagnosingbarren}, input states and measurements \cite{Ragone2024, diaz2023showcasingbarrenplateautheory} or even the noise in the quantum hardware \cite{Wang2021, StilckFrança2021}. Most recently, its origin has been discussed in terms of the so-called  \textit{curse of dimensionality} \cite{cerezo2023does}, which implies that the inner product between two exponentially large parametrized objects will be (on average) exponentially small and exhibit concentration. The statistical uncertainty in estimating such minuscule gradients makes it practically impossible to determine reliable descent directions without an exponentially large number of measurements \cite{larocca2025barren}. Therefore, what once was seen as an advantage (i.e., the exponentially large dimension of the Hilbert space) starts to seen as core of the BP phenomenon.}

{The BP problem has inpired extensive research into mitigation strategies for variational quantum computing \cite{cunningham2025investigating}. Promising approaches often involve constraining the quantum model's expressibility to avoid the concentration of measure in high-dimensional Hilbert spaces \cite{larocca2025barren}. This includes employing shallow circuits \cite{leone2022practical, cerezo2021cost, PhysRevLett.132.150603} or designing PQC with restricted dynamical Lie algebras \cite{PhysRevA.104.032401, PhysRevLett.129.070501}, which inherently limit the exploitable state space and prevent the exponential gradient decay. Similarly, variable-structure ansatzes \cite{Bilkis2023} and carefully crafted initialization strategies \cite{10.5555/3600270.3601622, PhysRevApplied.22.054005, park2024hardwareefficientansatzbarrenplateaus} have shown empirical success as practical heuristics. However, this approach introduces a critical trade-off: these same constraints often render the variational model efficiently classically simulable, thereby negating its potential for a quantum advantage \cite{cerezo2023does, bermejo2024quantumconvolutionalneuralnetworks, 10.1609/aaai.v37i6.25830, 3y65-f5w6}. In contrast, techniques such as merely switching classical optimizers \cite{Cerezo_2021, arrasmith2021effect,Thanasilp2023} or applying standard error mitigation \cite{StilckFrança2021, Wang2023, Quek2024} have been shown to offer no direct impact to avoid the BP problem itself.}

These aspects denote how the area of variational quantum computing studies still lacks a greater theoretical and empirical repertoire for further conclusions regarding the sought ``quantum advantage". The progress made in recent years has led to the development of numerous applications. This is because VQA and QML are characterized as a tool with an immense degree of diversification, whether for quantum simulations or optimization processes in general. In addition, works have already proven the universality of VQAs \cite{PhysRevA.103.L030401}, which makes their practicality applicable even in eras after NISQ \cite{zimboras2025myths}. {Accordingly, any promise of advantage is inherently problem- and regime-specific, hinging on ansatz structure, initialization, and depth budgets that avoid BP while preserving non-classical expressivity}

\section{Implementations} \label{secIV}

In this section, we present concrete implementations of VQAs for quantum simulation and organize them into four application pillars that will guide the discussion that follows: (i) ground- and excited-state problems, where these methods are employed to obtain molecular energies and spectra; (ii) quantum dynamics, covering both conservative (closed) evolutions and open-system scenarios, including variational treatments built upon Lindblad and quantum-state-diffusion formulations; (iii) finite-temperature physics, where variational thermalization and imaginary-time methods are used to prepare Gibbs states and study thermodynamic behavior; and (iv) a concise note in quantum neural networks when used in simulation-adjacent tasks. Table \textcolor{blue}{I} provides a compact summary of the principal algorithms referenced throughout this section, including their canonical use cases and representative cost functions, serving as a roadmap for the detailed subsections that follows.
\begin{figure*}
    \centering
    \includegraphics[width=\linewidth]{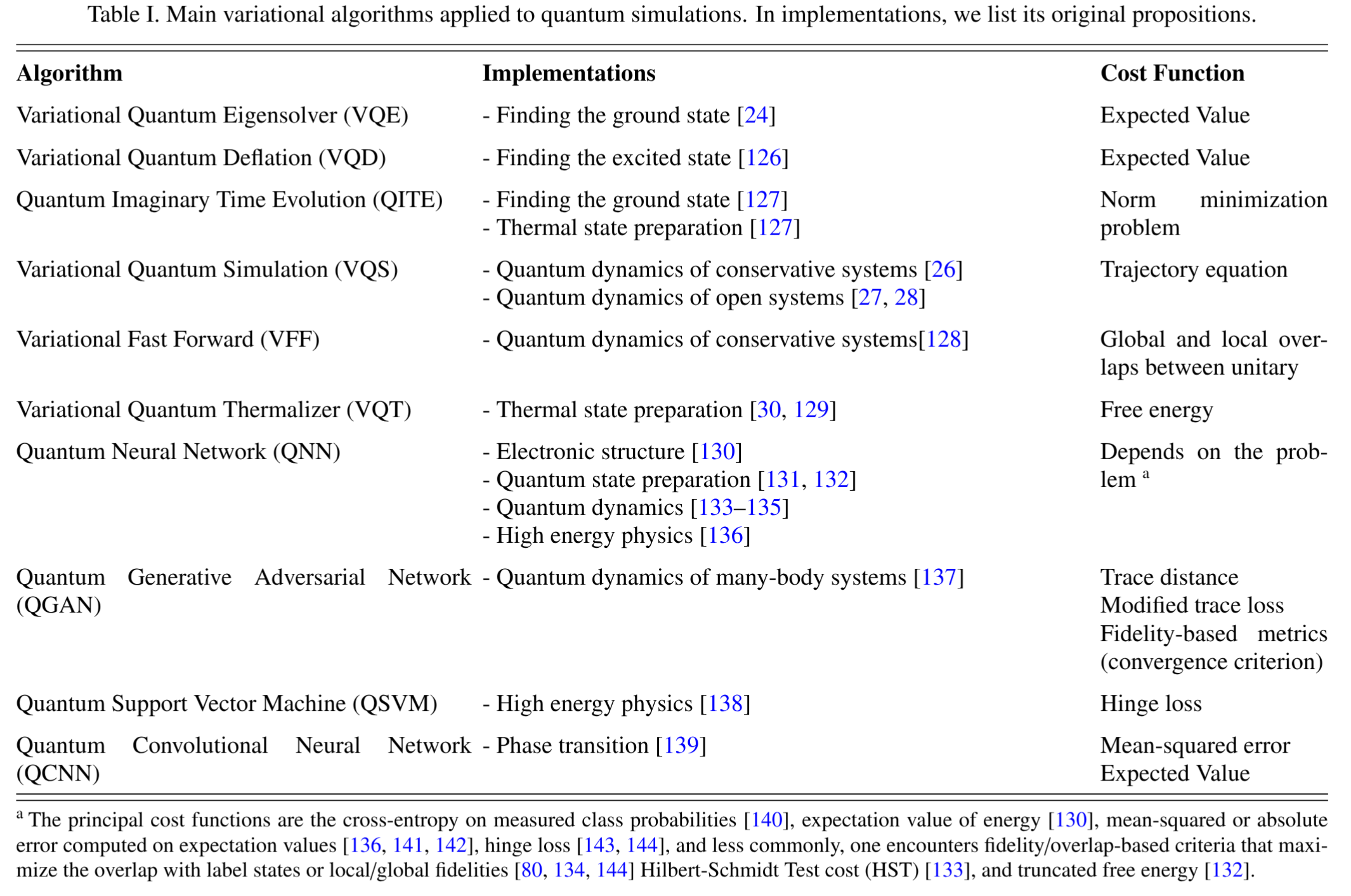}
    \label{tab:algoritmos_variacionais}
\end{figure*}

\subsection{Find the ground and excited states}

The task of finding the ground state of molecules is of fundamental importance, since it allows the determination of essential chemical properties, such as binding energy, electronic structure and reactivity \cite{kauzmann1970quantum, levine2014quantum, costain1958determination}. Computational chemistry methods, such as density functional theory, are widely used for this purpose, allowing results to be obtained with good accuracy and relatively low computational cost \cite{Parr1989}. However, for larger systems or those involving significant electronic correlation, the same problems regarding computational resources arise again (see Ex. \ref{ex:spin-1/2}).

In 1995, Kitaev developed a quantum algorithm capable of estimating the eigenvalues of Hermitian operators with logarithmic complexity using the Quantum Fourier transform (QFT) \cite{kitaev1995quantummeasurementsabelianstabilizer}. This proposal laid the foundation for the Quantum Phase Estimation (QPE) algorithm \cite{PhysRevLett.98.090501}. Years later, Lloyd and Abrams (1997; 1999) proved the exponential speedup of the algorithm by computing eigenvalues and eigenvectors of Hermitian operators \cite{PhysRevLett.79.2586, PhysRevLett.83.5162}. Inspired by these algorithms, Aspuru-Guzik \textit{et al.} (2005) proposed using these techniques to find the ground state of molecules \cite{doi:10.1126/science.1113479}, which set new limits for the field of quantum chemistry \cite{cao2019quantum}.

However, the approach faces similar problems to ref. \cite{doi:10.1073/pnas.0808245105}: while the number of qubits scales linearly, the number of gates needed to achieve a certain accuracy scales in $N^k$, where $k$ is a problem constant (for details see figure 2 in ref. \cite{doi:10.1073/pnas.0808245105}).
Considering these limitations, Peruzzo \textit{et al.} (2014) proposed the variational quantum eigensolver (VQE) for circuit depth reduction using classical hardware resources to optimize a cost function $C(\vec \theta)$ \cite{peruzzo2014variational}. In terms of structure, the VQE resembles the standard description of VQAs (Fig. \ref{fig:VQA}), but uses two components aimed at the ground state problem: the variational principle and specific classes of \textit{ans\"atze} \cite{TILLY20221, mcclean2016theory}. 

For this task, these structures are built to find the state $\ket{\Psi_{min}}$ that provides the lowest possible energy of the system, described by the Hamiltonian $H$, such that

\begin{equation}
E_{min} = \bra{\Psi_{min}} H \ket{\Psi_{min}} \,.
\end{equation}

This value is achieved during the optimization process by applying the variational theorem to ensure that

\begin{equation} \label{eq:variat}
E_{min} \leq E(\vec \theta) = \bra{\Psi(\vec \theta)} H \ket{\Psi(\vec \theta)} = Tr[O_k U^{\dagger}(\vec \theta )\rho U (\vec \theta)] \,,
\end{equation}
where $E_{min}$ corresponds to the mean of the observables $\langle O_k\rangle$ of interest. Eq. \eqref{eq:variat} denotes that the method consists of an approximation capable of ensuring that the optimization never exceeds the ground state limit \cite{peruzzo2014variational, TILLY20221, mcclean2016theory}.

Regarding \textit{ans\"atze}, HHAs always stand out when the issue is hardware optimization, with their main applications being concentrated in the \textit{Hardware Efficient Ansatz} (HEA) \cite{kandala2017hardware, doi:10.1021/acs.jctc.2c01057}. However, the emphasis of simulations in quantum chemistry is to use the characteristics of the molecule to ensure that eq. \eqref{eq:variat} approaches equality. In the literature, there is a certain consensus on the use of the quantum \textit{unitary coupled cluster} (qUCC) ansatz to achieve greater accuracy in these simulations, given its potential to exploit the symmetry of the molecule \cite{sennane2023calculating}.

This technique has its origins in classical quantum chemistry simulations, which define the wave function $\ket{\psi_{CC}}$ by applying the excitation operator $e^T$ to the reference state, usually represented by the Hartree-Fock state $\ket{HF}$, so that
\begin{equation}
\ket{\psi_{CC}} = e^T\ket{HF} \,,
\end{equation}
where $T=\sum_i T_i$, with $T_i$ representing an excitation of order $i$ \cite{RevModPhys.79.291, 10.1063/5.0161368}. However, the problem with this technique lies in the non-unitariness of the excitation operator, which becomes a problem when dealing with conservative quantum systems \cite{Romero_2019}. In this sense, we redefine $e^T \longrightarrow e^{T - T^\dagger}$ (UCC), thus ensuring the unitarity of the operator \cite{doi:10.1021/acs.jctc.8b01004}. Therefore,

\begin{equation}
\ket{\psi_{UCC}} = e^{T-T^{\dagger}}\ket{HF} \,,
\end{equation}
which fits into the class of universal quantum simulators described by ref. \cite{doi:10.1126/science.273.5278.1073}, and can be easily implemented in a quantum computer by applying logic gates (see Ex. \ref{ex:UCC}). 
Taking a similar approach, Quantum Imaginary Time Evolution (QITE) can be seen as a natural extension of VQE, in that both employ parameterized circuits and classical optimization to approximate the lowest energy state of a quantum system \cite{motta2020imaginary}. While VQE directly minimizes the expected energy \(\langle\psi(\vec\theta)|H|\psi(\vec\theta)\rangle\), QITE simulates the evolution in “imaginary time” \(e^{-H\tau}\) through McLachlan's variational principle \cite{McLachlan01011964}, adjusting \(\vec\theta\) to successively approximate the Gibbs state or the ground state from an arbitrary initial state \cite{mcardle2019variational}. Thus, QITE replaces the energy measurement steps of VQE with overlap measurements between nearby states in parameter space, but maintains the same hybrid quantum-classical structure and low-depth ansätze flexibility, making it equally suitable for implementation on NISQ devices \cite{nishi2021implementation, kamakari2022digital}.

There are also algorithms that use VQE as previous routines to find the excited state of molecules. In 2019, Higgott \textit{et al.} proposed the variational quantum deflation (VQD) algorithm \cite{Higgott2019variationalquantum}, using VQE to find the ground state and then applying the same principle by rewriting the Hamiltonian as
\begin{equation}
H'= H + a \ket{\psi_{min}}\bra{\psi_{min}} 
\end{equation}
ensuring that the energy of the state is $ E_{ex} \leq E(\vec \theta) = \bra{\Psi(\vec \theta)} H' \ket{\Psi(\vec \theta)}$ when the value of $a$ is greater than or equal to the distance between the first excited state and the ground state \cite{PhysRevA.75.012328}. This makes it possible to obtain an even more complete description of the molecule in terms of its stability and reactivity.

{In the vast majority of algorithms focused on quantum chemistry, the qUCC have been widely advocated for their ability to replicate the accuracy patterns of classical gold-standard methods like coupled cluster theory while maintaining physical interpretability\cite{sennane2023calculating, 10.1063/5.0019055, PhysRevA.95.020501}. In fact, PMA offer significant advantages in chemical accuracy and parameter efficiency compared to HHA, as they incorporate domain knowledge through physically motivated operator selection and initialization strategies \cite{Romero_2019, PhysRevX.6.031007}. However, recent rigorous analyses have revealed a more pessimistic outlook. Theoretical work by Mao \textit{et al.} (2024) demonstrates that even relaxed versions of unitary coupled cluster ansatzes exhibit exponential cost concentration when incorporating two-body excitations essential for chemical accuracy, with numerical evidence showing this BP behavior persists even at shallow depths \cite{Mao2024}. This fundamental trade-off between expressiveness and trainability suggests that the scalability advantages of PMA may be limited by the same BP phenomena that affect their hardware-efficient counterparts.}

\begin{examplebox}{Unitary Couple Cluster for $H_2$}{UCC}
First, the \emph{Hartree–Fock} reference state is chosen, which for $H_2$ corresponds to occupying the two lowest-energy orbitals (reduced basis), resulting in the state
\begin{equation}
\ket{\mathrm{HF}} =\ket{1100} \,,
\end{equation}
in four spin-orbitals.

Next, the cluster operator \(T\) is defined, restricted to single excitations (\textit{singles}), that is, to the movement of a single electron from an occupied orbital to a virtual orbital. In the case of $H_2$ there are exactly two possible excitations:
\[
\begin{aligned}
\theta_1 &: \quad a_{2}^{\dagger} a_{0}, \\
\theta_2 &: \quad a_{3}^{\dagger} a_{1}. \,,
\end{aligned}
\]
so that the excitation operator becomes
\begin{equation}
T(\theta_{1},\theta_{2})
= \theta_{1}\,a_{2}^{\dagger}a_{0}
+ \theta_{2}\,a_{3}^{\dagger}a_{1}\,,
\end{equation}
and the UCC ansatz is given by
\begin{equation}
\lvert \psi(\theta_{1},\theta_{2}) \rangle
= \exp\!\bigl(T - T^{\dagger}\bigr)\;\lvert \mathrm{HF} \rangle.
\end{equation}

To implement this operator on a quantum computer, one applies the Jordan–Wigner mapping (for example) to convert each fermionic term into a sum of products of Pauli matrices:
\begin{equation}
a_{p}^{\dagger}a_{q} - a_{q}^{\dagger}a_{p} \longrightarrow \frac{i}{2} (X_2Z_1Z_0Y_0 - Y_2 Z_1 Z_0X_0)
\end{equation}
which can ultimately be implemented on a quantum computer \cite{doi:10.1126/science.273.5278.1073}. 
\end{examplebox}

\subsection{Quantum dynamics}
\label{sec:quant_dyn}

The task of finding the ground state of molecules is characterized as a static problem, with a time-invariant solution. On the other hand, simulating the evolution of quantum systems requires understanding the dynamics of the problem, in order to characterize the behavior of the system in a time interval $t = [t_0, \ldots, T]$ \cite{RevModPhys.86.153}. In general, dynamic quantum simulations are classified into two categories: simulation of conservative quantum systems and simulation of open quantum systems. While the first considers closed systems \footnote{Isolated systems, which may or may not be influenced by external fields, as long as they do not interact with the particles generating the field \cite{sakurai2020modern}.}, the second considers dissipative dynamics from exchanges between the system and the environment \cite{breuer2002theory}. In the following subsections, the theoretical aspects of both categories are discussed.

\subsubsection{Conservative systems}

The evolution of a conservative quantum system is a topic already widely explored in the literature \cite{RevModPhys.86.153}. The Schr\"odinger equation for unitary evolutions provides a description of these systems from the relation
\begin{equation}
i \hbar \frac{\partial} {\partial t} \ket{\psi} = H \ket{\psi} \,,
\end{equation}
such that the evolution of the state $\ket{\psi}$ must guarantee that
\begin{equation}
\ket{\psi; t_0} \xrightarrow{\text{time evolution}} \ket{\psi(t)} \,,
\end{equation}
where both states are mapped by the time evolution operator
\begin{equation}
\mathbf{U}\ket{\psi; t_0} = \mathbf{e^{-iHt} }\ket{\psi; t_0} = \ket{\psi(t)} \ .
\end{equation}

As described previously, this operator can be implemented directly for the case of Hamiltonians with commuting operators (see Ex. \ref{ex:DQS}), or using the Trotter-Suziki formula \eqref{eq:trotter} for more general contexts. This procedure is known as trotterization, which maps the temporal evolution of the system by applying short steps, at the cost of greater circuit depth (see Ex. \ref{ex:trotter}).

\begin{examplebox}{{Trotterization}}{trotter}

Consider a qubit whose Hamiltonian can be written as the sum of two non-commuting terms,
\begin{equation}
H = \frac{\omega}{2}\,X + \frac{\Delta}{2}\,Z\,.
\end{equation}
The exact time evolution \(t\) is given by
\begin{equation}
U(t) = e^{-iHt},
\end{equation}
but we can approximate it by first-order trotterization in two steps, for example. To do this, we define the time step $\Delta t = t/2$ and write 
\begin{equation}
U(t) \approx \Bigl(e^{-\,iH_X\,\Delta t}\;e^{-\,iH_Z\,\Delta t}\Bigr)^{2}. \end{equation}
Each exponential corresponds to simple rotations:  
\begin{equation}
e^{-\,iH_X\,\Delta t}
= e^{-\,i(\omega/2)\,X\,\Delta t}
= R_X(\omega\,\Delta t),
\end{equation}
\begin{equation}
e^{-\,iH_Z\,\Delta t}
= e^{-\,i(\Delta/2)\,Z\,\Delta t}
= R_Z(\Delta\,\Delta t).
\end{equation}

Therefore, the resulting approximated circuit first applies a rotation in \(X\) by angle \(\omega\,\tfrac{t}{2}\) followed by a rotation in \(Z\) by angle \(\Delta\,\tfrac{t}{2}\), and repeating them a second time. The approximation error of this first-order scheme decreases as \(\mathcal{O}(t^2/4)\), and the error can be further reduced by increasing the number of steps or using higher-order formulas.
\end{examplebox}

Considering these aspects, VQAs emerge with the possibility of reducing the circuit depth, optimizing the result for different times. In 2017, Li and Benajamin proposed the variational quantum simulation (VQS) algorithm \cite{li2017efficient}, using McLachlan's variational principle to approximate the evolution of the simulated state $\ket{\Psi(\vec \theta)}$ to the state of interest, according to the equation below
\begin{equation}
\delta \left | \left | i \hbar\frac{\partial}{\partial t} - H \right | \right | \ket{\Psi(\theta)} = 0 \ .
\end{equation}

In sumary, the application of this variational principle can be represented by the calculation of the trajectory equation \cite{McLachlan01011964}, written as
\begin{equation} \label{eq:trajetory}
    \sum_j^l M_{i, j} \dot \theta_j = V_j \ ,
\end{equation}
where
\begin{equation} \label{eq:VQS_M}
    M_{i,j} = \operatorname{Re}\left(  \frac{\partial \bra{\Psi(\vec{\theta}(t))}}{\partial \theta_i} \frac{\partial \ket{\Psi(\vec{\theta}(t))}}{\partial \theta_j}  \right)
\end{equation}
and 
\begin{equation} \label{eq:VQS_V}
    V_j = -\operatorname{Im}\left( \frac{\partial \bra{\Psi(\vec{\theta}(t))}}{\partial \theta_j} H \ket{ \Psi(\vec{\Theta}(t))}    \right) \ . 
\end{equation}

In this case, both Eq. \eqref{eq:VQS_M} and \eqref{eq:VQS_V}, can be computed on quantum computers using the Hadamard transform \cite{yuan2019theory}, while Eq. \eqref{eq:trajetory} can be solved using the fourth-order Runge-Kutta method \cite{DORMAND198019}. In the original work, \citeauthor{li2017efficient} (2017) \cite{li2017efficient} demonstrated that this method can achieve the same efficiency as the trotterization technique using fewer layers to simulate the evolution of the system.

Taking a different approach, Cirstoiu \textit{et al.} proposed the variational fast forwarding (VFF) algorithm to study the temporal evolution of these systems \cite{cirstoiu2020variational}. In this case, instead of decomposing the propagator \(U(t)=e^{-iHt}\) into multiple Trotter steps, VFF employs a parameterized circuit $U(\vec \theta^*)$ that maps the state to the lowest energy eigenvectors and eigenvalues. This circuit can be acquired via optimization with VQE, resulting in the evolution
\begin{equation}
(e^{-iHt/n})^n \ket{\psi} \approx U(\vec \theta^*) (\tau(E, t))^n U^\dagger (\vec \theta ^*) 
\end{equation}
where
\begin{equation}
\tau(t) = \sum_j \exp{(-i E_j t)} \ket{\psi_j}\bra{\psi_j} \ .
\end{equation}

Note that the evolution is employed directly from $\tau(E, t)$, which does not depend on parameter optimization or circuit depth \cite{cerezo2021variational}. This makes it possible to reduce the computational cost in terms of iterations and/or implementation of logic gates.

\subsubsection{Open Systems}

Unlike closed quantum systems, open quantum systems deal with dissipation and absorption phenomena that make the evolution of the system's Hamiltonian a non-unitary process, and consequently, non-conservative \cite{breuer2002theory}. In this case, there are two primary theoretical tools for the description of open quantum systems: the representation in terms of Kraus operators and the Lindblad master equation. The first characterizes the dynamics through a completely positive and trace-preserving mapping, where the state of the system evolves according to 
\begin{equation}
\rho' = \sum_{k} K_{k}\,\rho\,K_{k}^\dagger\,,\quad \sum_{k}K_{k}^\dagger K_{k} = \mathbb{I}\,,
\end{equation}
with \(\{K_{k}\}\) being the Kraus operators that incorporate decoherence and dissipation effects in a discrete manner \cite{Wu_2007, vstelmachovivc2001dynamics}. The second describes the continuous evolution of the state via a Markovian differential equation\footnote{A Markovian process is a stochastic process that satisfies the Markov property, that is, whose conditional probability of future transition depends only on the present state and not on the previous history \cite{breuer2002theory}.}, given by 
\begin{equation} \label{eq:lind}
\frac{d\rho}{dt} \;=\; -\frac{i}{\hbar}[H,\rho] \;+\; \sum_{k}\Bigl(L_{k}\,\rho\,L_{k}^\dagger \;-\;\tfrac{1}{2}\{L_{k}^\dagger L_{k},\rho\}\Bigr)\,,
\end{equation}
where \(H\) is the Hamiltonian of the system and \(\{L_{k}\}\) are the jump operators that model the interaction with the environment in a continuous manner \cite{manzano2020short}.

In the context of quantum computing, there is a certain preference for characterizing open quantum systems from the Lindblad master equation, since it is possible to simulate the evolution of the system from the diffusivity of individual states, computing its average trajectory to reconstruct the density operator $\rho$ \cite{yuan2019theory, PhysRevA.101.012328, barison2021efficient}. In this sense, the quantum architecture deals with the first (conservative) term of Eq. \eqref{eq:lind}, while the second (dissipative) term is incorporated from {classical optimization techniques, which offers a potential advantage for open quantum systems (see Ex. \ref{ex:VQSContrast})}.

\begin{examplebox}{Variational quantum simulation advantage for open-systems}{VQSContrast}

{\textit{Classical limitations.} Exact classical methods, such as hierarchical equations of motion or tensor network approaches, face exponential memory growth with the number of bosonic modes and simulation time. This limits simulations to small environments or short-time dynamics. The exponential Hilbert space dimension makes studies of strong coupling or large baths intractable \cite{doi:10.1126/science.273.5278.1073}. \newline \newline
\textit{Variational quantum simulation advantage.} In VQS, the system is represented by a parameterized circuit state $|\psi(\vec{\theta}(t))\rangle$, with evolution governed by the time-dependent variational principle:
\begin{equation*}
M_{ij}\dot{\theta}_j = C_i,\qquad M_{ij}=\Re\langle\partial_i\psi|\partial_j\psi\rangle,\quad C_i=\Im\langle\partial_i\psi|H|\psi\rangle.
\end{equation*}
The quantum advantage arises from efficient state preparation and measurement: the cost scales polynomially with the number of parameters $p$ and inverse precision $1/\varepsilon$, independent of the Hilbert space dimension \cite{yuan2019theory, PhysRevLett.125.010501}. For an ansatz with $p = \operatorname{poly}(N)$, the per-time-step cost is $\operatorname{poly}(N, t, 1/\varepsilon)$, enabling simulations of large baths and long-time dynamics beyond classical reach \cite{PRXQuantum.3.010320}.}
\end{examplebox}

Following this approach, Endo \textit{et al.} proposed an expansion of VQS considering the description of the dynamics of open systems through the stochastic Schr\"odinger equation, whose evolution can be considered as an average of wave functions that undergo a continuous measurement induced by the environment \cite{PhysRevLett.125.010501}. To this end, the algorithm parameters take into account the state non-normalization factor $\alpha$, resulting in
\begin{equation} \label{eq:VQS-QSD}
\ket{\Psi(\vec \Theta)} = \alpha U(\vec \theta) \ket{\psi} \ ,
\end{equation}
so the parameters forwarded to Eq. \eqref{eq:trajetory} are redefined as $\vec \Theta = \{\alpha, \vec \theta \}$. With this, it is possible to define the effective Hamiltonian of the system as
\begin{equation} \label{eq:VQS-H}
H_{\mathrm{eff}} = -iH - \frac{1}{2}\sum_i J_k( L_i^\dagger L_i + \langle L^\dagger L_i \rangle ) ,
\end{equation}
which can be simulated via McLachlan's variational principle \cite{McLachlan01011964} in a similar way to the previous proposition. However, note that now the system suffers decohesion due to dissipative dynamics of the interaction with the environment, characterized in the non-normalization factor $\alpha$ of the state $\ket{\Psi(\vec \Theta)}$, according to Eq. \eqref{eq:VQS-QSD}.

Complementing this approach, Luo \textit{et al.} (2024) proposed the variational quantum simulation via quantum state diffusion algorithm (VQS-QSD) \cite{doi:10.1021/acs.jpclett.4c00576} by replacing the diffusivity term $\langle L^\dagger L \rangle$ in Eq. \eqref{eq:VQS-H} by a Gaussian noise \(z := \{z_k(t)\}\), referring to individual trajectories of the state in the environment. As a result, the stochastic evolution of the system is described as
\begin{equation} \label{eq:Heff}
\frac{\partial \ket{\psi_t}}{\partial t} = \left ( -iH - \frac{1}{2}\sum_k J_k L^{\dagger} L + \sum_k z_k^*(t) L_k \right ) \ket{\psi_k} 
\end{equation}
\begin{equation*}
    = -iH_{eff} \ket{\psi_t} \ ,
\end{equation*}
which again can be simulated using Mclaunch's variational principle \eqref{eq:trajetory}.

In addition to the aforementioned methodologies, a diverse family of variational algorithms has been developed, adhering to the core principles of hybrid quantum-classical optimization \cite{PhysRevA.101.012328, su2020quantum, Chen2024adaptivevariational}. These methods universally employ parameterized quantum circuits to approximate both dynamical evolution and target quantum states. A fundamental characteristic of this approach is its inherent capacity to describe mixed states, where \( \text{Tr}[\rho^2] < 1 \), making it particularly well-suited for simulating open quantum systems and finite-temperature ensembles. Consequently, the variational quantum simulation (VQS) framework has been successfully deployed across a wide spectrum of applications. This includes the study of strongly correlated electron dynamics in lattice models \cite{Kokail2019}, the investigation of electronic properties in periodic materials \cite{PhysRevResearch.4.013052}, and the simulation of nonnative dynamics \cite{galvão2025variationalquantumsimulationnonadditive}.

\subsection{Thermal state preparation}

Thermal state preparation is key to expanding the scope of quantum simulations to problems in quantum thermodynamics and nonequilibrium dynamics, as it allows to correctly model the statistics of these systems \cite{chen2023quantumthermalstatepreparation}. In NISQ devices, VQAs have shown particular promise for this task, by replacing long-time adiabatic heating protocols with shallow parameterized circuits that minimize the free energy or other thermodynamic metrics \cite{PhysRevLett.123.220502, PhysRevA.101.012328, verdon2019quantumhamiltonianbasedmodelsvariational}.

In 2019, Wu and Hsieh \textit{et al.} introduced a variational algorithm to prepare entangled Gibbs states between a system $A$ and ancillas, represented by state $B$ \cite{PhysRevLett.123.220502}. The protocol alternates evolutions under the Hamiltonian of interest \(H\) and under an entanglement Hamiltonian $H_{AB}$, which acts simultaneously on $A$ and $B$. For reduced-dimensional systems, it is possible to optimize the entire circuit on a classical computer and execute it in sequence in the quantum simulator. Thus, in a hybrid quantum-classical scheme, the free energy of the subsystem $A$ is defined as the cost function: 
\begin{equation}
F \;=\; Tr\bigl[\,\rho_A H \bigr] \;-\; T\,S(\rho_A)\,,
\end{equation}
where \(\rho_A\) is the reduced density of \(A\), $S(\rho_A) = -Tr\bigl[\rho_A\ln\rho_A\bigr]$ is the von Neumann entropy, and \(T\) is the effective temperature. In this case, minimization of $F$ ensures convergence to the Gibbs state of \(A\) \cite{PhysRevLett.123.220502} .

Inspired by this method, \cite{verdon2019quantumhamiltonianbasedmodelsvariational} proposed the Variational Quantum Thermalizer (VQT), which follows a similar protocol to the previous ones, considering the preparation of the thermal state $\rho(\vec \theta, \vec \phi)$ from the cost function defined from the free energy
\begin{equation} \label{eq:cost_VQT}
C(\vec \theta, \vec \phi) = \beta F((\vec \theta, \vec \phi))= \beta Tr[\rho(\vec \theta, \vec \phi) H] - S(\rho(\vec \theta, \vec \phi)) \,,
\end{equation}
where $\{\vec \theta, \vec \phi\}$ are parameters to be optimized in the circuit. Note also that the variational principle recovers the behavior of the VQE for zero temperature, since
\begin{equation}
C(\vec \theta, \vec \phi) \xrightarrow{\beta \rightarrow \infty} Tr[O_k \rho(\vec \theta, \vec \phi)] \,.
\end{equation}

In 2021, Sagastizabal \textit{et al.} demonstrated this technique experimentally using a gate-based quantum processor \cite{sagastizabal2021variational}. More recently, \citeauthor{PhysRevA.110.012445} (2024) \cite{PhysRevA.110.012445} also demonstrated in practice the faithful generation of Gibbs states for transverse-field and Heisenberg magnet models, both in state-vector simulations and on real \textit{hardware}, paving the way for applications in condensed matter physics.

In summary, variational quantum algorithms provide a powerful and flexible framework for the preparation of thermal states on near-term quantum devices. By minimizing the free energy of a parameterized ansatz, these hybrid quantum-classical methods circumvent the need for deep, coherent quantum circuits required by purely unitary approaches, making them particularly suitable for the NISQ era. The versatility of this methodology is evidenced by its growing number of applications, which extend beyond fundamental state preparation to include the modeling of quantum thermodynamic processes \cite{Silva_2024}. Notably, this algorithm has been successfully applied to problems such as work extraction \cite{6bm48ckl, PhysRevResearch.6.013038, PhysRevA.110.012443}. These applications underscore the potential of variational thermal state preparation to open new avenues for exploring finite-temperature quantum phenomena and quantum thermodynamics in experimentally relevant settings.

\subsection{Quantum Machine Learning}

Quantum Machine Learning (QML) compounds a set of methods and algorithms that aim to achieve computational advantages. Based on the pattern recognition capabilities of classical machine learning (ML), QML also utilizes quantum computing to achieve better results \cite{Biamonte2017}. Many QML algorithms are direct counterparts to well-established machine learning algorithms, the main difference being that they are now implemented in quantum circuits. {Similar to classical ML, the vast majority of QML algorithms focus on using classical data \cite{cerezo2022challenges}, so that the} many of the aforementioned algorithms fall beyond the scope of this work. In this section, the main application of QML presented is through the use of neural networks. 

Neural networks are exemplary ML models characterized by the application of layers with adjustable parameters. Originally proposed to {emulate} biological neural activity, the neural networks were rapidly adapted for applications in quantum computing \cite{schuld2014quest}. The quantum counterpart of classical neural networks, Quantum Neural Networks (QNN), have been explored in diverse areas and, {because of the variational approach, share a common structure with VQA \cite{jeswal2019recent}.} 


{
Quantum Machine Leaning, in general, is a versatile new area with multiple definitions and possible uses. Cerezo et al.(2022) identify four key applications for using QML: Quantum simulation, enhanced quantum computing, quantum machine perception, and classical data analysis \cite{cerezo2022challenges}. In this regard, the authors also predict the relationship between QML and quantum advantage in terms of sample or time complexity for quantum-mechanical processes. Even though there are only a few studies in the area, the range of applications for quantum simulation varies from chemistry to many-body dynamics. It is also expected that, in the fault-tolerant era of quantum computing, QML will play a central role in learning directly from quantum data produced by accurate quantum simulations \cite{cerezo2022challenges}.
}

{The extensive list of QML applications presented in ref. \cite{sajjan2022quantum} highlights the diversity of the area. On the other hand, the applications of QML in quantum simulation (the main focus of this paper) bring advantages not only in physics but also for material science and chemistry. Among the mentioned applications, the preparation of the quantum Gibbs state and Hamiltonians stands out due to the use of a variational quantum circuit with a Taylor series \cite{wang2021variational}.
The hybrid algorithm proposed by  Xia and Kais  \cite{xia2018quantum}, uses a restricted Boltzmann machine to calculate the energy levels of a quantum system. Using ($n + m$) qubits, where $n$ is the number of visible layers in the Boltzmann machine, while $m$ represents the hidden layers composed of auxiliary qubits, the simulations can be performed requiring only 13 qubits. The algorithm has a satisfactory performance (under the metrics of the study) in calculating the electronic structure of small molecular systems, opening up the perspective of material discovery through QML \cite{xia2018quantum}.}

{In addition,  Guan et al. review a series of QML applications in high-energy physics \cite{guan2021quantum}. Although all the applications presented being based on classical data, quantum simulation is presented as a future tool for simulating HEP, while QML can be used to analyze the obtained results. Another approach would be employing quantum objects, extracted from HEP experiments, directly in QML \cite{guan2021quantum}. The kernel QML algorithm for classification, Quantum Support Vector Machine (QSVM), was utilized by \citeauthor{wu2021application} in the analysis of Higgs boson production \cite{wu2021application}. Tests were performed on both quantum simulators and hardware \cite{wu2021application}. In this regard, the forthcoming work \cite{kan2026machine} discusses the employment of a quantum-mechanical harmonic oscillator in QML for the simulation of bosonic quantum states.}

{
Beyond the QML applications in quantum simulation discussed, there is also a possibility for quantum dynamical simulation \cite{gibbs2024dynamical}. A QNN is trained as a part of the Resource-Efficient Fast-Forwarding algorithm proposed by \citeauthor{gibbs2024dynamical}. When simulating the XY spin models, the algorithm achieved a high fidelity, but with times significantly longer compared to Trotterization \cite{gibbs2024dynamical}. QNNs have also been used in quantum simulation to study many-body quantum systems \cite{gardas2018quantum,long2024quantum,baul2025quantum,kim2024hamiltonian}. In ref. \cite{gardas2018quantum}, a variational Monte Carlo neural network is utilized in a D-Wave quantum sampler, while in \cite{long2024quantum}, the dissipative dynamics of many-body open quantum systems are simulated. Additionally, Baul et al. use a Quantum Convolutional Neural Network as a classifier to identify the phase transition region in the Hubbard Model \cite{baul2025quantum}. }

{Ref. \cite{kim2024hamiltonian} proposed a Hamiltonian Quantum Generative Adversarial Network (QGAN) using Quantum Optimal Control to learn and generate many-body quantum states. The QGAN framework, introduced by Goodfellow et al. \cite{goodfellow2014generative}, involves a generator that creates data resembling the original distribution and a discriminator that identifies real versus synthetic samples, with both improving through adversarial training. This concept extends to hybrid quantum-classical models, such as in \cite{romero2021variational}, where a quantum generator pairs with a classical or quantum discriminator, demonstrating the adaptation of classical machine learning for quantum simulation.}

{Therefore, while QML remains an emerging field, it has demonstrated remarkable potential for quantum simulation through variational quantum circuits. These approaches have shown applicability to preparation of quantum states \cite{kan2026machine,wang2021variational}, high energy physics \cite{guan2021quantum}, electronic structure \cite{xia2018quantum}, phase transition and quantum dynamics  \cite{gibbs2024dynamical,gardas2018quantum, long2024quantum, kim2024hamiltonian}. However, recent studies highlight persistent challenges including barren plateaus and classical simulability that must be addressed for scalable applications \cite{larocca2025barren, cerezo2023does}. Collectively, these advances suggest QML is expanding the frontiers of computational physics, materials science, and drug discovery while paving the way for practical quantum simulation across diverse scientific domains.}

\section{Concluding remarks and Outlooks} \label{secV}

Variational quantum simulations have demonstrated significant flexibility and robustness in solving complex problems on NISQ devices. Their effectiveness stems from the use of shallow parameterized circuits, which mitigate the detrimental effects of decoherence and noise by limiting circuit depth. This approach facilitates a tunable trade-off between computational precision and the hybrid quantum-classical resource cost. With well-designed strategies, VQAs have achieved competitive fidelities in diverse applications, including quantum chemistry, quantum thermodynamics, and system dynamics. Furthermore, they serve as a vital testbed for novel hardware architectures and calibration protocols, providing critical insights for the co-design of future quantum algorithms and technologies.

\begin{figure*}
    \centering
    \includegraphics[width=0.8\linewidth]{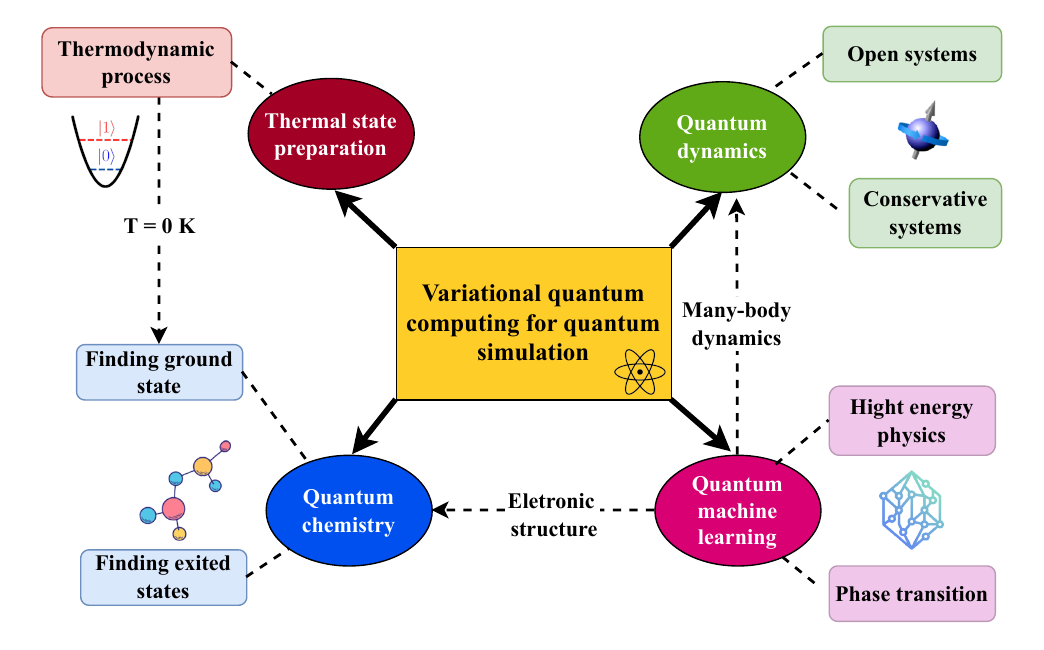}
    \caption{{Opportunities in variational quantum computing for quantum simulation. Variational quantum computing offers a versatile framework for quantum simulation, with multiple algorithms demonstrating diverse applications across quantum system characterization. The VQE enables ground state determination \cite{peruzzo2014variational}, while its extension VQD targets excited states \cite{Higgott2019variationalquantum}. For dynamical simulations, VQS handles both conservative \cite{li2017efficient} and open quantum systems \cite{PhysRevLett.125.010501, doi:10.1021/acs.jpclett.4c00576}, complemented by VFF for long-time evolution \cite{cirstoiu2020variational}. Thermal state preparation is addressed through both QITE \cite{motta2020imaginary} and VQT approaches \cite{verdon2019quantumhamiltonianbasedmodelsvariational, Selisko_2024}. Quantum machine learning further expands these capabilities, with QNNs applied to electronic structure \cite{xia2018quantum}, state preparation \cite{kan2026machine,wang2021variational}, dynamics \cite{gibbs2024dynamical,gardas2018quantum, long2024quantum}, and high-energy physics \cite{guan2021quantum}. Additional QML architectures include QGANs for many-body system dynamics \cite{kim2024hamiltonian}, QSVMs for classification tasks in high-energy physics \cite{wu2021application}, and QCNNs for phase transition detection \cite{baul2025quantum}. This diverse algorithmic landscape, employing cost functions ranging from energy expectation values to fidelity-based metrics, demonstrates the rich potential of variational methods for advancing quantum simulation across multiple scientific domains.}}
    \label{fig:VQC}
\end{figure*}

In this work, these aspects were explored across various quantum simulation contexts, encompassing the determination of molecular ground states, the preparation of thermal states, and the dynamics of quantum systems, {as illustrated in Fig. \ref{fig:VQC}}. Each application, focusing on a distinct domain, necessitates the development of tailored algorithms. Consequently, specific VQAs were formulated with cost functions meticulously aligned to the problem's physical characteristics and objectives, guiding the optimization towards the most relevant regions of the state space. The ansatz structure was carefully selected to efficiently represent the families of states of interest, ensuring a balance between expressibility and computational efficiency. This process was complemented by optimization strategies chosen for their compatibility with the ansatz, thereby promoting more effective convergence and enhancing the probability of successfully finding the desired solution.

This exploration reveals three fundamental challenges within current variational quantum simulation frameworks:
\begin{itemize}
    \item Achieving scalability and efficiency in modeling higher-dimensional quantum systems.
    \item Incorporating more complex temporal dynamics, including non-unitary interactions and system-environment correlation effects.
    \item Developing theoretical tools to generalize and unify variational approaches, thereby extending their scope and robustness.
\end{itemize}

{In principle, the primary obstacle to achieving practical near-term quantum utility remains the BP phenomenon, which necessitates problem classes where avoidance conditions (structured ansätze, shallow circuits, and informed initialization) can be satisfied without rendering the problem classically simulable. This challenge echoes historical developments in classical machine learning, where neural networks were once considered impractical until seminal work by Krizhevsky, Sutskever and the Nobel Prize laureate in Physics G. Hinton demonstrated their revival through architectural and training innovations for high-resolution images classification \cite{NIPS2012_c399862d}. Similarly, while BP analyses help identify fundamental limitations in scaling variational quantum models, they also guide research toward more promising avenues. By leveraging these insights and developing novel strategies including problem-inspired ansätze and noise-resilient optimization techniques, the field may ultimately realize the potential of variational quantum computing for quantum simulation, mirroring the transformative trajectory that reshaped classical deep learning.}

Therefore, a central focus for future work is to address these limitations by developing applications for systems closer to demonstrating a genuine quantum advantage. The field of variational quantum simulation remains a highly fertile ground for research. It is anticipated that this work will contribute to the advancement of the area by discussing and presenting practical problems across different contexts, thereby elucidating both the opportunities and the challenges that define the current scientific landscape.

\begin{acknowledgments}

C. Cruz {,} Lucas Q. Galvão {and Anna Beatriz M. de Souza} thank the Fundação de Amparo à Pesquisa do Estado da Bahia - FAPESB for its financial support (grant numbers APP0041/2023 {,} PPP0006/2024 {and BOL2154/2025}). This work has been funded by the project ``Master's and PhD in Quantum Technologies - QIN-FCRH-2025-5-1-1" and iNOVATeQ Lato Senso Especialização em Computação Quântica - Pesquisador, both supported by QuIIN - Quantum Industrial Innovation, EMBRAPII CIMATEC Competence Center in Quantum Technologies, with financial resources from the PPI IoT/Manufatura 4.0 of the MCTI grant number 053/2023, signed with EMBRAPII. Also, this work received partial financial support from CNPq (Grant Numbers: $305096/2022-2$ MAM).

\end{acknowledgments}

\section*{Funding}

This work has been partially funded by:  Fundação de Amparo à Pesquisa do Estado da Bahia - FAPESB (grant numbers APP0041/2023, PPP0006/2024) and BOL2154/2025; the project ``Master's and PhD in Quantum Technologies - QIN-FCRH-2025-5-1-1" and iNOVATeQ Lato Senso Especialização em Computação Quântica - Pesquisador, both supported by QuIIN - Quantum Industrial Innovation, EMBRAPII CIMATEC Competence Center in Quantum Technologies, with financial resources from the PPI IoT/Manufatura 4.0 of the MCTI grant number 053/2023, signed with EMBRAPII; and Conselho Nacional de Desenvolvimento Científico e Tecnológico - CNPq (Grant Numbers: $305096/2022-2$).

\section*{Data Availability}
This literature review did not generate or analyze any new datasets. All information is available in the published studies cited in the References.

\section*{Author Contributions Statement}
Conceptualization: Lucas Q. Galvão, Clebson Cruz. Methodology: Lucas Q. Galvão, Anna Beatriz M. de Souza, Clebson Cruz. Investigation: Lucas Q. Galvão, Anna Beatriz M. de Souza. Visualization: Lucas Q. Galvão, Anna Beatriz M. de Souza. Writing - original draft: Lucas Q. Galvão, Anna Beatriz M. de Souza. Writing - review \& editing: Clebson Cruz, Marcelo A. Moret. Supervision: Clebson Cruz, Marcelo A. Moret. Project administration: Clebson Cruz. Resources: Marcelo A. Moret,  Lucas Q. Galvão, Clebson Cruz. Funding acquisition: Clebson Cruz, Marcelo A. Moret. All authors read and approved the final version of the manuscript.

\bibliography{main}

\begin{thebibliography}{195}%
\makeatletter
\providecommand \@ifxundefined [1]{%
 \@ifx{#1\undefined}
}%
\providecommand \@ifnum [1]{%
 \ifnum #1\expandafter \@firstoftwo
 \else \expandafter \@secondoftwo
 \fi
}%
\providecommand \@ifx [1]{%
 \ifx #1\expandafter \@firstoftwo
 \else \expandafter \@secondoftwo
 \fi
}%
\providecommand \natexlab [1]{#1}%
\providecommand \enquote  [1]{``#1''}%
\providecommand \bibnamefont  [1]{#1}%
\providecommand \bibfnamefont [1]{#1}%
\providecommand \citenamefont [1]{#1}%
\providecommand \href@noop [0]{\@secondoftwo}%
\providecommand \href [0]{\begingroup \@sanitize@url \@href}%
\providecommand \@href[1]{\@@startlink{#1}\@@href}%
\providecommand \@@href[1]{\endgroup#1\@@endlink}%
\providecommand \@sanitize@url [0]{\catcode `\\12\catcode `\$12\catcode `\&12\catcode `\#12\catcode `\^12\catcode `\_12\catcode `\%12\relax}%
\providecommand \@@startlink[1]{}%
\providecommand \@@endlink[0]{}%
\providecommand \url  [0]{\begingroup\@sanitize@url \@url }%
\providecommand \@url [1]{\endgroup\@href {#1}{\urlprefix }}%
\providecommand \urlprefix  [0]{URL }%
\providecommand \Eprint [0]{\href }%
\providecommand \doibase [0]{http://dx.doi.org/}%
\providecommand \selectlanguage [0]{\@gobble}%
\providecommand \bibinfo  [0]{\@secondoftwo}%
\providecommand \bibfield  [0]{\@secondoftwo}%
\providecommand \translation [1]{[#1]}%
\providecommand \BibitemOpen [0]{}%
\providecommand \bibitemStop [0]{}%
\providecommand \bibitemNoStop [0]{.\EOS\space}%
\providecommand \EOS [0]{\spacefactor3000\relax}%
\providecommand \BibitemShut  [1]{\csname bibitem#1\endcsname}%
\let\auto@bib@innerbib\@empty
\bibitem [{\citenamefont {Rohrlich}(1990)}]{Rohrlich_1990}%
  \BibitemOpen
  \bibfield  {author} {\bibinfo {author} {\bibfnamefont {F.}~\bibnamefont {Rohrlich}},\ }\href {\doibase 10.1086/psaprocbienmeetp.1990.2.193094} {\bibfield  {journal} {\bibinfo  {journal} {PSA: Proceedings of the Biennial Meeting of the Philosophy of Science Association}\ }\textbf {\bibinfo {volume} {1990}},\ \bibinfo {pages} {507–518} (\bibinfo {year} {1990})}\BibitemShut {NoStop}%
\bibitem [{\citenamefont {Brookshear}\ \emph {et~al.}(2009)\citenamefont {Brookshear}, \citenamefont {Brylow},\ and\ \citenamefont {Manasa}}]{brookshear2009computer}%
  \BibitemOpen
  \bibfield  {author} {\bibinfo {author} {\bibfnamefont {J.~G.}\ \bibnamefont {Brookshear}}, \bibinfo {author} {\bibfnamefont {D.}~\bibnamefont {Brylow}}, \ and\ \bibinfo {author} {\bibfnamefont {S.}~\bibnamefont {Manasa}},\ }\href@noop {} {\  (\bibinfo {year} {2009})}\BibitemShut {NoStop}%
\bibitem [{\citenamefont {Feynman}(2018{\natexlab{a}})}]{feynman2018feynman}%
  \BibitemOpen
  \bibfield  {author} {\bibinfo {author} {\bibfnamefont {R.~P.}\ \bibnamefont {Feynman}},\ }\href@noop {} {\emph {\bibinfo {title} {Feynman lectures on computation}}}\ (\bibinfo  {publisher} {CRC Press},\ \bibinfo {year} {2018})\BibitemShut {NoStop}%
\bibitem [{\citenamefont {Feynman}(2018{\natexlab{b}})}]{feynman2018simulating}%
  \BibitemOpen
  \bibfield  {author} {\bibinfo {author} {\bibfnamefont {R.~P.}\ \bibnamefont {Feynman}},\ }in\ \href {https://www.taylorfrancis.com/chapters/edit/10.1201/9780429500459-11/simulating-physics-computers-richard-feynman} {\emph {\bibinfo {booktitle} {Feynman and computation}}}\ (\bibinfo  {publisher} {cRc Press},\ \bibinfo {year} {2018})\ pp.\ \bibinfo {pages} {133--153}\BibitemShut {NoStop}%
\bibitem [{\citenamefont {Nielsen}\ and\ \citenamefont {Chuang}(2010)}]{nielsen2010quantum}%
  \BibitemOpen
  \bibfield  {author} {\bibinfo {author} {\bibfnamefont {M.~A.}\ \bibnamefont {Nielsen}}\ and\ \bibinfo {author} {\bibfnamefont {I.~L.}\ \bibnamefont {Chuang}},\ }\href@noop {} {\emph {\bibinfo {title} {Quantum computation and quantum information}}}\ (\bibinfo  {publisher} {Cambridge university press},\ \bibinfo {year} {2010})\BibitemShut {NoStop}%
\bibitem [{\citenamefont {Troyer}\ and\ \citenamefont {Wiese}(2005)}]{PhysRevLett.94.170201}%
  \BibitemOpen
  \bibfield  {author} {\bibinfo {author} {\bibfnamefont {M.}~\bibnamefont {Troyer}}\ and\ \bibinfo {author} {\bibfnamefont {U.-J.}\ \bibnamefont {Wiese}},\ }\href {\doibase 10.1103/PhysRevLett.94.170201} {\bibfield  {journal} {\bibinfo  {journal} {Phys. Rev. Lett.}\ }\textbf {\bibinfo {volume} {94}},\ \bibinfo {pages} {170201} (\bibinfo {year} {2005})}\BibitemShut {NoStop}%
\bibitem [{\citenamefont {Qin}\ \emph {et~al.}(2022)\citenamefont {Qin}, \citenamefont {Schäfer}, \citenamefont {Andergassen}, \citenamefont {Corboz},\ and\ \citenamefont {Gull}}]{Qin2022}%
  \BibitemOpen
  \bibfield  {author} {\bibinfo {author} {\bibfnamefont {M.}~\bibnamefont {Qin}}, \bibinfo {author} {\bibfnamefont {T.}~\bibnamefont {Schäfer}}, \bibinfo {author} {\bibfnamefont {S.}~\bibnamefont {Andergassen}}, \bibinfo {author} {\bibfnamefont {P.}~\bibnamefont {Corboz}}, \ and\ \bibinfo {author} {\bibfnamefont {E.}~\bibnamefont {Gull}},\ }\href {\doibase https://doi.org/10.1146/annurev-conmatphys-090921-033948} {\bibfield  {journal} {\bibinfo  {journal} {Annual Review of Condensed Matter Physics}\ }\textbf {\bibinfo {volume} {13}},\ \bibinfo {pages} {275} (\bibinfo {year} {2022})}\BibitemShut {NoStop}%
\bibitem [{\citenamefont {Lloyd}(1996)}]{doi:10.1126/science.273.5278.1073}%
  \BibitemOpen
  \bibfield  {author} {\bibinfo {author} {\bibfnamefont {S.}~\bibnamefont {Lloyd}},\ }\href {\doibase 10.1126/science.273.5278.1073} {\bibfield  {journal} {\bibinfo  {journal} {Science}\ }\textbf {\bibinfo {volume} {273}},\ \bibinfo {pages} {1073} (\bibinfo {year} {1996})}\BibitemShut {NoStop}%
\bibitem [{\citenamefont {Georgescu}\ \emph {et~al.}(2014)\citenamefont {Georgescu}, \citenamefont {Ashhab},\ and\ \citenamefont {Nori}}]{RevModPhys.86.153}%
  \BibitemOpen
  \bibfield  {author} {\bibinfo {author} {\bibfnamefont {I.~M.}\ \bibnamefont {Georgescu}}, \bibinfo {author} {\bibfnamefont {S.}~\bibnamefont {Ashhab}}, \ and\ \bibinfo {author} {\bibfnamefont {F.}~\bibnamefont {Nori}},\ }\href {\doibase 10.1103/RevModPhys.86.153} {\bibfield  {journal} {\bibinfo  {journal} {Rev. Mod. Phys.}\ }\textbf {\bibinfo {volume} {86}},\ \bibinfo {pages} {153} (\bibinfo {year} {2014})}\BibitemShut {NoStop}%
\bibitem [{\citenamefont {Somaroo}\ \emph {et~al.}(1999)\citenamefont {Somaroo}, \citenamefont {Tseng}, \citenamefont {Havel}, \citenamefont {Laflamme},\ and\ \citenamefont {Cory}}]{PhysRevLett.82.5381}%
  \BibitemOpen
  \bibfield  {author} {\bibinfo {author} {\bibfnamefont {S.}~\bibnamefont {Somaroo}}, \bibinfo {author} {\bibfnamefont {C.~H.}\ \bibnamefont {Tseng}}, \bibinfo {author} {\bibfnamefont {T.~F.}\ \bibnamefont {Havel}}, \bibinfo {author} {\bibfnamefont {R.}~\bibnamefont {Laflamme}}, \ and\ \bibinfo {author} {\bibfnamefont {D.~G.}\ \bibnamefont {Cory}},\ }\href {\doibase 10.1103/PhysRevLett.82.5381} {\bibfield  {journal} {\bibinfo  {journal} {Phys. Rev. Lett.}\ }\textbf {\bibinfo {volume} {82}},\ \bibinfo {pages} {5381} (\bibinfo {year} {1999})}\BibitemShut {NoStop}%
\bibitem [{\citenamefont {Shor}(1999)}]{doi:10.1137/S0036144598347011}%
  \BibitemOpen
  \bibfield  {author} {\bibinfo {author} {\bibfnamefont {P.~W.}\ \bibnamefont {Shor}},\ }\href {\doibase 10.1137/S0036144598347011} {\bibfield  {journal} {\bibinfo  {journal} {SIAM Review}\ }\textbf {\bibinfo {volume} {41}},\ \bibinfo {pages} {303} (\bibinfo {year} {1999})}\BibitemShut {NoStop}%
\bibitem [{\citenamefont {Grover}(1996)}]{10.1145/237814.237866}%
  \BibitemOpen
  \bibfield  {author} {\bibinfo {author} {\bibfnamefont {L.~K.}\ \bibnamefont {Grover}},\ }in\ \href {\doibase 10.1145/237814.237866} {\emph {\bibinfo {booktitle} {Proceedings of the Twenty-Eighth Annual ACM Symposium on Theory of Computing}}},\ \bibinfo {series and number} {STOC '96}\ (\bibinfo  {publisher} {Association for Computing Machinery},\ \bibinfo {address} {New York, NY, USA},\ \bibinfo {year} {1996})\ p.\ \bibinfo {pages} {212–219}\BibitemShut {NoStop}%
\bibitem [{\citenamefont {Harrow}\ \emph {et~al.}(2009)\citenamefont {Harrow}, \citenamefont {Hassidim},\ and\ \citenamefont {Lloyd}}]{PhysRevLett.103.150502}%
  \BibitemOpen
  \bibfield  {author} {\bibinfo {author} {\bibfnamefont {A.~W.}\ \bibnamefont {Harrow}}, \bibinfo {author} {\bibfnamefont {A.}~\bibnamefont {Hassidim}}, \ and\ \bibinfo {author} {\bibfnamefont {S.}~\bibnamefont {Lloyd}},\ }\href {\doibase 10.1103/PhysRevLett.103.150502} {\bibfield  {journal} {\bibinfo  {journal} {Phys. Rev. Lett.}\ }\textbf {\bibinfo {volume} {103}},\ \bibinfo {pages} {150502} (\bibinfo {year} {2009})}\BibitemShut {NoStop}%
\bibitem [{\citenamefont {Biamonte}\ \emph {et~al.}(2017)\citenamefont {Biamonte}, \citenamefont {Wittek}, \citenamefont {Pancotti}, \citenamefont {Rebentrost}, \citenamefont {Wiebe},\ and\ \citenamefont {Lloyd}}]{Biamonte2017}%
  \BibitemOpen
  \bibfield  {author} {\bibinfo {author} {\bibfnamefont {J.}~\bibnamefont {Biamonte}}, \bibinfo {author} {\bibfnamefont {P.}~\bibnamefont {Wittek}}, \bibinfo {author} {\bibfnamefont {N.}~\bibnamefont {Pancotti}}, \bibinfo {author} {\bibfnamefont {P.}~\bibnamefont {Rebentrost}}, \bibinfo {author} {\bibfnamefont {N.}~\bibnamefont {Wiebe}}, \ and\ \bibinfo {author} {\bibfnamefont {S.}~\bibnamefont {Lloyd}},\ }\href@noop {} {\bibfield  {journal} {\bibinfo  {journal} {Nature}\ }\textbf {\bibinfo {volume} {549}},\ \bibinfo {pages} {195} (\bibinfo {year} {2017})}\BibitemShut {NoStop}%
\bibitem [{\citenamefont {Deutsch}(1985)}]{deutsch1985quantum}%
  \BibitemOpen
  \bibfield  {author} {\bibinfo {author} {\bibfnamefont {D.}~\bibnamefont {Deutsch}},\ }\href@noop {} {\bibfield  {journal} {\bibinfo  {journal} {Proceedings of the Royal Society of London. A. Mathematical and Physical Sciences}\ }\textbf {\bibinfo {volume} {400}},\ \bibinfo {pages} {97} (\bibinfo {year} {1985})}\BibitemShut {NoStop}%
\bibitem [{\citenamefont {Kassal}\ \emph {et~al.}(2008)\citenamefont {Kassal}, \citenamefont {Jordan}, \citenamefont {Love}, \citenamefont {Mohseni},\ and\ \citenamefont {Aspuru-Guzik}}]{doi:10.1073/pnas.0808245105}%
  \BibitemOpen
  \bibfield  {author} {\bibinfo {author} {\bibfnamefont {I.}~\bibnamefont {Kassal}}, \bibinfo {author} {\bibfnamefont {S.~P.}\ \bibnamefont {Jordan}}, \bibinfo {author} {\bibfnamefont {P.~J.}\ \bibnamefont {Love}}, \bibinfo {author} {\bibfnamefont {M.}~\bibnamefont {Mohseni}}, \ and\ \bibinfo {author} {\bibfnamefont {A.}~\bibnamefont {Aspuru-Guzik}},\ }\href {\doibase 10.1073/pnas.0808245105} {\bibfield  {journal} {\bibinfo  {journal} {Proceedings of the National Academy of Sciences}\ }\textbf {\bibinfo {volume} {105}},\ \bibinfo {pages} {18681} (\bibinfo {year} {2008})}\BibitemShut {NoStop}%
\bibitem [{\citenamefont {Cao}\ \emph {et~al.}(2019)\citenamefont {Cao}, \citenamefont {Romero}, \citenamefont {Olson}, \citenamefont {Degroote}, \citenamefont {Johnson}, \citenamefont {Kieferov{\'a}}, \citenamefont {Kivlichan}, \citenamefont {Menke}, \citenamefont {Peropadre}, \citenamefont {Sawaya} \emph {et~al.}}]{cao2019quantum}%
  \BibitemOpen
  \bibfield  {author} {\bibinfo {author} {\bibfnamefont {Y.}~\bibnamefont {Cao}}, \bibinfo {author} {\bibfnamefont {J.}~\bibnamefont {Romero}}, \bibinfo {author} {\bibfnamefont {J.~P.}\ \bibnamefont {Olson}}, \bibinfo {author} {\bibfnamefont {M.}~\bibnamefont {Degroote}}, \bibinfo {author} {\bibfnamefont {P.~D.}\ \bibnamefont {Johnson}}, \bibinfo {author} {\bibfnamefont {M.}~\bibnamefont {Kieferov{\'a}}}, \bibinfo {author} {\bibfnamefont {I.~D.}\ \bibnamefont {Kivlichan}}, \bibinfo {author} {\bibfnamefont {T.}~\bibnamefont {Menke}}, \bibinfo {author} {\bibfnamefont {B.}~\bibnamefont {Peropadre}}, \bibinfo {author} {\bibfnamefont {N.~P.}\ \bibnamefont {Sawaya}},  \emph {et~al.},\ }\href@noop {} {\bibfield  {journal} {\bibinfo  {journal} {Chemical reviews}\ }\textbf {\bibinfo {volume} {119}},\ \bibinfo {pages} {10856} (\bibinfo {year} {2019})}\BibitemShut {NoStop}%
\bibitem [{\citenamefont {Bravyi}\ \emph {et~al.}(2008)\citenamefont {Bravyi}, \citenamefont {DiVincenzo}, \citenamefont {Loss},\ and\ \citenamefont {Terhal}}]{PhysRevLett.101.070503}%
  \BibitemOpen
  \bibfield  {author} {\bibinfo {author} {\bibfnamefont {S.}~\bibnamefont {Bravyi}}, \bibinfo {author} {\bibfnamefont {D.~P.}\ \bibnamefont {DiVincenzo}}, \bibinfo {author} {\bibfnamefont {D.}~\bibnamefont {Loss}}, \ and\ \bibinfo {author} {\bibfnamefont {B.~M.}\ \bibnamefont {Terhal}},\ }\href {\doibase 10.1103/PhysRevLett.101.070503} {\bibfield  {journal} {\bibinfo  {journal} {Phys. Rev. Lett.}\ }\textbf {\bibinfo {volume} {101}},\ \bibinfo {pages} {070503} (\bibinfo {year} {2008})}\BibitemShut {NoStop}%
\bibitem [{\citenamefont {Chenu}\ \emph {et~al.}(2017)\citenamefont {Chenu}, \citenamefont {Beau}, \citenamefont {Cao},\ and\ \citenamefont {del Campo}}]{PhysRevLett.118.140403}%
  \BibitemOpen
  \bibfield  {author} {\bibinfo {author} {\bibfnamefont {A.}~\bibnamefont {Chenu}}, \bibinfo {author} {\bibfnamefont {M.}~\bibnamefont {Beau}}, \bibinfo {author} {\bibfnamefont {J.}~\bibnamefont {Cao}}, \ and\ \bibinfo {author} {\bibfnamefont {A.}~\bibnamefont {del Campo}},\ }\href {\doibase 10.1103/PhysRevLett.118.140403} {\bibfield  {journal} {\bibinfo  {journal} {Phys. Rev. Lett.}\ }\textbf {\bibinfo {volume} {118}},\ \bibinfo {pages} {140403} (\bibinfo {year} {2017})}\BibitemShut {NoStop}%
\bibitem [{\citenamefont {Preskill}(2025)}]{preskill2025battling}%
  \BibitemOpen
  \bibfield  {author} {\bibinfo {author} {\bibfnamefont {J.}~\bibnamefont {Preskill}},\ }\href {https://aip.brightspotcdn.com/PTO.v78.i7.42_1.online.pdf} {\bibfield  {journal} {\bibinfo  {journal} {Physics Today}\ }\textbf {\bibinfo {volume} {78}},\ \bibinfo {pages} {42} (\bibinfo {year} {2025})}\BibitemShut {NoStop}%
\bibitem [{\citenamefont {Preskill}(2018)}]{preskill2018quantum}%
  \BibitemOpen
  \bibfield  {author} {\bibinfo {author} {\bibfnamefont {J.}~\bibnamefont {Preskill}},\ }\href {https://doi.org/10.22331/q-2018-08-06-79} {\bibfield  {journal} {\bibinfo  {journal} {Quantum}\ }\textbf {\bibinfo {volume} {2}},\ \bibinfo {pages} {79} (\bibinfo {year} {2018})}\BibitemShut {NoStop}%
\bibitem [{\citenamefont {Cerezo}\ \emph {et~al.}(2021{\natexlab{a}})\citenamefont {Cerezo}, \citenamefont {Arrasmith}, \citenamefont {Babbush}, \citenamefont {Benjamin}, \citenamefont {Endo}, \citenamefont {Fujii}, \citenamefont {McClean}, \citenamefont {Mitarai}, \citenamefont {Yuan}, \citenamefont {Cincio} \emph {et~al.}}]{cerezo2021variational}%
  \BibitemOpen
  \bibfield  {author} {\bibinfo {author} {\bibfnamefont {M.}~\bibnamefont {Cerezo}}, \bibinfo {author} {\bibfnamefont {A.}~\bibnamefont {Arrasmith}}, \bibinfo {author} {\bibfnamefont {R.}~\bibnamefont {Babbush}}, \bibinfo {author} {\bibfnamefont {S.~C.}\ \bibnamefont {Benjamin}}, \bibinfo {author} {\bibfnamefont {S.}~\bibnamefont {Endo}}, \bibinfo {author} {\bibfnamefont {K.}~\bibnamefont {Fujii}}, \bibinfo {author} {\bibfnamefont {J.~R.}\ \bibnamefont {McClean}}, \bibinfo {author} {\bibfnamefont {K.}~\bibnamefont {Mitarai}}, \bibinfo {author} {\bibfnamefont {X.}~\bibnamefont {Yuan}}, \bibinfo {author} {\bibfnamefont {L.}~\bibnamefont {Cincio}},  \emph {et~al.},\ }\href@noop {} {\bibfield  {journal} {\bibinfo  {journal} {Nature Reviews Physics}\ }\textbf {\bibinfo {volume} {3}},\ \bibinfo {pages} {625} (\bibinfo {year} {2021}{\natexlab{a}})}\BibitemShut {NoStop}%
\bibitem [{\citenamefont {Bharti}\ \emph {et~al.}(2022)\citenamefont {Bharti}, \citenamefont {Cervera-Lierta}, \citenamefont {Kyaw}, \citenamefont {Haug}, \citenamefont {Alperin-Lea}, \citenamefont {Anand}, \citenamefont {Degroote}, \citenamefont {Heimonen}, \citenamefont {Kottmann}, \citenamefont {Menke}, \citenamefont {Mok}, \citenamefont {Sim}, \citenamefont {Kwek},\ and\ \citenamefont {Aspuru-Guzik}}]{RevModPhys.94.015004}%
  \BibitemOpen
  \bibfield  {author} {\bibinfo {author} {\bibfnamefont {K.}~\bibnamefont {Bharti}}, \bibinfo {author} {\bibfnamefont {A.}~\bibnamefont {Cervera-Lierta}}, \bibinfo {author} {\bibfnamefont {T.~H.}\ \bibnamefont {Kyaw}}, \bibinfo {author} {\bibfnamefont {T.}~\bibnamefont {Haug}}, \bibinfo {author} {\bibfnamefont {S.}~\bibnamefont {Alperin-Lea}}, \bibinfo {author} {\bibfnamefont {A.}~\bibnamefont {Anand}}, \bibinfo {author} {\bibfnamefont {M.}~\bibnamefont {Degroote}}, \bibinfo {author} {\bibfnamefont {H.}~\bibnamefont {Heimonen}}, \bibinfo {author} {\bibfnamefont {J.~S.}\ \bibnamefont {Kottmann}}, \bibinfo {author} {\bibfnamefont {T.}~\bibnamefont {Menke}}, \bibinfo {author} {\bibfnamefont {W.-K.}\ \bibnamefont {Mok}}, \bibinfo {author} {\bibfnamefont {S.}~\bibnamefont {Sim}}, \bibinfo {author} {\bibfnamefont {L.-C.}\ \bibnamefont {Kwek}}, \ and\ \bibinfo {author} {\bibfnamefont {A.}~\bibnamefont {Aspuru-Guzik}},\ }\href {\doibase 10.1103/RevModPhys.94.015004} {\bibfield  {journal} {\bibinfo  {journal} {Rev.
  Mod. Phys.}\ }\textbf {\bibinfo {volume} {94}},\ \bibinfo {pages} {015004} (\bibinfo {year} {2022})}\BibitemShut {NoStop}%
\bibitem [{\citenamefont {Peruzzo}\ \emph {et~al.}(2014)\citenamefont {Peruzzo}, \citenamefont {McClean}, \citenamefont {Shadbolt}, \citenamefont {Yung}, \citenamefont {Zhou}, \citenamefont {Love}, \citenamefont {Aspuru-Guzik},\ and\ \citenamefont {O'Brien}}]{peruzzo2014variational}%
  \BibitemOpen
  \bibfield  {author} {\bibinfo {author} {\bibfnamefont {A.}~\bibnamefont {Peruzzo}}, \bibinfo {author} {\bibfnamefont {J.}~\bibnamefont {McClean}}, \bibinfo {author} {\bibfnamefont {P.}~\bibnamefont {Shadbolt}}, \bibinfo {author} {\bibfnamefont {M.-H.}\ \bibnamefont {Yung}}, \bibinfo {author} {\bibfnamefont {X.-Q.}\ \bibnamefont {Zhou}}, \bibinfo {author} {\bibfnamefont {P.~J.}\ \bibnamefont {Love}}, \bibinfo {author} {\bibfnamefont {A.}~\bibnamefont {Aspuru-Guzik}}, \ and\ \bibinfo {author} {\bibfnamefont {J.~L.}\ \bibnamefont {O'Brien}},\ }\href@noop {} {\bibfield  {journal} {\bibinfo  {journal} {Nature Communications}\ }\textbf {\bibinfo {volume} {5}},\ \bibinfo {pages} {4213} (\bibinfo {year} {2014})}\BibitemShut {NoStop}%
\bibitem [{\citenamefont {Tilly}\ \emph {et~al.}(2022)\citenamefont {Tilly}, \citenamefont {Chen}, \citenamefont {Cao}, \citenamefont {Picozzi}, \citenamefont {Setia}, \citenamefont {Li}, \citenamefont {Grant}, \citenamefont {Wossnig}, \citenamefont {Rungger}, \citenamefont {Booth},\ and\ \citenamefont {Tennyson}}]{TILLY20221}%
  \BibitemOpen
  \bibfield  {author} {\bibinfo {author} {\bibfnamefont {J.}~\bibnamefont {Tilly}}, \bibinfo {author} {\bibfnamefont {H.}~\bibnamefont {Chen}}, \bibinfo {author} {\bibfnamefont {S.}~\bibnamefont {Cao}}, \bibinfo {author} {\bibfnamefont {D.}~\bibnamefont {Picozzi}}, \bibinfo {author} {\bibfnamefont {K.}~\bibnamefont {Setia}}, \bibinfo {author} {\bibfnamefont {Y.}~\bibnamefont {Li}}, \bibinfo {author} {\bibfnamefont {E.}~\bibnamefont {Grant}}, \bibinfo {author} {\bibfnamefont {L.}~\bibnamefont {Wossnig}}, \bibinfo {author} {\bibfnamefont {I.}~\bibnamefont {Rungger}}, \bibinfo {author} {\bibfnamefont {G.~H.}\ \bibnamefont {Booth}}, \ and\ \bibinfo {author} {\bibfnamefont {J.}~\bibnamefont {Tennyson}},\ }\href {\doibase https://doi.org/10.1016/j.physrep.2022.08.003} {\bibfield  {journal} {\bibinfo  {journal} {Physics Reports}\ }\textbf {\bibinfo {volume} {986}},\ \bibinfo {pages} {1} (\bibinfo {year} {2022})},\ \bibinfo {note} {the Variational Quantum Eigensolver: a review of methods and best
  practices}\BibitemShut {NoStop}%
\bibitem [{\citenamefont {Li}\ and\ \citenamefont {Benjamin}(2017)}]{li2017efficient}%
  \BibitemOpen
  \bibfield  {author} {\bibinfo {author} {\bibfnamefont {Y.}~\bibnamefont {Li}}\ and\ \bibinfo {author} {\bibfnamefont {S.~C.}\ \bibnamefont {Benjamin}},\ }\href@noop {} {\bibfield  {journal} {\bibinfo  {journal} {Physical Review X}\ }\textbf {\bibinfo {volume} {7}},\ \bibinfo {pages} {021050} (\bibinfo {year} {2017})}\BibitemShut {NoStop}%
\bibitem [{\citenamefont {Endo}\ \emph {et~al.}(2020)\citenamefont {Endo}, \citenamefont {Sun}, \citenamefont {Li}, \citenamefont {Benjamin},\ and\ \citenamefont {Yuan}}]{PhysRevLett.125.010501}%
  \BibitemOpen
  \bibfield  {author} {\bibinfo {author} {\bibfnamefont {S.}~\bibnamefont {Endo}}, \bibinfo {author} {\bibfnamefont {J.}~\bibnamefont {Sun}}, \bibinfo {author} {\bibfnamefont {Y.}~\bibnamefont {Li}}, \bibinfo {author} {\bibfnamefont {S.~C.}\ \bibnamefont {Benjamin}}, \ and\ \bibinfo {author} {\bibfnamefont {X.}~\bibnamefont {Yuan}},\ }\href {\doibase 10.1103/PhysRevLett.125.010501} {\bibfield  {journal} {\bibinfo  {journal} {Phys. Rev. Lett.}\ }\textbf {\bibinfo {volume} {125}},\ \bibinfo {pages} {010501} (\bibinfo {year} {2020})}\BibitemShut {NoStop}%
\bibitem [{\citenamefont {Luo}\ \emph {et~al.}(2024)\citenamefont {Luo}, \citenamefont {Lin},\ and\ \citenamefont {Gao}}]{doi:10.1021/acs.jpclett.4c00576}%
  \BibitemOpen
  \bibfield  {author} {\bibinfo {author} {\bibfnamefont {J.}~\bibnamefont {Luo}}, \bibinfo {author} {\bibfnamefont {K.}~\bibnamefont {Lin}}, \ and\ \bibinfo {author} {\bibfnamefont {X.}~\bibnamefont {Gao}},\ }\href {\doibase 10.1021/acs.jpclett.4c00576} {\bibfield  {journal} {\bibinfo  {journal} {The Journal of Physical Chemistry Letters}\ }\textbf {\bibinfo {volume} {15}},\ \bibinfo {pages} {3516} (\bibinfo {year} {2024})},\ \bibinfo {note} {pMID: 38517759}\BibitemShut {NoStop}%
\bibitem [{\citenamefont {Wu}\ and\ \citenamefont {Hsieh}(2019)}]{PhysRevLett.123.220502}%
  \BibitemOpen
  \bibfield  {author} {\bibinfo {author} {\bibfnamefont {J.}~\bibnamefont {Wu}}\ and\ \bibinfo {author} {\bibfnamefont {T.~H.}\ \bibnamefont {Hsieh}},\ }\href {\doibase 10.1103/PhysRevLett.123.220502} {\bibfield  {journal} {\bibinfo  {journal} {Phys. Rev. Lett.}\ }\textbf {\bibinfo {volume} {123}},\ \bibinfo {pages} {220502} (\bibinfo {year} {2019})}\BibitemShut {NoStop}%
\bibitem [{\citenamefont {Verdon}\ \emph {et~al.}(2019)\citenamefont {Verdon}, \citenamefont {Marks}, \citenamefont {Nanda}, \citenamefont {Leichenauer},\ and\ \citenamefont {Hidary}}]{verdon2019quantumhamiltonianbasedmodelsvariational}%
  \BibitemOpen
  \bibfield  {author} {\bibinfo {author} {\bibfnamefont {G.}~\bibnamefont {Verdon}}, \bibinfo {author} {\bibfnamefont {J.}~\bibnamefont {Marks}}, \bibinfo {author} {\bibfnamefont {S.}~\bibnamefont {Nanda}}, \bibinfo {author} {\bibfnamefont {S.}~\bibnamefont {Leichenauer}}, \ and\ \bibinfo {author} {\bibfnamefont {J.}~\bibnamefont {Hidary}},\ }\href {https://arxiv.org/abs/1910.02071} {\enquote {\bibinfo {title} {Quantum hamiltonian-based models and the variational quantum thermalizer algorithm},}\ } (\bibinfo {year} {2019})\BibitemShut {NoStop}%
\bibitem [{\citenamefont {Lee}\ \emph {et~al.}(2021{\natexlab{a}})\citenamefont {Lee}, \citenamefont {Patil}, \citenamefont {Zhang},\ and\ \citenamefont {Hsieh}}]{PhysRevResearch.3.023095}%
  \BibitemOpen
  \bibfield  {author} {\bibinfo {author} {\bibfnamefont {C.~K.}\ \bibnamefont {Lee}}, \bibinfo {author} {\bibfnamefont {P.}~\bibnamefont {Patil}}, \bibinfo {author} {\bibfnamefont {S.}~\bibnamefont {Zhang}}, \ and\ \bibinfo {author} {\bibfnamefont {C.~Y.}\ \bibnamefont {Hsieh}},\ }\href {\doibase 10.1103/PhysRevResearch.3.023095} {\bibfield  {journal} {\bibinfo  {journal} {Phys. Rev. Res.}\ }\textbf {\bibinfo {volume} {3}},\ \bibinfo {pages} {023095} (\bibinfo {year} {2021}{\natexlab{a}})}\BibitemShut {NoStop}%
\bibitem [{\citenamefont {Arrasmith}\ \emph {et~al.}(2019)\citenamefont {Arrasmith}, \citenamefont {Cincio}, \citenamefont {Sornborger}, \citenamefont {Zurek},\ and\ \citenamefont {Coles}}]{arrasmith2019variational}%
  \BibitemOpen
  \bibfield  {author} {\bibinfo {author} {\bibfnamefont {A.}~\bibnamefont {Arrasmith}}, \bibinfo {author} {\bibfnamefont {L.}~\bibnamefont {Cincio}}, \bibinfo {author} {\bibfnamefont {A.~T.}\ \bibnamefont {Sornborger}}, \bibinfo {author} {\bibfnamefont {W.~H.}\ \bibnamefont {Zurek}}, \ and\ \bibinfo {author} {\bibfnamefont {P.~J.}\ \bibnamefont {Coles}},\ }\href@noop {} {\bibfield  {journal} {\bibinfo  {journal} {Nature communications}\ }\textbf {\bibinfo {volume} {10}},\ \bibinfo {pages} {3438} (\bibinfo {year} {2019})}\BibitemShut {NoStop}%
\bibitem [{\citenamefont {Larocca}\ \emph {et~al.}(2025)\citenamefont {Larocca}, \citenamefont {Thanasilp}, \citenamefont {Wang}, \citenamefont {Sharma}, \citenamefont {Biamonte}, \citenamefont {Coles}, \citenamefont {Cincio}, \citenamefont {McClean}, \citenamefont {Holmes},\ and\ \citenamefont {Cerezo}}]{larocca2025barren}%
  \BibitemOpen
  \bibfield  {author} {\bibinfo {author} {\bibfnamefont {M.}~\bibnamefont {Larocca}}, \bibinfo {author} {\bibfnamefont {S.}~\bibnamefont {Thanasilp}}, \bibinfo {author} {\bibfnamefont {S.}~\bibnamefont {Wang}}, \bibinfo {author} {\bibfnamefont {K.}~\bibnamefont {Sharma}}, \bibinfo {author} {\bibfnamefont {J.}~\bibnamefont {Biamonte}}, \bibinfo {author} {\bibfnamefont {P.~J.}\ \bibnamefont {Coles}}, \bibinfo {author} {\bibfnamefont {L.}~\bibnamefont {Cincio}}, \bibinfo {author} {\bibfnamefont {J.~R.}\ \bibnamefont {McClean}}, \bibinfo {author} {\bibfnamefont {Z.}~\bibnamefont {Holmes}}, \ and\ \bibinfo {author} {\bibfnamefont {M.}~\bibnamefont {Cerezo}},\ }\href@noop {} {\bibfield  {journal} {\bibinfo  {journal} {Nature Reviews Physics}\ ,\ \bibinfo {pages} {1}} (\bibinfo {year} {2025})}\BibitemShut {NoStop}%
\bibitem [{\citenamefont {Zhang}\ \emph {et~al.}(2025{\natexlab{a}})\citenamefont {Zhang}, \citenamefont {Zhang}, \citenamefont {Sun}, \citenamefont {Lin}, \citenamefont {Huang}, \citenamefont {Lv},\ and\ \citenamefont {Yuan}}]{Zhang2025}%
  \BibitemOpen
  \bibfield  {author} {\bibinfo {author} {\bibfnamefont {Y.}~\bibnamefont {Zhang}}, \bibinfo {author} {\bibfnamefont {X.}~\bibnamefont {Zhang}}, \bibinfo {author} {\bibfnamefont {J.}~\bibnamefont {Sun}}, \bibinfo {author} {\bibfnamefont {H.}~\bibnamefont {Lin}}, \bibinfo {author} {\bibfnamefont {Y.}~\bibnamefont {Huang}}, \bibinfo {author} {\bibfnamefont {D.}~\bibnamefont {Lv}}, \ and\ \bibinfo {author} {\bibfnamefont {X.}~\bibnamefont {Yuan}},\ }\href {\doibase https://doi.org/10.1002/wcms.70020} {\bibfield  {journal} {\bibinfo  {journal} {WIREs Computational Molecular Science}\ }\textbf {\bibinfo {volume} {15}},\ \bibinfo {pages} {e70020} (\bibinfo {year} {2025}{\natexlab{a}})},\ \bibinfo {note} {e70020 CMS-1125.R1}\BibitemShut {NoStop}%
\bibitem [{\citenamefont {Bauer}\ \emph {et~al.}(2023)\citenamefont {Bauer}, \citenamefont {Davoudi}, \citenamefont {Klco},\ and\ \citenamefont {Savage}}]{Bauer2023}%
  \BibitemOpen
  \bibfield  {author} {\bibinfo {author} {\bibfnamefont {C.~W.}\ \bibnamefont {Bauer}}, \bibinfo {author} {\bibfnamefont {Z.}~\bibnamefont {Davoudi}}, \bibinfo {author} {\bibfnamefont {N.}~\bibnamefont {Klco}}, \ and\ \bibinfo {author} {\bibfnamefont {M.~J.}\ \bibnamefont {Savage}},\ }\href {\doibase 10.1038/s42254-023-00599-8} {\bibfield  {journal} {\bibinfo  {journal} {Nature Reviews Physics}\ }\textbf {\bibinfo {volume} {5}},\ \bibinfo {pages} {420} (\bibinfo {year} {2023})}\BibitemShut {NoStop}%
\bibitem [{\citenamefont {Bachelard}(1984)}]{bachelard1984new}%
  \BibitemOpen
  \bibfield  {author} {\bibinfo {author} {\bibfnamefont {G.}~\bibnamefont {Bachelard}},\ }\href {https://www.amazon.com.br/New-Scientific-Spirit-Gaston-Bachelard/dp/0807015008} {\emph {\bibinfo {title} {The new scientific spirit}}}\ (\bibinfo  {publisher} {Beacon Press},\ \bibinfo {year} {1984})\BibitemShut {NoStop}%
\bibitem [{\citenamefont {Wolfram}(2002)}]{wolfram2002new}%
  \BibitemOpen
  \bibfield  {author} {\bibinfo {author} {\bibfnamefont {S.}~\bibnamefont {Wolfram}},\ }\href {https://books.google.com.br/books?id=dw_vAAAAMAAJ} {\emph {\bibinfo {title} {A New Kind of Science}}}\ (\bibinfo  {publisher} {Wolfram Media},\ \bibinfo {year} {2002})\BibitemShut {NoStop}%
\bibitem [{\citenamefont {Saha}\ and\ \citenamefont {Mukherjee}(2021)}]{simulations}%
  \BibitemOpen
  \bibfield  {author} {\bibinfo {author} {\bibfnamefont {S.~K.}\ \bibnamefont {Saha}}\ and\ \bibinfo {author} {\bibfnamefont {M.}~\bibnamefont {Mukherjee}},\ }\href {https://link.springer.com/book/10.1007/978-981-15-8315-5} {\emph {\bibinfo {title} {Recent Advances in Computational Mechanics and Simulations}}}\ (\bibinfo  {publisher} {Springer},\ \bibinfo {year} {2021})\BibitemShut {NoStop}%
\bibitem [{\citenamefont {Press}(2007)}]{press2007numerical}%
  \BibitemOpen
  \bibfield  {author} {\bibinfo {author} {\bibfnamefont {W.~H.}\ \bibnamefont {Press}},\ }\href@noop {} {\emph {\bibinfo {title} {Numerical recipes 3rd edition: The art of scientific computing}}}\ (\bibinfo  {publisher} {Cambridge university press},\ \bibinfo {year} {2007})\BibitemShut {NoStop}%
\bibitem [{\citenamefont {Hilbert}\ and\ \citenamefont {López}(2011)}]{doi:10.1126/science.1200970}%
  \BibitemOpen
  \bibfield  {author} {\bibinfo {author} {\bibfnamefont {M.}~\bibnamefont {Hilbert}}\ and\ \bibinfo {author} {\bibfnamefont {P.}~\bibnamefont {López}},\ }\href {\doibase 10.1126/science.1200970} {\bibfield  {journal} {\bibinfo  {journal} {Science}\ }\textbf {\bibinfo {volume} {332}},\ \bibinfo {pages} {60} (\bibinfo {year} {2011})}\BibitemShut {NoStop}%
\bibitem [{\citenamefont {Bartley}(2024)}]{bartley2023big}%
  \BibitemOpen
  \bibfield  {author} {\bibinfo {author} {\bibfnamefont {K.}~\bibnamefont {Bartley}},\ }\href@noop {} {\bibfield  {journal} {\bibinfo  {journal} {URL https://rivery. io/blog/big-data-statistics-how-much-data-is-there-in-the-world}\ } (\bibinfo {year} {2024})}\BibitemShut {NoStop}%
\bibitem [{\citenamefont {Benioff}(1982)}]{PhysRevLett.48.1581}%
  \BibitemOpen
  \bibfield  {author} {\bibinfo {author} {\bibfnamefont {P.}~\bibnamefont {Benioff}},\ }\href {\doibase 10.1103/PhysRevLett.48.1581} {\bibfield  {journal} {\bibinfo  {journal} {Phys. Rev. Lett.}\ }\textbf {\bibinfo {volume} {48}},\ \bibinfo {pages} {1581} (\bibinfo {year} {1982})}\BibitemShut {NoStop}%
\bibitem [{\citenamefont {Kluber}(2023)}]{kluber2023trotterization}%
  \BibitemOpen
  \bibfield  {author} {\bibinfo {author} {\bibfnamefont {G.}~\bibnamefont {Kluber}},\ }\href@noop {} {\bibfield  {journal} {\bibinfo  {journal} {arXiv preprint arXiv:2310.13296}\ } (\bibinfo {year} {2023})}\BibitemShut {NoStop}%
\bibitem [{\citenamefont {Kalos}(2012)}]{kalos2012monte}%
  \BibitemOpen
  \bibfield  {author} {\bibinfo {author} {\bibfnamefont {M.~H.}\ \bibnamefont {Kalos}},\ }\href@noop {} {\emph {\bibinfo {title} {Monte Carlo methods in quantum problems}}},\ Vol.\ \bibinfo {volume} {125}\ (\bibinfo  {publisher} {Springer Science \& Business Media},\ \bibinfo {year} {2012})\BibitemShut {NoStop}%
\bibitem [{\citenamefont {Poulin}\ \emph {et~al.}(2011)\citenamefont {Poulin}, \citenamefont {Qarry}, \citenamefont {Somma},\ and\ \citenamefont {Verstraete}}]{PhysRevLett.106.170501}%
  \BibitemOpen
  \bibfield  {author} {\bibinfo {author} {\bibfnamefont {D.}~\bibnamefont {Poulin}}, \bibinfo {author} {\bibfnamefont {A.}~\bibnamefont {Qarry}}, \bibinfo {author} {\bibfnamefont {R.}~\bibnamefont {Somma}}, \ and\ \bibinfo {author} {\bibfnamefont {F.}~\bibnamefont {Verstraete}},\ }\href {\doibase 10.1103/PhysRevLett.106.170501} {\bibfield  {journal} {\bibinfo  {journal} {Phys. Rev. Lett.}\ }\textbf {\bibinfo {volume} {106}},\ \bibinfo {pages} {170501} (\bibinfo {year} {2011})}\BibitemShut {NoStop}%
\bibitem [{\citenamefont {Heyl}\ \emph {et~al.}(2019)\citenamefont {Heyl}, \citenamefont {Hauke},\ and\ \citenamefont {Zoller}}]{doi:10.1126/sciadv.aau8342}%
  \BibitemOpen
  \bibfield  {author} {\bibinfo {author} {\bibfnamefont {M.}~\bibnamefont {Heyl}}, \bibinfo {author} {\bibfnamefont {P.}~\bibnamefont {Hauke}}, \ and\ \bibinfo {author} {\bibfnamefont {P.}~\bibnamefont {Zoller}},\ }\href {\doibase 10.1126/sciadv.aau8342} {\bibfield  {journal} {\bibinfo  {journal} {Science Advances}\ }\textbf {\bibinfo {volume} {5}},\ \bibinfo {pages} {eaau8342} (\bibinfo {year} {2019})}\BibitemShut {NoStop}%
\bibitem [{\citenamefont {Sakurai}\ and\ \citenamefont {Napolitano}(2020)}]{sakurai2020modern}%
  \BibitemOpen
  \bibfield  {author} {\bibinfo {author} {\bibfnamefont {J.~J.}\ \bibnamefont {Sakurai}}\ and\ \bibinfo {author} {\bibfnamefont {J.}~\bibnamefont {Napolitano}},\ }\href@noop {} {\emph {\bibinfo {title} {Modern quantum mechanics}}}\ (\bibinfo  {publisher} {Cambridge University Press},\ \bibinfo {year} {2020})\BibitemShut {NoStop}%
\bibitem [{\citenamefont {Somhorst}\ \emph {et~al.}(2023)\citenamefont {Somhorst}, \citenamefont {van~der Meer}, \citenamefont {Correa~Anguita}, \citenamefont {Schadow}, \citenamefont {Snijders}, \citenamefont {de~Goede}, \citenamefont {Kassenberg}, \citenamefont {Venderbosch}, \citenamefont {Taballione}, \citenamefont {Epping} \emph {et~al.}}]{somhorst2023quantum}%
  \BibitemOpen
  \bibfield  {author} {\bibinfo {author} {\bibfnamefont {F.~H.}\ \bibnamefont {Somhorst}}, \bibinfo {author} {\bibfnamefont {R.}~\bibnamefont {van~der Meer}}, \bibinfo {author} {\bibfnamefont {M.}~\bibnamefont {Correa~Anguita}}, \bibinfo {author} {\bibfnamefont {R.}~\bibnamefont {Schadow}}, \bibinfo {author} {\bibfnamefont {H.~J.}\ \bibnamefont {Snijders}}, \bibinfo {author} {\bibfnamefont {M.}~\bibnamefont {de~Goede}}, \bibinfo {author} {\bibfnamefont {B.}~\bibnamefont {Kassenberg}}, \bibinfo {author} {\bibfnamefont {P.}~\bibnamefont {Venderbosch}}, \bibinfo {author} {\bibfnamefont {C.}~\bibnamefont {Taballione}}, \bibinfo {author} {\bibfnamefont {J.}~\bibnamefont {Epping}},  \emph {et~al.},\ }\href@noop {} {\bibfield  {journal} {\bibinfo  {journal} {Nature communications}\ }\textbf {\bibinfo {volume} {14}},\ \bibinfo {pages} {3895} (\bibinfo {year} {2023})}\BibitemShut {NoStop}%
\bibitem [{\citenamefont {Cornish}\ \emph {et~al.}(2024)\citenamefont {Cornish}, \citenamefont {Tarbutt},\ and\ \citenamefont {Hazzard}}]{cornish2024quantum}%
  \BibitemOpen
  \bibfield  {author} {\bibinfo {author} {\bibfnamefont {S.~L.}\ \bibnamefont {Cornish}}, \bibinfo {author} {\bibfnamefont {M.~R.}\ \bibnamefont {Tarbutt}}, \ and\ \bibinfo {author} {\bibfnamefont {K.~R.}\ \bibnamefont {Hazzard}},\ }\href@noop {} {\bibfield  {journal} {\bibinfo  {journal} {Nature Physics}\ }\textbf {\bibinfo {volume} {20}},\ \bibinfo {pages} {730} (\bibinfo {year} {2024})}\BibitemShut {NoStop}%
\bibitem [{\citenamefont {Huang}\ \emph {et~al.}(2022)\citenamefont {Huang}, \citenamefont {Broughton}, \citenamefont {Cotler}, \citenamefont {Chen}, \citenamefont {Li}, \citenamefont {Mohseni}, \citenamefont {Neven}, \citenamefont {Babbush}, \citenamefont {Kueng}, \citenamefont {Preskill},\ and\ \citenamefont {McClean}}]{doi:10.1126/science.abn7293}%
  \BibitemOpen
  \bibfield  {author} {\bibinfo {author} {\bibfnamefont {H.-Y.}\ \bibnamefont {Huang}}, \bibinfo {author} {\bibfnamefont {M.}~\bibnamefont {Broughton}}, \bibinfo {author} {\bibfnamefont {J.}~\bibnamefont {Cotler}}, \bibinfo {author} {\bibfnamefont {S.}~\bibnamefont {Chen}}, \bibinfo {author} {\bibfnamefont {J.}~\bibnamefont {Li}}, \bibinfo {author} {\bibfnamefont {M.}~\bibnamefont {Mohseni}}, \bibinfo {author} {\bibfnamefont {H.}~\bibnamefont {Neven}}, \bibinfo {author} {\bibfnamefont {R.}~\bibnamefont {Babbush}}, \bibinfo {author} {\bibfnamefont {R.}~\bibnamefont {Kueng}}, \bibinfo {author} {\bibfnamefont {J.}~\bibnamefont {Preskill}}, \ and\ \bibinfo {author} {\bibfnamefont {J.~R.}\ \bibnamefont {McClean}},\ }\href {\doibase 10.1126/science.abn7293} {\bibfield  {journal} {\bibinfo  {journal} {Science}\ }\textbf {\bibinfo {volume} {376}},\ \bibinfo {pages} {1182} (\bibinfo {year} {2022})}\BibitemShut {NoStop}%
\bibitem [{\citenamefont {Daley}\ \emph {et~al.}(2022{\natexlab{a}})\citenamefont {Daley}, \citenamefont {Bloch}, \citenamefont {Kokail}, \citenamefont {Flannigan}, \citenamefont {Pearson}, \citenamefont {Troyer},\ and\ \citenamefont {Zoller}}]{daley2022practical}%
  \BibitemOpen
  \bibfield  {author} {\bibinfo {author} {\bibfnamefont {A.~J.}\ \bibnamefont {Daley}}, \bibinfo {author} {\bibfnamefont {I.}~\bibnamefont {Bloch}}, \bibinfo {author} {\bibfnamefont {C.}~\bibnamefont {Kokail}}, \bibinfo {author} {\bibfnamefont {S.}~\bibnamefont {Flannigan}}, \bibinfo {author} {\bibfnamefont {N.}~\bibnamefont {Pearson}}, \bibinfo {author} {\bibfnamefont {M.}~\bibnamefont {Troyer}}, \ and\ \bibinfo {author} {\bibfnamefont {P.}~\bibnamefont {Zoller}},\ }\href@noop {} {\bibfield  {journal} {\bibinfo  {journal} {Nature}\ }\textbf {\bibinfo {volume} {607}},\ \bibinfo {pages} {667} (\bibinfo {year} {2022}{\natexlab{a}})}\BibitemShut {NoStop}%
\bibitem [{\citenamefont {Verstraete}\ \emph {et~al.}(2009)\citenamefont {Verstraete}, \citenamefont {Cirac},\ and\ \citenamefont {Latorre}}]{PhysRevA.79.032316}%
  \BibitemOpen
  \bibfield  {author} {\bibinfo {author} {\bibfnamefont {F.}~\bibnamefont {Verstraete}}, \bibinfo {author} {\bibfnamefont {J.~I.}\ \bibnamefont {Cirac}}, \ and\ \bibinfo {author} {\bibfnamefont {J.~I.}\ \bibnamefont {Latorre}},\ }\href {\doibase 10.1103/PhysRevA.79.032316} {\bibfield  {journal} {\bibinfo  {journal} {Phys. Rev. A}\ }\textbf {\bibinfo {volume} {79}},\ \bibinfo {pages} {032316} (\bibinfo {year} {2009})}\BibitemShut {NoStop}%
\bibitem [{\citenamefont {van Oudenaarden}\ and\ \citenamefont {Mooij}(1996)}]{PhysRevLett.76.4947}%
  \BibitemOpen
  \bibfield  {author} {\bibinfo {author} {\bibfnamefont {A.}~\bibnamefont {van Oudenaarden}}\ and\ \bibinfo {author} {\bibfnamefont {J.~E.}\ \bibnamefont {Mooij}},\ }\href {\doibase 10.1103/PhysRevLett.76.4947} {\bibfield  {journal} {\bibinfo  {journal} {Phys. Rev. Lett.}\ }\textbf {\bibinfo {volume} {76}},\ \bibinfo {pages} {4947} (\bibinfo {year} {1996})}\BibitemShut {NoStop}%
\bibitem [{\citenamefont {Benenti}\ and\ \citenamefont {Strini}(2008)}]{benenti2008quantum}%
  \BibitemOpen
  \bibfield  {author} {\bibinfo {author} {\bibfnamefont {G.}~\bibnamefont {Benenti}}\ and\ \bibinfo {author} {\bibfnamefont {G.}~\bibnamefont {Strini}},\ }\href@noop {} {\bibfield  {journal} {\bibinfo  {journal} {American Journal of Physics}\ }\textbf {\bibinfo {volume} {76}},\ \bibinfo {pages} {657} (\bibinfo {year} {2008})}\BibitemShut {NoStop}%
\bibitem [{\citenamefont {Nielsen}\ \emph {et~al.}(2006)\citenamefont {Nielsen}, \citenamefont {Dowling}, \citenamefont {Gu},\ and\ \citenamefont {Doherty}}]{doi:10.1126/science.1121541}%
  \BibitemOpen
  \bibfield  {author} {\bibinfo {author} {\bibfnamefont {M.~A.}\ \bibnamefont {Nielsen}}, \bibinfo {author} {\bibfnamefont {M.~R.}\ \bibnamefont {Dowling}}, \bibinfo {author} {\bibfnamefont {M.}~\bibnamefont {Gu}}, \ and\ \bibinfo {author} {\bibfnamefont {A.~C.}\ \bibnamefont {Doherty}},\ }\href {\doibase 10.1126/science.1121541} {\bibfield  {journal} {\bibinfo  {journal} {Science}\ }\textbf {\bibinfo {volume} {311}},\ \bibinfo {pages} {1133} (\bibinfo {year} {2006})}\BibitemShut {NoStop}%
\bibitem [{\citenamefont {Alves}\ \emph {et~al.}(2020)\citenamefont {Alves}, \citenamefont {Gomes}, \citenamefont {Santana},\ and\ \citenamefont {Santos}}]{alves2020simulating}%
  \BibitemOpen
  \bibfield  {author} {\bibinfo {author} {\bibfnamefont {{\'E}.~M.}\ \bibnamefont {Alves}}, \bibinfo {author} {\bibfnamefont {F.~D.}\ \bibnamefont {Gomes}}, \bibinfo {author} {\bibfnamefont {H.~S.}\ \bibnamefont {Santana}}, \ and\ \bibinfo {author} {\bibfnamefont {A.~C.}\ \bibnamefont {Santos}},\ }\href {\doibase https://doi.org/10.1590/1806-9126-RBEF-2019-0299} {\bibfield  {journal} {\bibinfo  {journal} {Revista Brasileira de Ensino de F{\'\i}sica}\ }\textbf {\bibinfo {volume} {42}},\ \bibinfo {pages} {e20190299} (\bibinfo {year} {2020})}\BibitemShut {NoStop}%
\bibitem [{\citenamefont {Zhang}\ \emph {et~al.}(2025{\natexlab{b}})\citenamefont {Zhang}, \citenamefont {Zhang}, \citenamefont {Sun}, \citenamefont {Lin}, \citenamefont {Huang}, \citenamefont {Lv},\ and\ \citenamefont {Yuan}}]{https://doi.org/10.1002/wcms.70020}%
  \BibitemOpen
  \bibfield  {author} {\bibinfo {author} {\bibfnamefont {Y.}~\bibnamefont {Zhang}}, \bibinfo {author} {\bibfnamefont {X.}~\bibnamefont {Zhang}}, \bibinfo {author} {\bibfnamefont {J.}~\bibnamefont {Sun}}, \bibinfo {author} {\bibfnamefont {H.}~\bibnamefont {Lin}}, \bibinfo {author} {\bibfnamefont {Y.}~\bibnamefont {Huang}}, \bibinfo {author} {\bibfnamefont {D.}~\bibnamefont {Lv}}, \ and\ \bibinfo {author} {\bibfnamefont {X.}~\bibnamefont {Yuan}},\ }\href {\doibase https://doi.org/10.1002/wcms.70020} {\bibfield  {journal} {\bibinfo  {journal} {WIREs Computational Molecular Science}\ }\textbf {\bibinfo {volume} {15}},\ \bibinfo {pages} {e70020} (\bibinfo {year} {2025}{\natexlab{b}})},\ \bibinfo {note} {e70020 CMS-1125.R1}\BibitemShut {NoStop}%
\bibitem [{\citenamefont {Cerezo}\ \emph {et~al.}(2022)\citenamefont {Cerezo}, \citenamefont {Verdon}, \citenamefont {Huang}, \citenamefont {Cincio},\ and\ \citenamefont {Coles}}]{cerezo2022challenges}%
  \BibitemOpen
  \bibfield  {author} {\bibinfo {author} {\bibfnamefont {M.}~\bibnamefont {Cerezo}}, \bibinfo {author} {\bibfnamefont {G.}~\bibnamefont {Verdon}}, \bibinfo {author} {\bibfnamefont {H.-Y.}\ \bibnamefont {Huang}}, \bibinfo {author} {\bibfnamefont {L.}~\bibnamefont {Cincio}}, \ and\ \bibinfo {author} {\bibfnamefont {P.~J.}\ \bibnamefont {Coles}},\ }\href {https://www.nature.com/articles/s43588-022-00311-3} {\bibfield  {journal} {\bibinfo  {journal} {Nature computational science}\ }\textbf {\bibinfo {volume} {2}},\ \bibinfo {pages} {567} (\bibinfo {year} {2022})}\BibitemShut {NoStop}%
\bibitem [{\citenamefont {Childs}\ \emph {et~al.}(2018)\citenamefont {Childs}, \citenamefont {Maslov}, \citenamefont {Nam}, \citenamefont {Ross},\ and\ \citenamefont {Su}}]{doi:10.1073/pnas.1801723115}%
  \BibitemOpen
  \bibfield  {author} {\bibinfo {author} {\bibfnamefont {A.~M.}\ \bibnamefont {Childs}}, \bibinfo {author} {\bibfnamefont {D.}~\bibnamefont {Maslov}}, \bibinfo {author} {\bibfnamefont {Y.}~\bibnamefont {Nam}}, \bibinfo {author} {\bibfnamefont {N.~J.}\ \bibnamefont {Ross}}, \ and\ \bibinfo {author} {\bibfnamefont {Y.}~\bibnamefont {Su}},\ }\href {\doibase 10.1073/pnas.1801723115} {\bibfield  {journal} {\bibinfo  {journal} {Proceedings of the National Academy of Sciences}\ }\textbf {\bibinfo {volume} {115}},\ \bibinfo {pages} {9456} (\bibinfo {year} {2018})}\BibitemShut {NoStop}%
\bibitem [{\citenamefont {Daley}\ \emph {et~al.}(2022{\natexlab{b}})\citenamefont {Daley}, \citenamefont {Bloch}, \citenamefont {Kokail}, \citenamefont {Flannigan}, \citenamefont {Pearson}, \citenamefont {Troyer},\ and\ \citenamefont {Zoller}}]{Daley2022}%
  \BibitemOpen
  \bibfield  {author} {\bibinfo {author} {\bibfnamefont {A.~J.}\ \bibnamefont {Daley}}, \bibinfo {author} {\bibfnamefont {I.}~\bibnamefont {Bloch}}, \bibinfo {author} {\bibfnamefont {C.}~\bibnamefont {Kokail}}, \bibinfo {author} {\bibfnamefont {S.}~\bibnamefont {Flannigan}}, \bibinfo {author} {\bibfnamefont {N.}~\bibnamefont {Pearson}}, \bibinfo {author} {\bibfnamefont {M.}~\bibnamefont {Troyer}}, \ and\ \bibinfo {author} {\bibfnamefont {P.}~\bibnamefont {Zoller}},\ }\href {\doibase 10.1038/s41586-022-04940-6} {\bibfield  {journal} {\bibinfo  {journal} {Nature}\ }\textbf {\bibinfo {volume} {607}},\ \bibinfo {pages} {667} (\bibinfo {year} {2022}{\natexlab{b}})}\BibitemShut {NoStop}%
\bibitem [{\citenamefont {{IBM}}(2016)}]{IBM2016QuantumCloud}%
  \BibitemOpen
  \bibfield  {author} {\bibinfo {author} {\bibnamefont {{IBM}}},\ }\href {https://uk.newsroom.ibm.com/2016-May-04-IBM-Makes-Quantum-Computing-Available-on-IBM-Cloud-to-Accelerate-Innovation} {\enquote {\bibinfo {title} {{IBM Makes Quantum Computing Available on IBM Cloud to Accelerate Innovation}},}\ }\bibinfo {howpublished} {IBM Newsroom} (\bibinfo {year} {2016}),\ \bibinfo {note} {acesso em: 10 de jan. 2025}\BibitemShut {NoStop}%
\bibitem [{\citenamefont {Gebheim}(2021)}]{gebheim2021state}%
  \BibitemOpen
  \bibfield  {author} {\bibinfo {author} {\bibfnamefont {M.}~\bibnamefont {Gebheim}},\ }\href {https://www.bench.com/setting-the-benchmark/the-state-of-the-qubit-and-quantum-computing} {\enquote {\bibinfo {title} {The state of the qubit and quantum computing},}\ } (\bibinfo {year} {2021}),\ \bibinfo {note} {accessed: 2025-05-01}\BibitemShut {NoStop}%
\bibitem [{\citenamefont {SPINQ}(2025)}]{spinquanta2025top}%
  \BibitemOpen
  \bibfield  {author} {\bibinfo {author} {\bibnamefont {SPINQ}},\ }\href {https://www.spinquanta.com/news-detail/top-companies-working-on-quantum-computers-updated20250114101321} {\enquote {\bibinfo {title} {Top 11 companies working on quantum computers [2025 updated]},}\ } (\bibinfo {year} {2025}),\ \bibinfo {note} {accessed: 2025-05-01}\BibitemShut {NoStop}%
\bibitem [{\citenamefont {PRESKILL}()}]{PRESKILL1998}%
  \BibitemOpen
  \bibfield  {author} {\bibinfo {author} {\bibfnamefont {J.}~\bibnamefont {PRESKILL}},\ }\enquote {\bibinfo {title} {Fault-tolerant quantum computation},}\ in\ \href {\doibase 10.1142/9789812385253_0008} {\emph {\bibinfo {booktitle} {Introduction to Quantum Computation and Information}}},\ pp.\ \bibinfo {pages} {213--269}\BibitemShut {NoStop}%
\bibitem [{\citenamefont {Shor}(1996)}]{548464}%
  \BibitemOpen
  \bibfield  {author} {\bibinfo {author} {\bibfnamefont {P.}~\bibnamefont {Shor}},\ }in\ \href {\doibase 10.1109/SFCS.1996.548464} {\emph {\bibinfo {booktitle} {Proceedings of 37th Conference on Foundations of Computer Science}}}\ (\bibinfo {year} {1996})\ pp.\ \bibinfo {pages} {56--65}\BibitemShut {NoStop}%
\bibitem [{\citenamefont {Aharonov}\ and\ \citenamefont {Ben-Or}(1997)}]{aharonov1997fault}%
  \BibitemOpen
  \bibfield  {author} {\bibinfo {author} {\bibfnamefont {D.}~\bibnamefont {Aharonov}}\ and\ \bibinfo {author} {\bibfnamefont {M.}~\bibnamefont {Ben-Or}},\ }in\ \href@noop {} {\emph {\bibinfo {booktitle} {Proceedings of the twenty-ninth annual ACM symposium on Theory of computing}}}\ (\bibinfo {year} {1997})\ pp.\ \bibinfo {pages} {176--188}\BibitemShut {NoStop}%
\bibitem [{\citenamefont {Gottesman}(1998)}]{PhysRevA.57.127}%
  \BibitemOpen
  \bibfield  {author} {\bibinfo {author} {\bibfnamefont {D.}~\bibnamefont {Gottesman}},\ }\href {\doibase 10.1103/PhysRevA.57.127} {\bibfield  {journal} {\bibinfo  {journal} {Phys. Rev. A}\ }\textbf {\bibinfo {volume} {57}},\ \bibinfo {pages} {127} (\bibinfo {year} {1998})}\BibitemShut {NoStop}%
\bibitem [{\citenamefont {Fellous-Asiani}\ \emph {et~al.}(2021)\citenamefont {Fellous-Asiani}, \citenamefont {Chai}, \citenamefont {Whitney}, \citenamefont {Auff\`eves},\ and\ \citenamefont {Ng}}]{PRXQuantum.2.040335}%
  \BibitemOpen
  \bibfield  {author} {\bibinfo {author} {\bibfnamefont {M.}~\bibnamefont {Fellous-Asiani}}, \bibinfo {author} {\bibfnamefont {J.~H.}\ \bibnamefont {Chai}}, \bibinfo {author} {\bibfnamefont {R.~S.}\ \bibnamefont {Whitney}}, \bibinfo {author} {\bibfnamefont {A.}~\bibnamefont {Auff\`eves}}, \ and\ \bibinfo {author} {\bibfnamefont {H.~K.}\ \bibnamefont {Ng}},\ }\href {\doibase 10.1103/PRXQuantum.2.040335} {\bibfield  {journal} {\bibinfo  {journal} {PRX Quantum}\ }\textbf {\bibinfo {volume} {2}},\ \bibinfo {pages} {040335} (\bibinfo {year} {2021})}\BibitemShut {NoStop}%
\bibitem [{\citenamefont {King}\ \emph {et~al.}(2025)\citenamefont {King}, \citenamefont {Nocera}, \citenamefont {Rams}, \citenamefont {Dziarmaga}, \citenamefont {Wiersema}, \citenamefont {Bernoudy}, \citenamefont {Raymond}, \citenamefont {Kaushal}, \citenamefont {Heinsdorf}, \citenamefont {Harris}, \citenamefont {Boothby}, \citenamefont {Altomare}, \citenamefont {Asad}, \citenamefont {Berkley}, \citenamefont {Boschnak}, \citenamefont {Chern}, \citenamefont {Christiani}, \citenamefont {Cibere}, \citenamefont {Connor}, \citenamefont {Dehn}, \citenamefont {Deshpande}, \citenamefont {Ejtemaee}, \citenamefont {Farre}, \citenamefont {Hamer}, \citenamefont {Hoskinson}, \citenamefont {Huang}, \citenamefont {Johnson}, \citenamefont {Kortas}, \citenamefont {Ladizinsky}, \citenamefont {Lanting}, \citenamefont {Lai}, \citenamefont {Li}, \citenamefont {MacDonald}, \citenamefont {Marsden}, \citenamefont {McGeoch}, \citenamefont {Molavi}, \citenamefont {Oh}, \citenamefont {Neufeld}, \citenamefont {Norouzpour},
  \citenamefont {Pasvolsky}, \citenamefont {Poitras}, \citenamefont {Poulin-Lamarre}, \citenamefont {Prescott}, \citenamefont {Reis}, \citenamefont {Rich}, \citenamefont {Samani}, \citenamefont {Sheldan}, \citenamefont {Smirnov}, \citenamefont {Sterpka}, \citenamefont {Clavera}, \citenamefont {Tsai}, \citenamefont {Volkmann}, \citenamefont {Whiticar}, \citenamefont {Whittaker}, \citenamefont {Wilkinson}, \citenamefont {Yao}, \citenamefont {Yi}, \citenamefont {Sandvik}, \citenamefont {Alvarez}, \citenamefont {Melko}, \citenamefont {Carrasquilla}, \citenamefont {Franz},\ and\ \citenamefont {Amin}}]{doi:10.1126/science.ado6285}%
  \BibitemOpen
  \bibfield  {author} {\bibinfo {author} {\bibfnamefont {A.~D.}\ \bibnamefont {King}}, \bibinfo {author} {\bibfnamefont {A.}~\bibnamefont {Nocera}}, \bibinfo {author} {\bibfnamefont {M.~M.}\ \bibnamefont {Rams}}, \bibinfo {author} {\bibfnamefont {J.}~\bibnamefont {Dziarmaga}}, \bibinfo {author} {\bibfnamefont {R.}~\bibnamefont {Wiersema}}, \bibinfo {author} {\bibfnamefont {W.}~\bibnamefont {Bernoudy}}, \bibinfo {author} {\bibfnamefont {J.}~\bibnamefont {Raymond}}, \bibinfo {author} {\bibfnamefont {N.}~\bibnamefont {Kaushal}}, \bibinfo {author} {\bibfnamefont {N.}~\bibnamefont {Heinsdorf}}, \bibinfo {author} {\bibfnamefont {R.}~\bibnamefont {Harris}}, \bibinfo {author} {\bibfnamefont {K.}~\bibnamefont {Boothby}}, \bibinfo {author} {\bibfnamefont {F.}~\bibnamefont {Altomare}}, \bibinfo {author} {\bibfnamefont {M.}~\bibnamefont {Asad}}, \bibinfo {author} {\bibfnamefont {A.~J.}\ \bibnamefont {Berkley}}, \bibinfo {author} {\bibfnamefont {M.}~\bibnamefont {Boschnak}}, \bibinfo {author} {\bibfnamefont
  {K.}~\bibnamefont {Chern}}, \bibinfo {author} {\bibfnamefont {H.}~\bibnamefont {Christiani}}, \bibinfo {author} {\bibfnamefont {S.}~\bibnamefont {Cibere}}, \bibinfo {author} {\bibfnamefont {J.}~\bibnamefont {Connor}}, \bibinfo {author} {\bibfnamefont {M.~H.}\ \bibnamefont {Dehn}}, \bibinfo {author} {\bibfnamefont {R.}~\bibnamefont {Deshpande}}, \bibinfo {author} {\bibfnamefont {S.}~\bibnamefont {Ejtemaee}}, \bibinfo {author} {\bibfnamefont {P.}~\bibnamefont {Farre}}, \bibinfo {author} {\bibfnamefont {K.}~\bibnamefont {Hamer}}, \bibinfo {author} {\bibfnamefont {E.}~\bibnamefont {Hoskinson}}, \bibinfo {author} {\bibfnamefont {S.}~\bibnamefont {Huang}}, \bibinfo {author} {\bibfnamefont {M.~W.}\ \bibnamefont {Johnson}}, \bibinfo {author} {\bibfnamefont {S.}~\bibnamefont {Kortas}}, \bibinfo {author} {\bibfnamefont {E.}~\bibnamefont {Ladizinsky}}, \bibinfo {author} {\bibfnamefont {T.}~\bibnamefont {Lanting}}, \bibinfo {author} {\bibfnamefont {T.}~\bibnamefont {Lai}}, \bibinfo {author} {\bibfnamefont
  {R.}~\bibnamefont {Li}}, \bibinfo {author} {\bibfnamefont {A.~J.~R.}\ \bibnamefont {MacDonald}}, \bibinfo {author} {\bibfnamefont {G.}~\bibnamefont {Marsden}}, \bibinfo {author} {\bibfnamefont {C.~C.}\ \bibnamefont {McGeoch}}, \bibinfo {author} {\bibfnamefont {R.}~\bibnamefont {Molavi}}, \bibinfo {author} {\bibfnamefont {T.}~\bibnamefont {Oh}}, \bibinfo {author} {\bibfnamefont {R.}~\bibnamefont {Neufeld}}, \bibinfo {author} {\bibfnamefont {M.}~\bibnamefont {Norouzpour}}, \bibinfo {author} {\bibfnamefont {J.}~\bibnamefont {Pasvolsky}}, \bibinfo {author} {\bibfnamefont {P.}~\bibnamefont {Poitras}}, \bibinfo {author} {\bibfnamefont {G.}~\bibnamefont {Poulin-Lamarre}}, \bibinfo {author} {\bibfnamefont {T.}~\bibnamefont {Prescott}}, \bibinfo {author} {\bibfnamefont {M.}~\bibnamefont {Reis}}, \bibinfo {author} {\bibfnamefont {C.}~\bibnamefont {Rich}}, \bibinfo {author} {\bibfnamefont {M.}~\bibnamefont {Samani}}, \bibinfo {author} {\bibfnamefont {B.}~\bibnamefont {Sheldan}}, \bibinfo {author} {\bibfnamefont
  {A.}~\bibnamefont {Smirnov}}, \bibinfo {author} {\bibfnamefont {E.}~\bibnamefont {Sterpka}}, \bibinfo {author} {\bibfnamefont {B.~T.}\ \bibnamefont {Clavera}}, \bibinfo {author} {\bibfnamefont {N.}~\bibnamefont {Tsai}}, \bibinfo {author} {\bibfnamefont {M.}~\bibnamefont {Volkmann}}, \bibinfo {author} {\bibfnamefont {A.~M.}\ \bibnamefont {Whiticar}}, \bibinfo {author} {\bibfnamefont {J.~D.}\ \bibnamefont {Whittaker}}, \bibinfo {author} {\bibfnamefont {W.}~\bibnamefont {Wilkinson}}, \bibinfo {author} {\bibfnamefont {J.}~\bibnamefont {Yao}}, \bibinfo {author} {\bibfnamefont {T.~J.}\ \bibnamefont {Yi}}, \bibinfo {author} {\bibfnamefont {A.~W.}\ \bibnamefont {Sandvik}}, \bibinfo {author} {\bibfnamefont {G.}~\bibnamefont {Alvarez}}, \bibinfo {author} {\bibfnamefont {R.~G.}\ \bibnamefont {Melko}}, \bibinfo {author} {\bibfnamefont {J.}~\bibnamefont {Carrasquilla}}, \bibinfo {author} {\bibfnamefont {M.}~\bibnamefont {Franz}}, \ and\ \bibinfo {author} {\bibfnamefont {M.~H.}\ \bibnamefont {Amin}},\ }\href {\doibase
  10.1126/science.ado6285} {\bibfield  {journal} {\bibinfo  {journal} {Science}\ }\textbf {\bibinfo {volume} {388}},\ \bibinfo {pages} {199} (\bibinfo {year} {2025})},\ \Eprint {http://arxiv.org/abs/https://www.science.org/doi/pdf/10.1126/science.ado6285} {https://www.science.org/doi/pdf/10.1126/science.ado6285} \BibitemShut {NoStop}%
\bibitem [{\citenamefont {Combarro}\ \emph {et~al.}(2023)\citenamefont {Combarro}, \citenamefont {Gonz{\'a}lez-Castillo},\ and\ \citenamefont {Di~Meglio}}]{combarro2023practical}%
  \BibitemOpen
  \bibfield  {author} {\bibinfo {author} {\bibfnamefont {E.~F.}\ \bibnamefont {Combarro}}, \bibinfo {author} {\bibfnamefont {S.}~\bibnamefont {Gonz{\'a}lez-Castillo}}, \ and\ \bibinfo {author} {\bibfnamefont {A.}~\bibnamefont {Di~Meglio}},\ }\href@noop {} {\emph {\bibinfo {title} {A practical guide to quantum machine learning and quantum optimization: Hands-on approach to modern quantum algorithms}}}\ (\bibinfo  {publisher} {Packt Publishing Ltd},\ \bibinfo {year} {2023})\BibitemShut {NoStop}%
\bibitem [{\citenamefont {Yuan}\ \emph {et~al.}(2019)\citenamefont {Yuan}, \citenamefont {Endo}, \citenamefont {Zhao}, \citenamefont {Li},\ and\ \citenamefont {Benjamin}}]{yuan2019theory}%
  \BibitemOpen
  \bibfield  {author} {\bibinfo {author} {\bibfnamefont {X.}~\bibnamefont {Yuan}}, \bibinfo {author} {\bibfnamefont {S.}~\bibnamefont {Endo}}, \bibinfo {author} {\bibfnamefont {Q.}~\bibnamefont {Zhao}}, \bibinfo {author} {\bibfnamefont {Y.}~\bibnamefont {Li}}, \ and\ \bibinfo {author} {\bibfnamefont {S.~C.}\ \bibnamefont {Benjamin}},\ }\href@noop {} {\bibfield  {journal} {\bibinfo  {journal} {Quantum}\ }\textbf {\bibinfo {volume} {3}},\ \bibinfo {pages} {191} (\bibinfo {year} {2019})}\BibitemShut {NoStop}%
\bibitem [{\citenamefont {Holmes}\ \emph {et~al.}(2022)\citenamefont {Holmes}, \citenamefont {Sharma}, \citenamefont {Cerezo},\ and\ \citenamefont {Coles}}]{PRXQuantum.3.010313}%
  \BibitemOpen
  \bibfield  {author} {\bibinfo {author} {\bibfnamefont {Z.}~\bibnamefont {Holmes}}, \bibinfo {author} {\bibfnamefont {K.}~\bibnamefont {Sharma}}, \bibinfo {author} {\bibfnamefont {M.}~\bibnamefont {Cerezo}}, \ and\ \bibinfo {author} {\bibfnamefont {P.~J.}\ \bibnamefont {Coles}},\ }\href {\doibase 10.1103/PRXQuantum.3.010313} {\bibfield  {journal} {\bibinfo  {journal} {PRX Quantum}\ }\textbf {\bibinfo {volume} {3}},\ \bibinfo {pages} {010313} (\bibinfo {year} {2022})}\BibitemShut {NoStop}%
\bibitem [{\citenamefont {Choquette}\ \emph {et~al.}(2021)\citenamefont {Choquette}, \citenamefont {Di~Paolo}, \citenamefont {Barkoutsos}, \citenamefont {S\'en\'echal}, \citenamefont {Tavernelli},\ and\ \citenamefont {Blais}}]{PhysRevResearch.3.023092}%
  \BibitemOpen
  \bibfield  {author} {\bibinfo {author} {\bibfnamefont {A.}~\bibnamefont {Choquette}}, \bibinfo {author} {\bibfnamefont {A.}~\bibnamefont {Di~Paolo}}, \bibinfo {author} {\bibfnamefont {P.~K.}\ \bibnamefont {Barkoutsos}}, \bibinfo {author} {\bibfnamefont {D.}~\bibnamefont {S\'en\'echal}}, \bibinfo {author} {\bibfnamefont {I.}~\bibnamefont {Tavernelli}}, \ and\ \bibinfo {author} {\bibfnamefont {A.}~\bibnamefont {Blais}},\ }\href {\doibase 10.1103/PhysRevResearch.3.023092} {\bibfield  {journal} {\bibinfo  {journal} {Phys. Rev. Res.}\ }\textbf {\bibinfo {volume} {3}},\ \bibinfo {pages} {023092} (\bibinfo {year} {2021})}\BibitemShut {NoStop}%
\bibitem [{\citenamefont {Lee}\ \emph {et~al.}(2018)\citenamefont {Lee}, \citenamefont {Huggins}, \citenamefont {Head-Gordon},\ and\ \citenamefont {Whaley}}]{lee2018generalized}%
  \BibitemOpen
  \bibfield  {author} {\bibinfo {author} {\bibfnamefont {J.}~\bibnamefont {Lee}}, \bibinfo {author} {\bibfnamefont {W.~J.}\ \bibnamefont {Huggins}}, \bibinfo {author} {\bibfnamefont {M.}~\bibnamefont {Head-Gordon}}, \ and\ \bibinfo {author} {\bibfnamefont {K.~B.}\ \bibnamefont {Whaley}},\ }\href@noop {} {\bibfield  {journal} {\bibinfo  {journal} {Journal of chemical theory and computation}\ }\textbf {\bibinfo {volume} {15}},\ \bibinfo {pages} {311} (\bibinfo {year} {2018})}\BibitemShut {NoStop}%
\bibitem [{\citenamefont {Wiersema}\ \emph {et~al.}(2020)\citenamefont {Wiersema}, \citenamefont {Zhou}, \citenamefont {de~Sereville}, \citenamefont {Carrasquilla}, \citenamefont {Kim},\ and\ \citenamefont {Yuen}}]{PRXQuantum.1.020319}%
  \BibitemOpen
  \bibfield  {author} {\bibinfo {author} {\bibfnamefont {R.}~\bibnamefont {Wiersema}}, \bibinfo {author} {\bibfnamefont {C.}~\bibnamefont {Zhou}}, \bibinfo {author} {\bibfnamefont {Y.}~\bibnamefont {de~Sereville}}, \bibinfo {author} {\bibfnamefont {J.~F.}\ \bibnamefont {Carrasquilla}}, \bibinfo {author} {\bibfnamefont {Y.~B.}\ \bibnamefont {Kim}}, \ and\ \bibinfo {author} {\bibfnamefont {H.}~\bibnamefont {Yuen}},\ }\href {\doibase 10.1103/PRXQuantum.1.020319} {\bibfield  {journal} {\bibinfo  {journal} {PRX Quantum}\ }\textbf {\bibinfo {volume} {1}},\ \bibinfo {pages} {020319} (\bibinfo {year} {2020})}\BibitemShut {NoStop}%
\bibitem [{\citenamefont {Park}\ and\ \citenamefont {Killoran}(2024)}]{park2024hamiltonian}%
  \BibitemOpen
  \bibfield  {author} {\bibinfo {author} {\bibfnamefont {C.-Y.}\ \bibnamefont {Park}}\ and\ \bibinfo {author} {\bibfnamefont {N.}~\bibnamefont {Killoran}},\ }\href@noop {} {\bibfield  {journal} {\bibinfo  {journal} {Quantum}\ }\textbf {\bibinfo {volume} {8}},\ \bibinfo {pages} {1239} (\bibinfo {year} {2024})}\BibitemShut {NoStop}%
\bibitem [{\citenamefont {Leone}\ \emph {et~al.}(2024)\citenamefont {Leone}, \citenamefont {Oliviero}, \citenamefont {Cincio},\ and\ \citenamefont {Cerezo}}]{leone2022practical}%
  \BibitemOpen
  \bibfield  {author} {\bibinfo {author} {\bibfnamefont {L.}~\bibnamefont {Leone}}, \bibinfo {author} {\bibfnamefont {S.~F.}\ \bibnamefont {Oliviero}}, \bibinfo {author} {\bibfnamefont {L.}~\bibnamefont {Cincio}}, \ and\ \bibinfo {author} {\bibfnamefont {M.}~\bibnamefont {Cerezo}},\ }\href {\doibase 10.22331/q-2024-07-03-1395} {\bibfield  {journal} {\bibinfo  {journal} {{Quantum}}\ }\textbf {\bibinfo {volume} {8}},\ \bibinfo {pages} {1395} (\bibinfo {year} {2024})}\BibitemShut {NoStop}%
\bibitem [{\citenamefont {Kandala}\ \emph {et~al.}(2017)\citenamefont {Kandala}, \citenamefont {Mezzacapo}, \citenamefont {Temme}, \citenamefont {Takita}, \citenamefont {Brink}, \citenamefont {Chow},\ and\ \citenamefont {Gambetta}}]{kandala2017hardware}%
  \BibitemOpen
  \bibfield  {author} {\bibinfo {author} {\bibfnamefont {A.}~\bibnamefont {Kandala}}, \bibinfo {author} {\bibfnamefont {A.}~\bibnamefont {Mezzacapo}}, \bibinfo {author} {\bibfnamefont {K.}~\bibnamefont {Temme}}, \bibinfo {author} {\bibfnamefont {M.}~\bibnamefont {Takita}}, \bibinfo {author} {\bibfnamefont {M.}~\bibnamefont {Brink}}, \bibinfo {author} {\bibfnamefont {J.~M.}\ \bibnamefont {Chow}}, \ and\ \bibinfo {author} {\bibfnamefont {J.~M.}\ \bibnamefont {Gambetta}},\ }\href@noop {} {\bibfield  {journal} {\bibinfo  {journal} {nature}\ }\textbf {\bibinfo {volume} {549}},\ \bibinfo {pages} {242} (\bibinfo {year} {2017})}\BibitemShut {NoStop}%
\bibitem [{\citenamefont {Qin}(2023)}]{Qin_2023}%
  \BibitemOpen
  \bibfield  {author} {\bibinfo {author} {\bibfnamefont {J.}~\bibnamefont {Qin}},\ }\href {\doibase 10.1088/1742-6596/2634/1/012043} {\bibfield  {journal} {\bibinfo  {journal} {Journal of Physics: Conference Series}\ }\textbf {\bibinfo {volume} {2634}},\ \bibinfo {pages} {012043} (\bibinfo {year} {2023})}\BibitemShut {NoStop}%
\bibitem [{\citenamefont {Cerezo}\ \emph {et~al.}(2021{\natexlab{b}})\citenamefont {Cerezo}, \citenamefont {Sone}, \citenamefont {Volkoff}, \citenamefont {Cincio},\ and\ \citenamefont {Coles}}]{cerezo2021cost}%
  \BibitemOpen
  \bibfield  {author} {\bibinfo {author} {\bibfnamefont {M.}~\bibnamefont {Cerezo}}, \bibinfo {author} {\bibfnamefont {A.}~\bibnamefont {Sone}}, \bibinfo {author} {\bibfnamefont {T.}~\bibnamefont {Volkoff}}, \bibinfo {author} {\bibfnamefont {L.}~\bibnamefont {Cincio}}, \ and\ \bibinfo {author} {\bibfnamefont {P.~J.}\ \bibnamefont {Coles}},\ }\href@noop {} {\bibfield  {journal} {\bibinfo  {journal} {Nature communications}\ }\textbf {\bibinfo {volume} {12}},\ \bibinfo {pages} {1791} (\bibinfo {year} {2021}{\natexlab{b}})}\BibitemShut {NoStop}%
\bibitem [{\citenamefont {Guerreschi}\ and\ \citenamefont {Smelyanskiy}(2017)}]{guerreschi2017practical}%
  \BibitemOpen
  \bibfield  {author} {\bibinfo {author} {\bibfnamefont {G.~G.}\ \bibnamefont {Guerreschi}}\ and\ \bibinfo {author} {\bibfnamefont {M.}~\bibnamefont {Smelyanskiy}},\ }\href@noop {} {\bibfield  {journal} {\bibinfo  {journal} {arXiv preprint arXiv:1701.01450}\ } (\bibinfo {year} {2017})}\BibitemShut {NoStop}%
\bibitem [{\citenamefont {Mitarai}\ \emph {et~al.}(2018)\citenamefont {Mitarai}, \citenamefont {Negoro}, \citenamefont {Kitagawa},\ and\ \citenamefont {Fujii}}]{mitarai2018quantum}%
  \BibitemOpen
  \bibfield  {author} {\bibinfo {author} {\bibfnamefont {K.}~\bibnamefont {Mitarai}}, \bibinfo {author} {\bibfnamefont {M.}~\bibnamefont {Negoro}}, \bibinfo {author} {\bibfnamefont {M.}~\bibnamefont {Kitagawa}}, \ and\ \bibinfo {author} {\bibfnamefont {K.}~\bibnamefont {Fujii}},\ }\href@noop {} {\bibfield  {journal} {\bibinfo  {journal} {Physical Review A}\ }\textbf {\bibinfo {volume} {98}},\ \bibinfo {pages} {032309} (\bibinfo {year} {2018})}\BibitemShut {NoStop}%
\bibitem [{\citenamefont {Schuld}\ \emph {et~al.}(2019)\citenamefont {Schuld}, \citenamefont {Bergholm}, \citenamefont {Gogolin}, \citenamefont {Izaac},\ and\ \citenamefont {Killoran}}]{schuld2019evaluating}%
  \BibitemOpen
  \bibfield  {author} {\bibinfo {author} {\bibfnamefont {M.}~\bibnamefont {Schuld}}, \bibinfo {author} {\bibfnamefont {V.}~\bibnamefont {Bergholm}}, \bibinfo {author} {\bibfnamefont {C.}~\bibnamefont {Gogolin}}, \bibinfo {author} {\bibfnamefont {J.}~\bibnamefont {Izaac}}, \ and\ \bibinfo {author} {\bibfnamefont {N.}~\bibnamefont {Killoran}},\ }\href@noop {} {\bibfield  {journal} {\bibinfo  {journal} {Physical Review A}\ }\textbf {\bibinfo {volume} {99}},\ \bibinfo {pages} {032331} (\bibinfo {year} {2019})}\BibitemShut {NoStop}%
\bibitem [{\citenamefont {Kölle}\ \emph {et~al.}(2024)\citenamefont {Kölle}, \citenamefont {Witter}, \citenamefont {Rohe}, \citenamefont {Stenzel}, \citenamefont {Altmann},\ and\ \citenamefont {Gabor}}]{10646539}%
  \BibitemOpen
  \bibfield  {author} {\bibinfo {author} {\bibfnamefont {M.}~\bibnamefont {Kölle}}, \bibinfo {author} {\bibfnamefont {T.}~\bibnamefont {Witter}}, \bibinfo {author} {\bibfnamefont {T.}~\bibnamefont {Rohe}}, \bibinfo {author} {\bibfnamefont {G.}~\bibnamefont {Stenzel}}, \bibinfo {author} {\bibfnamefont {P.}~\bibnamefont {Altmann}}, \ and\ \bibinfo {author} {\bibfnamefont {T.}~\bibnamefont {Gabor}},\ }in\ \href {\doibase 10.1109/QSW62656.2024.00031} {\emph {\bibinfo {booktitle} {2024 IEEE International Conference on Quantum Software (QSW)}}}\ (\bibinfo {year} {2024})\ pp.\ \bibinfo {pages} {157--167}\BibitemShut {NoStop}%
\bibitem [{\citenamefont {Bressert}(2012)}]{bressert2012scipy}%
  \BibitemOpen
  \bibfield  {author} {\bibinfo {author} {\bibfnamefont {E.}~\bibnamefont {Bressert}},\ }\href@noop {} {\  (\bibinfo {year} {2012})}\BibitemShut {NoStop}%
\bibitem [{\citenamefont {Powell}(1994)}]{powell1994advances}%
  \BibitemOpen
  \bibfield  {author} {\bibinfo {author} {\bibfnamefont {M.}~\bibnamefont {Powell}},\ }in\ \href@noop {} {\emph {\bibinfo {booktitle} {Proceeding of the 6th Workshop on Optimization and Numerical Analysis}}}\ (\bibinfo {year} {1994})\ pp.\ \bibinfo {pages} {5--67}\BibitemShut {NoStop}%
\bibitem [{\citenamefont {Cheng}\ \emph {et~al.}(2024)\citenamefont {Cheng}, \citenamefont {Chen}, \citenamefont {Zhang},\ and\ \citenamefont {Zhang}}]{cheng2024quantum}%
  \BibitemOpen
  \bibfield  {author} {\bibinfo {author} {\bibfnamefont {L.}~\bibnamefont {Cheng}}, \bibinfo {author} {\bibfnamefont {Y.-Q.}\ \bibnamefont {Chen}}, \bibinfo {author} {\bibfnamefont {S.-X.}\ \bibnamefont {Zhang}}, \ and\ \bibinfo {author} {\bibfnamefont {S.}~\bibnamefont {Zhang}},\ }\href@noop {} {\bibfield  {journal} {\bibinfo  {journal} {Communications Physics}\ }\textbf {\bibinfo {volume} {7}},\ \bibinfo {pages} {83} (\bibinfo {year} {2024})}\BibitemShut {NoStop}%
\bibitem [{\citenamefont {Singh}\ \emph {et~al.}(2023)\citenamefont {Singh}, \citenamefont {Majumder},\ and\ \citenamefont {Mishra}}]{singh2023benchmarking}%
  \BibitemOpen
  \bibfield  {author} {\bibinfo {author} {\bibfnamefont {H.}~\bibnamefont {Singh}}, \bibinfo {author} {\bibfnamefont {S.}~\bibnamefont {Majumder}}, \ and\ \bibinfo {author} {\bibfnamefont {S.}~\bibnamefont {Mishra}},\ }\href@noop {} {\bibfield  {journal} {\bibinfo  {journal} {The Journal of Chemical Physics}\ }\textbf {\bibinfo {volume} {159}} (\bibinfo {year} {2023})}\BibitemShut {NoStop}%
\bibitem [{\citenamefont {Pellow-Jarman}\ \emph {et~al.}(2021)\citenamefont {Pellow-Jarman}, \citenamefont {Sinayskiy}, \citenamefont {Pillay},\ and\ \citenamefont {Petruccione}}]{pellow2021comparison}%
  \BibitemOpen
  \bibfield  {author} {\bibinfo {author} {\bibfnamefont {A.}~\bibnamefont {Pellow-Jarman}}, \bibinfo {author} {\bibfnamefont {I.}~\bibnamefont {Sinayskiy}}, \bibinfo {author} {\bibfnamefont {A.}~\bibnamefont {Pillay}}, \ and\ \bibinfo {author} {\bibfnamefont {F.}~\bibnamefont {Petruccione}},\ }\href@noop {} {\bibfield  {journal} {\bibinfo  {journal} {Quantum Information Processing}\ }\textbf {\bibinfo {volume} {20}},\ \bibinfo {pages} {202} (\bibinfo {year} {2021})}\BibitemShut {NoStop}%
\bibitem [{\citenamefont {Xie}\ \emph {et~al.}(2019)\citenamefont {Xie}, \citenamefont {Byrd},\ and\ \citenamefont {Nocedal}}]{Xie2019Analysis}%
  \BibitemOpen
  \bibfield  {author} {\bibinfo {author} {\bibfnamefont {Y.}~\bibnamefont {Xie}}, \bibinfo {author} {\bibfnamefont {R.}~\bibnamefont {Byrd}}, \ and\ \bibinfo {author} {\bibfnamefont {J.}~\bibnamefont {Nocedal}},\ }\href {\doibase 10.1137/19m1240794} {\bibfield  {journal} {\bibinfo  {journal} {SIAM J. Optim.}\ }\textbf {\bibinfo {volume} {30}},\ \bibinfo {pages} {182} (\bibinfo {year} {2019})}\BibitemShut {NoStop}%
\bibitem [{\citenamefont {Berahas}\ and\ \citenamefont {Takác}(2017)}]{Berahas2017A}%
  \BibitemOpen
  \bibfield  {author} {\bibinfo {author} {\bibfnamefont {A.}~\bibnamefont {Berahas}}\ and\ \bibinfo {author} {\bibfnamefont {M.}~\bibnamefont {Takác}},\ }\href {\doibase 10.1080/10556788.2019.1658107} {\bibfield  {journal} {\bibinfo  {journal} {Optimization Methods and Software}\ }\textbf {\bibinfo {volume} {35}},\ \bibinfo {pages} {191 } (\bibinfo {year} {2017})}\BibitemShut {NoStop}%
\bibitem [{\citenamefont {Sim}\ \emph {et~al.}(2019)\citenamefont {Sim}, \citenamefont {Johnson},\ and\ \citenamefont {Aspuru-Guzik}}]{Sim2019}%
  \BibitemOpen
  \bibfield  {author} {\bibinfo {author} {\bibfnamefont {S.}~\bibnamefont {Sim}}, \bibinfo {author} {\bibfnamefont {P.~D.}\ \bibnamefont {Johnson}}, \ and\ \bibinfo {author} {\bibfnamefont {A.}~\bibnamefont {Aspuru-Guzik}},\ }\href {\doibase https://doi.org/10.1002/qute.201900070} {\bibfield  {journal} {\bibinfo  {journal} {Advanced Quantum Technologies}\ }\textbf {\bibinfo {volume} {2}},\ \bibinfo {pages} {1900070} (\bibinfo {year} {2019})}\BibitemShut {NoStop}%
\bibitem [{\citenamefont {Akshay}\ \emph {et~al.}(2020)\citenamefont {Akshay}, \citenamefont {Philathong}, \citenamefont {Morales},\ and\ \citenamefont {Biamonte}}]{PhysRevLett.124.090504}%
  \BibitemOpen
  \bibfield  {author} {\bibinfo {author} {\bibfnamefont {V.}~\bibnamefont {Akshay}}, \bibinfo {author} {\bibfnamefont {H.}~\bibnamefont {Philathong}}, \bibinfo {author} {\bibfnamefont {M.~E.~S.}\ \bibnamefont {Morales}}, \ and\ \bibinfo {author} {\bibfnamefont {J.~D.}\ \bibnamefont {Biamonte}},\ }\href {\doibase 10.1103/PhysRevLett.124.090504} {\bibfield  {journal} {\bibinfo  {journal} {Phys. Rev. Lett.}\ }\textbf {\bibinfo {volume} {124}},\ \bibinfo {pages} {090504} (\bibinfo {year} {2020})}\BibitemShut {NoStop}%
\bibitem [{\citenamefont {Bittel}\ and\ \citenamefont {Kliesch}(2021)}]{PhysRevLett.127.120502}%
  \BibitemOpen
  \bibfield  {author} {\bibinfo {author} {\bibfnamefont {L.}~\bibnamefont {Bittel}}\ and\ \bibinfo {author} {\bibfnamefont {M.}~\bibnamefont {Kliesch}},\ }\href {\doibase 10.1103/PhysRevLett.127.120502} {\bibfield  {journal} {\bibinfo  {journal} {Phys. Rev. Lett.}\ }\textbf {\bibinfo {volume} {127}},\ \bibinfo {pages} {120502} (\bibinfo {year} {2021})}\BibitemShut {NoStop}%
\bibitem [{\citenamefont {Anschuetz}\ and\ \citenamefont {Kiani}(2022)}]{Anschuetz2022}%
  \BibitemOpen
  \bibfield  {author} {\bibinfo {author} {\bibfnamefont {E.~R.}\ \bibnamefont {Anschuetz}}\ and\ \bibinfo {author} {\bibfnamefont {B.~T.}\ \bibnamefont {Kiani}},\ }\href {\doibase 10.1038/s41467-022-35364-5} {\bibfield  {journal} {\bibinfo  {journal} {Nature Communications}\ }\textbf {\bibinfo {volume} {13}},\ \bibinfo {pages} {7760} (\bibinfo {year} {2022})}\BibitemShut {NoStop}%
\bibitem [{\citenamefont {Fontana}\ \emph {et~al.}(2022)\citenamefont {Fontana}, \citenamefont {Cerezo}, \citenamefont {Arrasmith}, \citenamefont {Rungger},\ and\ \citenamefont {Coles}}]{Fontana2022nontrivial}%
  \BibitemOpen
  \bibfield  {author} {\bibinfo {author} {\bibfnamefont {E.}~\bibnamefont {Fontana}}, \bibinfo {author} {\bibfnamefont {M.}~\bibnamefont {Cerezo}}, \bibinfo {author} {\bibfnamefont {A.}~\bibnamefont {Arrasmith}}, \bibinfo {author} {\bibfnamefont {I.}~\bibnamefont {Rungger}}, \ and\ \bibinfo {author} {\bibfnamefont {P.~J.}\ \bibnamefont {Coles}},\ }\href {\doibase 10.22331/q-2022-09-15-804} {\bibfield  {journal} {\bibinfo  {journal} {{Quantum}}\ }\textbf {\bibinfo {volume} {6}},\ \bibinfo {pages} {804} (\bibinfo {year} {2022})}\BibitemShut {NoStop}%
\bibitem [{\citenamefont {Uvarov}\ and\ \citenamefont {Biamonte}(2021)}]{uvarov2021barren}%
  \BibitemOpen
  \bibfield  {author} {\bibinfo {author} {\bibfnamefont {A.}~\bibnamefont {Uvarov}}\ and\ \bibinfo {author} {\bibfnamefont {J.~D.}\ \bibnamefont {Biamonte}},\ }\href@noop {} {\bibfield  {journal} {\bibinfo  {journal} {Journal of Physics A: Mathematical and Theoretical}\ }\textbf {\bibinfo {volume} {54}},\ \bibinfo {pages} {245301} (\bibinfo {year} {2021})}\BibitemShut {NoStop}%
\bibitem [{\citenamefont {Fontana}\ \emph {et~al.}(2024)\citenamefont {Fontana}, \citenamefont {Herman}, \citenamefont {Chakrabarti}, \citenamefont {Kumar}, \citenamefont {Yalovetzky}, \citenamefont {Heredge}, \citenamefont {Sureshbabu},\ and\ \citenamefont {Pistoia}}]{Fontana2024}%
  \BibitemOpen
  \bibfield  {author} {\bibinfo {author} {\bibfnamefont {E.}~\bibnamefont {Fontana}}, \bibinfo {author} {\bibfnamefont {D.}~\bibnamefont {Herman}}, \bibinfo {author} {\bibfnamefont {S.}~\bibnamefont {Chakrabarti}}, \bibinfo {author} {\bibfnamefont {N.}~\bibnamefont {Kumar}}, \bibinfo {author} {\bibfnamefont {R.}~\bibnamefont {Yalovetzky}}, \bibinfo {author} {\bibfnamefont {J.}~\bibnamefont {Heredge}}, \bibinfo {author} {\bibfnamefont {S.~H.}\ \bibnamefont {Sureshbabu}}, \ and\ \bibinfo {author} {\bibfnamefont {M.}~\bibnamefont {Pistoia}},\ }\href {\doibase 10.1038/s41467-024-49910-w} {\bibfield  {journal} {\bibinfo  {journal} {Nature Communications}\ }\textbf {\bibinfo {volume} {15}},\ \bibinfo {pages} {7171} (\bibinfo {year} {2024})}\BibitemShut {NoStop}%
\bibitem [{\citenamefont {Qi}\ \emph {et~al.}(2023)\citenamefont {Qi}, \citenamefont {Wang}, \citenamefont {Zhu}, \citenamefont {Gani},\ and\ \citenamefont {Gong}}]{Qi2023}%
  \BibitemOpen
  \bibfield  {author} {\bibinfo {author} {\bibfnamefont {H.}~\bibnamefont {Qi}}, \bibinfo {author} {\bibfnamefont {L.}~\bibnamefont {Wang}}, \bibinfo {author} {\bibfnamefont {H.}~\bibnamefont {Zhu}}, \bibinfo {author} {\bibfnamefont {A.}~\bibnamefont {Gani}}, \ and\ \bibinfo {author} {\bibfnamefont {C.}~\bibnamefont {Gong}},\ }\href {\doibase 10.1007/s11128-023-04188-7} {\bibfield  {journal} {\bibinfo  {journal} {Quantum Information Processing}\ }\textbf {\bibinfo {volume} {22}},\ \bibinfo {pages} {435} (\bibinfo {year} {2023})}\BibitemShut {NoStop}%
\bibitem [{\citenamefont {McClean}\ \emph {et~al.}(2018)\citenamefont {McClean}, \citenamefont {Boixo}, \citenamefont {Smelyanskiy}, \citenamefont {Babbush},\ and\ \citenamefont {Neven}}]{mcclean2018barren}%
  \BibitemOpen
  \bibfield  {author} {\bibinfo {author} {\bibfnamefont {J.~R.}\ \bibnamefont {McClean}}, \bibinfo {author} {\bibfnamefont {S.}~\bibnamefont {Boixo}}, \bibinfo {author} {\bibfnamefont {V.~N.}\ \bibnamefont {Smelyanskiy}}, \bibinfo {author} {\bibfnamefont {R.}~\bibnamefont {Babbush}}, \ and\ \bibinfo {author} {\bibfnamefont {H.}~\bibnamefont {Neven}},\ }\href@noop {} {\bibfield  {journal} {\bibinfo  {journal} {Nature communications}\ }\textbf {\bibinfo {volume} {9}},\ \bibinfo {pages} {4812} (\bibinfo {year} {2018})}\BibitemShut {NoStop}%
\bibitem [{\citenamefont {Casella}\ and\ \citenamefont {Berger}(2024)}]{casella2024statistical}%
  \BibitemOpen
  \bibfield  {author} {\bibinfo {author} {\bibfnamefont {G.}~\bibnamefont {Casella}}\ and\ \bibinfo {author} {\bibfnamefont {R.}~\bibnamefont {Berger}},\ }\href {https://books.google.com.br/books?id=cqUIEQAAQBAJ&lpg=PP1&ots=BSfrODDEEU&dq=Statistical%20Inference&lr&hl=pt-BR&pg=PP1#v=onepage&q=Statistical%20Inference&f=false} {\emph {\bibinfo {title} {Statistical inference}}}\ (\bibinfo  {publisher} {CRC press},\ \bibinfo {year} {2024})\BibitemShut {NoStop}%
\bibitem [{\citenamefont {Larocca}\ \emph {et~al.}(2022)\citenamefont {Larocca}, \citenamefont {Czarnik}, \citenamefont {Sharma}, \citenamefont {Muraleedharan}, \citenamefont {Coles},\ and\ \citenamefont {Cerezo}}]{Larocca2022diagnosingbarren}%
  \BibitemOpen
  \bibfield  {author} {\bibinfo {author} {\bibfnamefont {M.}~\bibnamefont {Larocca}}, \bibinfo {author} {\bibfnamefont {P.}~\bibnamefont {Czarnik}}, \bibinfo {author} {\bibfnamefont {K.}~\bibnamefont {Sharma}}, \bibinfo {author} {\bibfnamefont {G.}~\bibnamefont {Muraleedharan}}, \bibinfo {author} {\bibfnamefont {P.~J.}\ \bibnamefont {Coles}}, \ and\ \bibinfo {author} {\bibfnamefont {M.}~\bibnamefont {Cerezo}},\ }\href {\doibase 10.22331/q-2022-09-29-824} {\bibfield  {journal} {\bibinfo  {journal} {{Quantum}}\ }\textbf {\bibinfo {volume} {6}},\ \bibinfo {pages} {824} (\bibinfo {year} {2022})}\BibitemShut {NoStop}%
\bibitem [{\citenamefont {Ragone}\ \emph {et~al.}(2024)\citenamefont {Ragone}, \citenamefont {Bakalov}, \citenamefont {Sauvage}, \citenamefont {Kemper}, \citenamefont {Ortiz~Marrero}, \citenamefont {Larocca},\ and\ \citenamefont {Cerezo}}]{Ragone2024}%
  \BibitemOpen
  \bibfield  {author} {\bibinfo {author} {\bibfnamefont {M.}~\bibnamefont {Ragone}}, \bibinfo {author} {\bibfnamefont {B.~N.}\ \bibnamefont {Bakalov}}, \bibinfo {author} {\bibfnamefont {F.}~\bibnamefont {Sauvage}}, \bibinfo {author} {\bibfnamefont {A.~F.}\ \bibnamefont {Kemper}}, \bibinfo {author} {\bibfnamefont {C.}~\bibnamefont {Ortiz~Marrero}}, \bibinfo {author} {\bibfnamefont {M.}~\bibnamefont {Larocca}}, \ and\ \bibinfo {author} {\bibfnamefont {M.}~\bibnamefont {Cerezo}},\ }\href {\doibase 10.1038/s41467-024-49909-3} {\bibfield  {journal} {\bibinfo  {journal} {Nature Communications}\ }\textbf {\bibinfo {volume} {15}},\ \bibinfo {pages} {7172} (\bibinfo {year} {2024})}\BibitemShut {NoStop}%
\bibitem [{\citenamefont {Diaz}\ \emph {et~al.}(2023)\citenamefont {Diaz}, \citenamefont {García-Martín}, \citenamefont {Kazi}, \citenamefont {Larocca},\ and\ \citenamefont {Cerezo}}]{diaz2023showcasingbarrenplateautheory}%
  \BibitemOpen
  \bibfield  {author} {\bibinfo {author} {\bibfnamefont {N.~L.}\ \bibnamefont {Diaz}}, \bibinfo {author} {\bibfnamefont {D.}~\bibnamefont {García-Martín}}, \bibinfo {author} {\bibfnamefont {S.}~\bibnamefont {Kazi}}, \bibinfo {author} {\bibfnamefont {M.}~\bibnamefont {Larocca}}, \ and\ \bibinfo {author} {\bibfnamefont {M.}~\bibnamefont {Cerezo}},\ }\href {https://arxiv.org/abs/2310.11505} {\enquote {\bibinfo {title} {Showcasing a barren plateau theory beyond the dynamical lie algebra},}\ } (\bibinfo {year} {2023}),\ \Eprint {http://arxiv.org/abs/2310.11505} {arXiv:2310.11505 [quant-ph]} \BibitemShut {NoStop}%
\bibitem [{\citenamefont {Wang}\ \emph {et~al.}(2021{\natexlab{a}})\citenamefont {Wang}, \citenamefont {Fontana}, \citenamefont {Cerezo}, \citenamefont {Sharma}, \citenamefont {Sone}, \citenamefont {Cincio},\ and\ \citenamefont {Coles}}]{Wang2021}%
  \BibitemOpen
  \bibfield  {author} {\bibinfo {author} {\bibfnamefont {S.}~\bibnamefont {Wang}}, \bibinfo {author} {\bibfnamefont {E.}~\bibnamefont {Fontana}}, \bibinfo {author} {\bibfnamefont {M.}~\bibnamefont {Cerezo}}, \bibinfo {author} {\bibfnamefont {K.}~\bibnamefont {Sharma}}, \bibinfo {author} {\bibfnamefont {A.}~\bibnamefont {Sone}}, \bibinfo {author} {\bibfnamefont {L.}~\bibnamefont {Cincio}}, \ and\ \bibinfo {author} {\bibfnamefont {P.~J.}\ \bibnamefont {Coles}},\ }\href {\doibase 10.1038/s41467-021-27045-6} {\bibfield  {journal} {\bibinfo  {journal} {Nature Communications}\ }\textbf {\bibinfo {volume} {12}},\ \bibinfo {pages} {6961} (\bibinfo {year} {2021}{\natexlab{a}})}\BibitemShut {NoStop}%
\bibitem [{\citenamefont {Stilck~Fran{\c{c}}a}\ and\ \citenamefont {Garc{\'i}a-Patr{\'o}n}(2021)}]{StilckFrança2021}%
  \BibitemOpen
  \bibfield  {author} {\bibinfo {author} {\bibfnamefont {D.}~\bibnamefont {Stilck~Fran{\c{c}}a}}\ and\ \bibinfo {author} {\bibfnamefont {R.}~\bibnamefont {Garc{\'i}a-Patr{\'o}n}},\ }\href {\doibase 10.1038/s41567-021-01356-3} {\bibfield  {journal} {\bibinfo  {journal} {Nature Physics}\ }\textbf {\bibinfo {volume} {17}},\ \bibinfo {pages} {1221} (\bibinfo {year} {2021})}\BibitemShut {NoStop}%
\bibitem [{\citenamefont {Cerezo}\ \emph {et~al.}(2025)\citenamefont {Cerezo}, \citenamefont {Larocca}, \citenamefont {Garc{\'i}a-Mart{\'i}n}, \citenamefont {Diaz}, \citenamefont {Braccia}, \citenamefont {Fontana}, \citenamefont {Rudolph}, \citenamefont {Bermejo}, \citenamefont {Ijaz}, \citenamefont {Thanasilp}, \citenamefont {Anschuetz},\ and\ \citenamefont {Holmes}}]{cerezo2023does}%
  \BibitemOpen
  \bibfield  {author} {\bibinfo {author} {\bibfnamefont {M.}~\bibnamefont {Cerezo}}, \bibinfo {author} {\bibfnamefont {M.}~\bibnamefont {Larocca}}, \bibinfo {author} {\bibfnamefont {D.}~\bibnamefont {Garc{\'i}a-Mart{\'i}n}}, \bibinfo {author} {\bibfnamefont {N.~L.}\ \bibnamefont {Diaz}}, \bibinfo {author} {\bibfnamefont {P.}~\bibnamefont {Braccia}}, \bibinfo {author} {\bibfnamefont {E.}~\bibnamefont {Fontana}}, \bibinfo {author} {\bibfnamefont {M.~S.}\ \bibnamefont {Rudolph}}, \bibinfo {author} {\bibfnamefont {P.}~\bibnamefont {Bermejo}}, \bibinfo {author} {\bibfnamefont {A.}~\bibnamefont {Ijaz}}, \bibinfo {author} {\bibfnamefont {S.}~\bibnamefont {Thanasilp}}, \bibinfo {author} {\bibfnamefont {E.~R.}\ \bibnamefont {Anschuetz}}, \ and\ \bibinfo {author} {\bibfnamefont {Z.}~\bibnamefont {Holmes}},\ }\href {\doibase 10.1038/s41467-025-63099-6} {\bibfield  {journal} {\bibinfo  {journal} {Nature Communications}\ }\textbf {\bibinfo {volume} {16}},\ \bibinfo {pages} {7907} (\bibinfo {year} {2025})}\BibitemShut
  {NoStop}%
\bibitem [{\citenamefont {Cunningham}\ and\ \citenamefont {Zhuang}(2025)}]{cunningham2025investigating}%
  \BibitemOpen
  \bibfield  {author} {\bibinfo {author} {\bibfnamefont {J.}~\bibnamefont {Cunningham}}\ and\ \bibinfo {author} {\bibfnamefont {J.}~\bibnamefont {Zhuang}},\ }\href@noop {} {\bibfield  {journal} {\bibinfo  {journal} {Quantum Information Processing}\ }\textbf {\bibinfo {volume} {24}},\ \bibinfo {pages} {1} (\bibinfo {year} {2025})}\BibitemShut {NoStop}%
\bibitem [{\citenamefont {Zhang}\ \emph {et~al.}(2024)\citenamefont {Zhang}, \citenamefont {Liu},\ and\ \citenamefont {Zhang}}]{PhysRevLett.132.150603}%
  \BibitemOpen
  \bibfield  {author} {\bibinfo {author} {\bibfnamefont {H.-K.}\ \bibnamefont {Zhang}}, \bibinfo {author} {\bibfnamefont {S.}~\bibnamefont {Liu}}, \ and\ \bibinfo {author} {\bibfnamefont {S.-X.}\ \bibnamefont {Zhang}},\ }\href {\doibase 10.1103/PhysRevLett.132.150603} {\bibfield  {journal} {\bibinfo  {journal} {Phys. Rev. Lett.}\ }\textbf {\bibinfo {volume} {132}},\ \bibinfo {pages} {150603} (\bibinfo {year} {2024})}\BibitemShut {NoStop}%
\bibitem [{\citenamefont {Lee}\ \emph {et~al.}(2021{\natexlab{b}})\citenamefont {Lee}, \citenamefont {Magann}, \citenamefont {Rabitz},\ and\ \citenamefont {Arenz}}]{PhysRevA.104.032401}%
  \BibitemOpen
  \bibfield  {author} {\bibinfo {author} {\bibfnamefont {J.}~\bibnamefont {Lee}}, \bibinfo {author} {\bibfnamefont {A.~B.}\ \bibnamefont {Magann}}, \bibinfo {author} {\bibfnamefont {H.~A.}\ \bibnamefont {Rabitz}}, \ and\ \bibinfo {author} {\bibfnamefont {C.}~\bibnamefont {Arenz}},\ }\href {\doibase 10.1103/PhysRevA.104.032401} {\bibfield  {journal} {\bibinfo  {journal} {Phys. Rev. A}\ }\textbf {\bibinfo {volume} {104}},\ \bibinfo {pages} {032401} (\bibinfo {year} {2021}{\natexlab{b}})}\BibitemShut {NoStop}%
\bibitem [{\citenamefont {K\"okc\"u}\ \emph {et~al.}(2022)\citenamefont {K\"okc\"u}, \citenamefont {Steckmann}, \citenamefont {Wang}, \citenamefont {Freericks}, \citenamefont {Dumitrescu},\ and\ \citenamefont {Kemper}}]{PhysRevLett.129.070501}%
  \BibitemOpen
  \bibfield  {author} {\bibinfo {author} {\bibfnamefont {E.}~\bibnamefont {K\"okc\"u}}, \bibinfo {author} {\bibfnamefont {T.}~\bibnamefont {Steckmann}}, \bibinfo {author} {\bibfnamefont {Y.}~\bibnamefont {Wang}}, \bibinfo {author} {\bibfnamefont {J.~K.}\ \bibnamefont {Freericks}}, \bibinfo {author} {\bibfnamefont {E.~F.}\ \bibnamefont {Dumitrescu}}, \ and\ \bibinfo {author} {\bibfnamefont {A.~F.}\ \bibnamefont {Kemper}},\ }\href {\doibase 10.1103/PhysRevLett.129.070501} {\bibfield  {journal} {\bibinfo  {journal} {Phys. Rev. Lett.}\ }\textbf {\bibinfo {volume} {129}},\ \bibinfo {pages} {070501} (\bibinfo {year} {2022})}\BibitemShut {NoStop}%
\bibitem [{\citenamefont {Bilkis}\ \emph {et~al.}(2023)\citenamefont {Bilkis}, \citenamefont {Cerezo}, \citenamefont {Verdon}, \citenamefont {Coles},\ and\ \citenamefont {Cincio}}]{Bilkis2023}%
  \BibitemOpen
  \bibfield  {author} {\bibinfo {author} {\bibfnamefont {M.}~\bibnamefont {Bilkis}}, \bibinfo {author} {\bibfnamefont {M.}~\bibnamefont {Cerezo}}, \bibinfo {author} {\bibfnamefont {G.}~\bibnamefont {Verdon}}, \bibinfo {author} {\bibfnamefont {P.~J.}\ \bibnamefont {Coles}}, \ and\ \bibinfo {author} {\bibfnamefont {L.}~\bibnamefont {Cincio}},\ }\href {\doibase 10.1007/s42484-023-00132-1} {\bibfield  {journal} {\bibinfo  {journal} {Quantum Machine Intelligence}\ }\textbf {\bibinfo {volume} {5}},\ \bibinfo {pages} {43} (\bibinfo {year} {2023})}\BibitemShut {NoStop}%
\bibitem [{\citenamefont {Zhang}\ \emph {et~al.}(2022)\citenamefont {Zhang}, \citenamefont {Liu}, \citenamefont {Hsieh},\ and\ \citenamefont {Tao}}]{10.5555/3600270.3601622}%
  \BibitemOpen
  \bibfield  {author} {\bibinfo {author} {\bibfnamefont {K.}~\bibnamefont {Zhang}}, \bibinfo {author} {\bibfnamefont {L.}~\bibnamefont {Liu}}, \bibinfo {author} {\bibfnamefont {M.-H.}\ \bibnamefont {Hsieh}}, \ and\ \bibinfo {author} {\bibfnamefont {D.}~\bibnamefont {Tao}},\ }in\ \href@noop {} {\emph {\bibinfo {booktitle} {Proceedings of the 36th International Conference on Neural Information Processing Systems}}},\ \bibinfo {series and number} {NIPS '22}\ (\bibinfo  {publisher} {Curran Associates Inc.},\ \bibinfo {address} {Red Hook, NY, USA},\ \bibinfo {year} {2022})\BibitemShut {NoStop}%
\bibitem [{\citenamefont {Wang}\ \emph {et~al.}(2024)\citenamefont {Wang}, \citenamefont {Qi}, \citenamefont {Ferrie},\ and\ \citenamefont {Dong}}]{PhysRevApplied.22.054005}%
  \BibitemOpen
  \bibfield  {author} {\bibinfo {author} {\bibfnamefont {Y.}~\bibnamefont {Wang}}, \bibinfo {author} {\bibfnamefont {B.}~\bibnamefont {Qi}}, \bibinfo {author} {\bibfnamefont {C.}~\bibnamefont {Ferrie}}, \ and\ \bibinfo {author} {\bibfnamefont {D.}~\bibnamefont {Dong}},\ }\href {\doibase 10.1103/PhysRevApplied.22.054005} {\bibfield  {journal} {\bibinfo  {journal} {Phys. Rev. Appl.}\ }\textbf {\bibinfo {volume} {22}},\ \bibinfo {pages} {054005} (\bibinfo {year} {2024})}\BibitemShut {NoStop}%
\bibitem [{\citenamefont {Park}\ \emph {et~al.}(2024)\citenamefont {Park}, \citenamefont {Kang},\ and\ \citenamefont {Huh}}]{park2024hardwareefficientansatzbarrenplateaus}%
  \BibitemOpen
  \bibfield  {author} {\bibinfo {author} {\bibfnamefont {C.-Y.}\ \bibnamefont {Park}}, \bibinfo {author} {\bibfnamefont {M.}~\bibnamefont {Kang}}, \ and\ \bibinfo {author} {\bibfnamefont {J.}~\bibnamefont {Huh}},\ }\href {https://arxiv.org/abs/2403.04844} {\enquote {\bibinfo {title} {Hardware-efficient ansatz without barren plateaus in any depth},}\ } (\bibinfo {year} {2024}),\ \Eprint {http://arxiv.org/abs/2403.04844} {arXiv:2403.04844 [quant-ph]} \BibitemShut {NoStop}%
\bibitem [{\citenamefont {Bermejo}\ \emph {et~al.}(2024)\citenamefont {Bermejo}, \citenamefont {Braccia}, \citenamefont {Rudolph}, \citenamefont {Holmes}, \citenamefont {Cincio},\ and\ \citenamefont {Cerezo}}]{bermejo2024quantumconvolutionalneuralnetworks}%
  \BibitemOpen
  \bibfield  {author} {\bibinfo {author} {\bibfnamefont {P.}~\bibnamefont {Bermejo}}, \bibinfo {author} {\bibfnamefont {P.}~\bibnamefont {Braccia}}, \bibinfo {author} {\bibfnamefont {M.~S.}\ \bibnamefont {Rudolph}}, \bibinfo {author} {\bibfnamefont {Z.}~\bibnamefont {Holmes}}, \bibinfo {author} {\bibfnamefont {L.}~\bibnamefont {Cincio}}, \ and\ \bibinfo {author} {\bibfnamefont {M.}~\bibnamefont {Cerezo}},\ }\href {https://arxiv.org/abs/2408.12739} {\enquote {\bibinfo {title} {Quantum convolutional neural networks are (effectively) classically simulable},}\ } (\bibinfo {year} {2024}),\ \Eprint {http://arxiv.org/abs/2408.12739} {arXiv:2408.12739 [quant-ph]} \BibitemShut {NoStop}%
\bibitem [{\citenamefont {Basheer}\ \emph {et~al.}(2023)\citenamefont {Basheer}, \citenamefont {Feng}, \citenamefont {Ferrie},\ and\ \citenamefont {Li}}]{10.1609/aaai.v37i6.25830}%
  \BibitemOpen
  \bibfield  {author} {\bibinfo {author} {\bibfnamefont {A.}~\bibnamefont {Basheer}}, \bibinfo {author} {\bibfnamefont {Y.}~\bibnamefont {Feng}}, \bibinfo {author} {\bibfnamefont {C.}~\bibnamefont {Ferrie}}, \ and\ \bibinfo {author} {\bibfnamefont {S.}~\bibnamefont {Li}},\ }in\ \href {\doibase 10.1609/aaai.v37i6.25830} {\emph {\bibinfo {booktitle} {Proceedings of the Thirty-Seventh AAAI Conference on Artificial Intelligence and Thirty-Fifth Conference on Innovative Applications of Artificial Intelligence and Thirteenth Symposium on Educational Advances in Artificial Intelligence}}},\ \bibinfo {series and number} {AAAI'23/IAAI'23/EAAI'23}\ (\bibinfo  {publisher} {AAAI Press},\ \bibinfo {year} {2023})\BibitemShut {NoStop}%
\bibitem [{\citenamefont {Goh}\ \emph {et~al.}(2025)\citenamefont {Goh}, \citenamefont {Larocca}, \citenamefont {Cincio}, \citenamefont {Cerezo},\ and\ \citenamefont {Sauvage}}]{3y65-f5w6}%
  \BibitemOpen
  \bibfield  {author} {\bibinfo {author} {\bibfnamefont {M.~L.}\ \bibnamefont {Goh}}, \bibinfo {author} {\bibfnamefont {M.}~\bibnamefont {Larocca}}, \bibinfo {author} {\bibfnamefont {L.}~\bibnamefont {Cincio}}, \bibinfo {author} {\bibfnamefont {M.}~\bibnamefont {Cerezo}}, \ and\ \bibinfo {author} {\bibfnamefont {F.}~\bibnamefont {Sauvage}},\ }\href {\doibase 10.1103/3y65-f5w6} {\bibfield  {journal} {\bibinfo  {journal} {Phys. Rev. Res.}\ }\textbf {\bibinfo {volume} {7}},\ \bibinfo {pages} {033266} (\bibinfo {year} {2025})}\BibitemShut {NoStop}%
\bibitem [{\citenamefont {Cerezo}\ and\ \citenamefont {Coles}(2021)}]{Cerezo_2021}%
  \BibitemOpen
  \bibfield  {author} {\bibinfo {author} {\bibfnamefont {M.}~\bibnamefont {Cerezo}}\ and\ \bibinfo {author} {\bibfnamefont {P.~J.}\ \bibnamefont {Coles}},\ }\href {\doibase 10.1088/2058-9565/abf51a} {\bibfield  {journal} {\bibinfo  {journal} {Quantum Science and Technology}\ }\textbf {\bibinfo {volume} {6}},\ \bibinfo {pages} {035006} (\bibinfo {year} {2021})}\BibitemShut {NoStop}%
\bibitem [{\citenamefont {Arrasmith}\ \emph {et~al.}(2021)\citenamefont {Arrasmith}, \citenamefont {Cerezo}, \citenamefont {Czarnik}, \citenamefont {Cincio},\ and\ \citenamefont {Coles}}]{arrasmith2021effect}%
  \BibitemOpen
  \bibfield  {author} {\bibinfo {author} {\bibfnamefont {A.}~\bibnamefont {Arrasmith}}, \bibinfo {author} {\bibfnamefont {M.}~\bibnamefont {Cerezo}}, \bibinfo {author} {\bibfnamefont {P.}~\bibnamefont {Czarnik}}, \bibinfo {author} {\bibfnamefont {L.}~\bibnamefont {Cincio}}, \ and\ \bibinfo {author} {\bibfnamefont {P.~J.}\ \bibnamefont {Coles}},\ }\href@noop {} {\bibfield  {journal} {\bibinfo  {journal} {Quantum}\ }\textbf {\bibinfo {volume} {5}},\ \bibinfo {pages} {558} (\bibinfo {year} {2021})}\BibitemShut {NoStop}%
\bibitem [{\citenamefont {Thanasilp}\ \emph {et~al.}(2023)\citenamefont {Thanasilp}, \citenamefont {Wang}, \citenamefont {Nghiem}, \citenamefont {Coles},\ and\ \citenamefont {Cerezo}}]{Thanasilp2023}%
  \BibitemOpen
  \bibfield  {author} {\bibinfo {author} {\bibfnamefont {S.}~\bibnamefont {Thanasilp}}, \bibinfo {author} {\bibfnamefont {S.}~\bibnamefont {Wang}}, \bibinfo {author} {\bibfnamefont {N.~A.}\ \bibnamefont {Nghiem}}, \bibinfo {author} {\bibfnamefont {P.}~\bibnamefont {Coles}}, \ and\ \bibinfo {author} {\bibfnamefont {M.}~\bibnamefont {Cerezo}},\ }\href {\doibase 10.1007/s42484-023-00103-6} {\bibfield  {journal} {\bibinfo  {journal} {Quantum Machine Intelligence}\ }\textbf {\bibinfo {volume} {5}},\ \bibinfo {pages} {21} (\bibinfo {year} {2023})}\BibitemShut {NoStop}%
\bibitem [{\citenamefont {Wang}\ \emph {et~al.}(2023)\citenamefont {Wang}, \citenamefont {Chen},\ and\ \citenamefont {Wang}}]{Wang2023}%
  \BibitemOpen
  \bibfield  {author} {\bibinfo {author} {\bibfnamefont {K.}~\bibnamefont {Wang}}, \bibinfo {author} {\bibfnamefont {Y.-A.}\ \bibnamefont {Chen}}, \ and\ \bibinfo {author} {\bibfnamefont {X.}~\bibnamefont {Wang}},\ }\href {\doibase 10.1007/s11432-023-3786-1} {\bibfield  {journal} {\bibinfo  {journal} {Science China Information Sciences}\ }\textbf {\bibinfo {volume} {66}},\ \bibinfo {pages} {180508} (\bibinfo {year} {2023})}\BibitemShut {NoStop}%
\bibitem [{\citenamefont {Quek}\ \emph {et~al.}(2024)\citenamefont {Quek}, \citenamefont {Stilck~Fran{\c{c}}a}, \citenamefont {Khatri}, \citenamefont {Meyer},\ and\ \citenamefont {Eisert}}]{Quek2024}%
  \BibitemOpen
  \bibfield  {author} {\bibinfo {author} {\bibfnamefont {Y.}~\bibnamefont {Quek}}, \bibinfo {author} {\bibfnamefont {D.}~\bibnamefont {Stilck~Fran{\c{c}}a}}, \bibinfo {author} {\bibfnamefont {S.}~\bibnamefont {Khatri}}, \bibinfo {author} {\bibfnamefont {J.~J.}\ \bibnamefont {Meyer}}, \ and\ \bibinfo {author} {\bibfnamefont {J.}~\bibnamefont {Eisert}},\ }\href {\doibase 10.1038/s41567-024-02536-7} {\bibfield  {journal} {\bibinfo  {journal} {Nature Physics}\ }\textbf {\bibinfo {volume} {20}},\ \bibinfo {pages} {1648} (\bibinfo {year} {2024})}\BibitemShut {NoStop}%
\bibitem [{\citenamefont {Biamonte}(2021)}]{PhysRevA.103.L030401}%
  \BibitemOpen
  \bibfield  {author} {\bibinfo {author} {\bibfnamefont {J.}~\bibnamefont {Biamonte}},\ }\href {\doibase 10.1103/PhysRevA.103.L030401} {\bibfield  {journal} {\bibinfo  {journal} {Phys. Rev. A}\ }\textbf {\bibinfo {volume} {103}},\ \bibinfo {pages} {L030401} (\bibinfo {year} {2021})}\BibitemShut {NoStop}%
\bibitem [{\citenamefont {Zimbor{\'a}s}\ \emph {et~al.}(2025)\citenamefont {Zimbor{\'a}s}, \citenamefont {Koczor}, \citenamefont {Holmes}, \citenamefont {Borrelli}, \citenamefont {Gily{\'e}n}, \citenamefont {Huang}, \citenamefont {Cai}, \citenamefont {Ac{\'\i}n}, \citenamefont {Aolita}, \citenamefont {Banchi} \emph {et~al.}}]{zimboras2025myths}%
  \BibitemOpen
  \bibfield  {author} {\bibinfo {author} {\bibfnamefont {Z.}~\bibnamefont {Zimbor{\'a}s}}, \bibinfo {author} {\bibfnamefont {B.}~\bibnamefont {Koczor}}, \bibinfo {author} {\bibfnamefont {Z.}~\bibnamefont {Holmes}}, \bibinfo {author} {\bibfnamefont {E.-M.}\ \bibnamefont {Borrelli}}, \bibinfo {author} {\bibfnamefont {A.}~\bibnamefont {Gily{\'e}n}}, \bibinfo {author} {\bibfnamefont {H.-Y.}\ \bibnamefont {Huang}}, \bibinfo {author} {\bibfnamefont {Z.}~\bibnamefont {Cai}}, \bibinfo {author} {\bibfnamefont {A.}~\bibnamefont {Ac{\'\i}n}}, \bibinfo {author} {\bibfnamefont {L.}~\bibnamefont {Aolita}}, \bibinfo {author} {\bibfnamefont {L.}~\bibnamefont {Banchi}},  \emph {et~al.},\ }\href@noop {} {\bibfield  {journal} {\bibinfo  {journal} {arXiv preprint arXiv:2501.05694}\ } (\bibinfo {year} {2025})}\BibitemShut {NoStop}%
\bibitem [{\citenamefont {Higgott}\ \emph {et~al.}(2019)\citenamefont {Higgott}, \citenamefont {Wang},\ and\ \citenamefont {Brierley}}]{Higgott2019variationalquantum}%
  \BibitemOpen
  \bibfield  {author} {\bibinfo {author} {\bibfnamefont {O.}~\bibnamefont {Higgott}}, \bibinfo {author} {\bibfnamefont {D.}~\bibnamefont {Wang}}, \ and\ \bibinfo {author} {\bibfnamefont {S.}~\bibnamefont {Brierley}},\ }\href {\doibase 10.22331/q-2019-07-01-156} {\bibfield  {journal} {\bibinfo  {journal} {{Quantum}}\ }\textbf {\bibinfo {volume} {3}},\ \bibinfo {pages} {156} (\bibinfo {year} {2019})}\BibitemShut {NoStop}%
\bibitem [{\citenamefont {Motta}\ \emph {et~al.}(2020)\citenamefont {Motta}, \citenamefont {Sun}, \citenamefont {Tan}, \citenamefont {O’Rourke}, \citenamefont {Ye}, \citenamefont {Minnich}, \citenamefont {McClean},\ and\ \citenamefont {Chan}}]{motta2020imaginary}%
  \BibitemOpen
  \bibfield  {author} {\bibinfo {author} {\bibfnamefont {M.}~\bibnamefont {Motta}}, \bibinfo {author} {\bibfnamefont {C.}~\bibnamefont {Sun}}, \bibinfo {author} {\bibfnamefont {A.~T.~K.}\ \bibnamefont {Tan}}, \bibinfo {author} {\bibfnamefont {M.~J.}\ \bibnamefont {O’Rourke}}, \bibinfo {author} {\bibfnamefont {E.}~\bibnamefont {Ye}}, \bibinfo {author} {\bibfnamefont {A.~J.}\ \bibnamefont {Minnich}}, \bibinfo {author} {\bibfnamefont {J.~R.}\ \bibnamefont {McClean}}, \ and\ \bibinfo {author} {\bibfnamefont {G.~K.-L.}\ \bibnamefont {Chan}},\ }\href@noop {} {\bibfield  {journal} {\bibinfo  {journal} {Nature Physics}\ }\textbf {\bibinfo {volume} {16}},\ \bibinfo {pages} {205} (\bibinfo {year} {2020})}\BibitemShut {NoStop}%
\bibitem [{\citenamefont {Cirstoiu}\ \emph {et~al.}(2020)\citenamefont {Cirstoiu}, \citenamefont {Wang}, \citenamefont {Brierley}, \citenamefont {Kashefi},\ and\ \citenamefont {Ekert}}]{cirstoiu2020variational}%
  \BibitemOpen
  \bibfield  {author} {\bibinfo {author} {\bibfnamefont {C.}~\bibnamefont {Cirstoiu}}, \bibinfo {author} {\bibfnamefont {X.}~\bibnamefont {Wang}}, \bibinfo {author} {\bibfnamefont {S.}~\bibnamefont {Brierley}}, \bibinfo {author} {\bibfnamefont {E.}~\bibnamefont {Kashefi}}, \ and\ \bibinfo {author} {\bibfnamefont {A.}~\bibnamefont {Ekert}},\ }\href {\doibase https://doi.org/10.1038/s41534-020-00302-0} {\bibfield  {journal} {\bibinfo  {journal} {npj Quantum Information}\ }\textbf {\bibinfo {volume} {6}},\ \bibinfo {pages} {82} (\bibinfo {year} {2020})}\BibitemShut {NoStop}%
\bibitem [{\citenamefont {Selisko}\ \emph {et~al.}(2023)\citenamefont {Selisko}, \citenamefont {Amsler}, \citenamefont {Hammerschmidt}, \citenamefont {Drautz},\ and\ \citenamefont {Eckl}}]{Selisko_2024}%
  \BibitemOpen
  \bibfield  {author} {\bibinfo {author} {\bibfnamefont {J.}~\bibnamefont {Selisko}}, \bibinfo {author} {\bibfnamefont {M.}~\bibnamefont {Amsler}}, \bibinfo {author} {\bibfnamefont {T.}~\bibnamefont {Hammerschmidt}}, \bibinfo {author} {\bibfnamefont {R.}~\bibnamefont {Drautz}}, \ and\ \bibinfo {author} {\bibfnamefont {T.}~\bibnamefont {Eckl}},\ }\href {\doibase 10.1088/2058-9565/ad1340} {\bibfield  {journal} {\bibinfo  {journal} {Quantum Science and Technology}\ }\textbf {\bibinfo {volume} {9}},\ \bibinfo {pages} {015026} (\bibinfo {year} {2023})}\BibitemShut {NoStop}%
\bibitem [{\citenamefont {Xia}\ and\ \citenamefont {Kais}(2018)}]{xia2018quantum}%
  \BibitemOpen
  \bibfield  {author} {\bibinfo {author} {\bibfnamefont {R.}~\bibnamefont {Xia}}\ and\ \bibinfo {author} {\bibfnamefont {S.}~\bibnamefont {Kais}},\ }\href {https://www.nature.com/articles/s41467-018-06598-z} {\bibfield  {journal} {\bibinfo  {journal} {Nature communications}\ }\textbf {\bibinfo {volume} {9}},\ \bibinfo {pages} {4195} (\bibinfo {year} {2018})}\BibitemShut {NoStop}%
\bibitem [{\citenamefont {Kan}\ and\ \citenamefont {Mao}(2026)}]{kan2026machine}%
  \BibitemOpen
  \bibfield  {author} {\bibinfo {author} {\bibfnamefont {S.}~\bibnamefont {Kan}}\ and\ \bibinfo {author} {\bibfnamefont {Y.}~\bibnamefont {Mao}},\ }in\ \href {https://www.sciencedirect.com/science/article/abs/pii/B9780443302596000177} {\emph {\bibinfo {booktitle} {Quantum Computational AI}}}\ (\bibinfo  {publisher} {Elsevier},\ \bibinfo {year} {2026})\ pp.\ \bibinfo {pages} {153--170}\BibitemShut {NoStop}%
\bibitem [{\citenamefont {Wang}\ \emph {et~al.}(2021{\natexlab{b}})\citenamefont {Wang}, \citenamefont {Li},\ and\ \citenamefont {Wang}}]{wang2021variational}%
  \BibitemOpen
  \bibfield  {author} {\bibinfo {author} {\bibfnamefont {Y.}~\bibnamefont {Wang}}, \bibinfo {author} {\bibfnamefont {G.}~\bibnamefont {Li}}, \ and\ \bibinfo {author} {\bibfnamefont {X.}~\bibnamefont {Wang}},\ }\href {https://journals.aps.org/prapplied/abstract/10.1103/PhysRevApplied.16.054035} {\bibfield  {journal} {\bibinfo  {journal} {Physical Review Applied}\ }\textbf {\bibinfo {volume} {16}},\ \bibinfo {pages} {054035} (\bibinfo {year} {2021}{\natexlab{b}})}\BibitemShut {NoStop}%
\bibitem [{\citenamefont {Gibbs}\ \emph {et~al.}(2024)\citenamefont {Gibbs}, \citenamefont {Holmes}, \citenamefont {Caro}, \citenamefont {Ezzell}, \citenamefont {Huang}, \citenamefont {Cincio}, \citenamefont {Sornborger},\ and\ \citenamefont {Coles}}]{gibbs2024dynamical}%
  \BibitemOpen
  \bibfield  {author} {\bibinfo {author} {\bibfnamefont {J.}~\bibnamefont {Gibbs}}, \bibinfo {author} {\bibfnamefont {Z.}~\bibnamefont {Holmes}}, \bibinfo {author} {\bibfnamefont {M.~C.}\ \bibnamefont {Caro}}, \bibinfo {author} {\bibfnamefont {N.}~\bibnamefont {Ezzell}}, \bibinfo {author} {\bibfnamefont {H.-Y.}\ \bibnamefont {Huang}}, \bibinfo {author} {\bibfnamefont {L.}~\bibnamefont {Cincio}}, \bibinfo {author} {\bibfnamefont {A.~T.}\ \bibnamefont {Sornborger}}, \ and\ \bibinfo {author} {\bibfnamefont {P.~J.}\ \bibnamefont {Coles}},\ }\href {https://journals.aps.org/prresearch/abstract/10.1103/PhysRevResearch.6.013241} {\bibfield  {journal} {\bibinfo  {journal} {Physical Review Research}\ }\textbf {\bibinfo {volume} {6}},\ \bibinfo {pages} {013241} (\bibinfo {year} {2024})}\BibitemShut {NoStop}%
\bibitem [{\citenamefont {Gardas}\ \emph {et~al.}(2018)\citenamefont {Gardas}, \citenamefont {Rams},\ and\ \citenamefont {Dziarmaga}}]{gardas2018quantum}%
  \BibitemOpen
  \bibfield  {author} {\bibinfo {author} {\bibfnamefont {B.}~\bibnamefont {Gardas}}, \bibinfo {author} {\bibfnamefont {M.~M.}\ \bibnamefont {Rams}}, \ and\ \bibinfo {author} {\bibfnamefont {J.}~\bibnamefont {Dziarmaga}},\ }\href {https://journals.aps.org/prb/abstract/10.1103/PhysRevB.98.184304} {\bibfield  {journal} {\bibinfo  {journal} {Physical Review B}\ }\textbf {\bibinfo {volume} {98}},\ \bibinfo {pages} {184304} (\bibinfo {year} {2018})}\BibitemShut {NoStop}%
\bibitem [{\citenamefont {Long}\ \emph {et~al.}(2024)\citenamefont {Long}, \citenamefont {Cao}, \citenamefont {Ge}, \citenamefont {Li}, \citenamefont {Yan}, \citenamefont {Xu}, \citenamefont {Wang},\ and\ \citenamefont {Zheng}}]{long2024quantum}%
  \BibitemOpen
  \bibfield  {author} {\bibinfo {author} {\bibfnamefont {C.}~\bibnamefont {Long}}, \bibinfo {author} {\bibfnamefont {L.}~\bibnamefont {Cao}}, \bibinfo {author} {\bibfnamefont {L.}~\bibnamefont {Ge}}, \bibinfo {author} {\bibfnamefont {Q.-X.}\ \bibnamefont {Li}}, \bibinfo {author} {\bibfnamefont {Y.}~\bibnamefont {Yan}}, \bibinfo {author} {\bibfnamefont {R.-X.}\ \bibnamefont {Xu}}, \bibinfo {author} {\bibfnamefont {Y.}~\bibnamefont {Wang}}, \ and\ \bibinfo {author} {\bibfnamefont {X.}~\bibnamefont {Zheng}},\ }\href {https://pubs.aip.org/aip/jcp/article/161/8/084105/3309338} {\bibfield  {journal} {\bibinfo  {journal} {The Journal of Chemical Physics}\ }\textbf {\bibinfo {volume} {161}} (\bibinfo {year} {2024})}\BibitemShut {NoStop}%
\bibitem [{\citenamefont {Guan}\ \emph {et~al.}(2021)\citenamefont {Guan}, \citenamefont {Perdue}, \citenamefont {Pesah}, \citenamefont {Schuld}, \citenamefont {Terashi}, \citenamefont {Vallecorsa},\ and\ \citenamefont {Vlimant}}]{guan2021quantum}%
  \BibitemOpen
  \bibfield  {author} {\bibinfo {author} {\bibfnamefont {W.}~\bibnamefont {Guan}}, \bibinfo {author} {\bibfnamefont {G.}~\bibnamefont {Perdue}}, \bibinfo {author} {\bibfnamefont {A.}~\bibnamefont {Pesah}}, \bibinfo {author} {\bibfnamefont {M.}~\bibnamefont {Schuld}}, \bibinfo {author} {\bibfnamefont {K.}~\bibnamefont {Terashi}}, \bibinfo {author} {\bibfnamefont {S.}~\bibnamefont {Vallecorsa}}, \ and\ \bibinfo {author} {\bibfnamefont {J.-R.}\ \bibnamefont {Vlimant}},\ }\href {https://iopscience.iop.org/article/10.1088/2632-2153/abc17d/meta} {\bibfield  {journal} {\bibinfo  {journal} {Machine Learning: Science and Technology}\ }\textbf {\bibinfo {volume} {2}},\ \bibinfo {pages} {011003} (\bibinfo {year} {2021})}\BibitemShut {NoStop}%
\bibitem [{\citenamefont {Kim}\ \emph {et~al.}(2024)\citenamefont {Kim}, \citenamefont {Lloyd},\ and\ \citenamefont {Marvian}}]{kim2024hamiltonian}%
  \BibitemOpen
  \bibfield  {author} {\bibinfo {author} {\bibfnamefont {L.}~\bibnamefont {Kim}}, \bibinfo {author} {\bibfnamefont {S.}~\bibnamefont {Lloyd}}, \ and\ \bibinfo {author} {\bibfnamefont {M.}~\bibnamefont {Marvian}},\ }\href {https://journals.aps.org/prresearch/abstract/10.1103/PhysRevResearch.6.033019} {\bibfield  {journal} {\bibinfo  {journal} {Physical Review Research}\ }\textbf {\bibinfo {volume} {6}},\ \bibinfo {pages} {033019} (\bibinfo {year} {2024})}\BibitemShut {NoStop}%
\bibitem [{\citenamefont {Wu}\ \emph {et~al.}(2021)\citenamefont {Wu}, \citenamefont {Sun}, \citenamefont {Guan}, \citenamefont {Zhou}, \citenamefont {Chan}, \citenamefont {Cheng}, \citenamefont {Pham}, \citenamefont {Qian}, \citenamefont {Wang}, \citenamefont {Zhang} \emph {et~al.}}]{wu2021application}%
  \BibitemOpen
  \bibfield  {author} {\bibinfo {author} {\bibfnamefont {S.~L.}\ \bibnamefont {Wu}}, \bibinfo {author} {\bibfnamefont {S.}~\bibnamefont {Sun}}, \bibinfo {author} {\bibfnamefont {W.}~\bibnamefont {Guan}}, \bibinfo {author} {\bibfnamefont {C.}~\bibnamefont {Zhou}}, \bibinfo {author} {\bibfnamefont {J.}~\bibnamefont {Chan}}, \bibinfo {author} {\bibfnamefont {C.~L.}\ \bibnamefont {Cheng}}, \bibinfo {author} {\bibfnamefont {T.}~\bibnamefont {Pham}}, \bibinfo {author} {\bibfnamefont {Y.}~\bibnamefont {Qian}}, \bibinfo {author} {\bibfnamefont {A.~Z.}\ \bibnamefont {Wang}}, \bibinfo {author} {\bibfnamefont {R.}~\bibnamefont {Zhang}},  \emph {et~al.},\ }\href {https://journals.aps.org/prresearch/abstract/10.1103/PhysRevResearch.3.033221} {\bibfield  {journal} {\bibinfo  {journal} {Physical Review Research}\ }\textbf {\bibinfo {volume} {3}},\ \bibinfo {pages} {033221} (\bibinfo {year} {2021})}\BibitemShut {NoStop}%
\bibitem [{\citenamefont {Baul}\ \emph {et~al.}(2025)\citenamefont {Baul}, \citenamefont {Fotso}, \citenamefont {Terletska}, \citenamefont {Tam},\ and\ \citenamefont {Moreno}}]{baul2025quantum}%
  \BibitemOpen
  \bibfield  {author} {\bibinfo {author} {\bibfnamefont {A.}~\bibnamefont {Baul}}, \bibinfo {author} {\bibfnamefont {H.}~\bibnamefont {Fotso}}, \bibinfo {author} {\bibfnamefont {H.}~\bibnamefont {Terletska}}, \bibinfo {author} {\bibfnamefont {K.-M.}\ \bibnamefont {Tam}}, \ and\ \bibinfo {author} {\bibfnamefont {J.}~\bibnamefont {Moreno}},\ }\href {https://www.mdpi.com/2624-960X/7/2/18} {\bibfield  {journal} {\bibinfo  {journal} {Quantum Reports}\ }\textbf {\bibinfo {volume} {7}},\ \bibinfo {pages} {18} (\bibinfo {year} {2025})}\BibitemShut {NoStop}%
\bibitem [{\citenamefont {Mahmud}\ \emph {et~al.}(2024)\citenamefont {Mahmud}, \citenamefont {Mashtura}, \citenamefont {Fattah},\ and\ \citenamefont {Saquib}}]{mahmud2024quantum}%
  \BibitemOpen
  \bibfield  {author} {\bibinfo {author} {\bibfnamefont {J.}~\bibnamefont {Mahmud}}, \bibinfo {author} {\bibfnamefont {R.}~\bibnamefont {Mashtura}}, \bibinfo {author} {\bibfnamefont {S.~A.}\ \bibnamefont {Fattah}}, \ and\ \bibinfo {author} {\bibfnamefont {M.}~\bibnamefont {Saquib}},\ }\href@noop {} {\bibfield  {journal} {\bibinfo  {journal} {Quantum Machine Intelligence}\ }\textbf {\bibinfo {volume} {6}},\ \bibinfo {pages} {11} (\bibinfo {year} {2024})}\BibitemShut {NoStop}%
\bibitem [{\citenamefont {de~Jesus}\ \emph {et~al.}(2025)\citenamefont {de~Jesus}, \citenamefont {da~Silva}, \citenamefont {Pires}, \citenamefont {da~Silva}, \citenamefont {dos Santos~Cruz},\ and\ \citenamefont {da~Silva}}]{de2025exploring}%
  \BibitemOpen
  \bibfield  {author} {\bibinfo {author} {\bibfnamefont {G.~F.}\ \bibnamefont {de~Jesus}}, \bibinfo {author} {\bibfnamefont {M.~H.~F.}\ \bibnamefont {da~Silva}}, \bibinfo {author} {\bibfnamefont {O.~M.}\ \bibnamefont {Pires}}, \bibinfo {author} {\bibfnamefont {L.~C.}\ \bibnamefont {da~Silva}}, \bibinfo {author} {\bibfnamefont {C.}~\bibnamefont {dos Santos~Cruz}}, \ and\ \bibinfo {author} {\bibfnamefont {V.~L.}\ \bibnamefont {da~Silva}},\ }\href@noop {} {\bibfield  {journal} {\bibinfo  {journal} {Entropy}\ }\textbf {\bibinfo {volume} {27}},\ \bibinfo {pages} {490} (\bibinfo {year} {2025})}\BibitemShut {NoStop}%
\bibitem [{\citenamefont {Ogur}\ and\ \citenamefont {Yilmaz}(2023)}]{ogur2023effect}%
  \BibitemOpen
  \bibfield  {author} {\bibinfo {author} {\bibfnamefont {B.}~\bibnamefont {Ogur}}\ and\ \bibinfo {author} {\bibfnamefont {I.}~\bibnamefont {Yilmaz}},\ }\href@noop {} {\bibfield  {journal} {\bibinfo  {journal} {Quantum Inf. Comput.}\ }\textbf {\bibinfo {volume} {23}},\ \bibinfo {pages} {181} (\bibinfo {year} {2023})}\BibitemShut {NoStop}%
\bibitem [{\citenamefont {Schatzki}\ \emph {et~al.}(2024)\citenamefont {Schatzki}, \citenamefont {Larocca}, \citenamefont {Nguyen}, \citenamefont {Sauvage},\ and\ \citenamefont {Cerezo}}]{schatzki2024theoretical}%
  \BibitemOpen
  \bibfield  {author} {\bibinfo {author} {\bibfnamefont {L.}~\bibnamefont {Schatzki}}, \bibinfo {author} {\bibfnamefont {M.}~\bibnamefont {Larocca}}, \bibinfo {author} {\bibfnamefont {Q.~T.}\ \bibnamefont {Nguyen}}, \bibinfo {author} {\bibfnamefont {F.}~\bibnamefont {Sauvage}}, \ and\ \bibinfo {author} {\bibfnamefont {M.}~\bibnamefont {Cerezo}},\ }\href@noop {} {\bibfield  {journal} {\bibinfo  {journal} {npj Quantum Information}\ }\textbf {\bibinfo {volume} {10}},\ \bibinfo {pages} {12} (\bibinfo {year} {2024})}\BibitemShut {NoStop}%
\bibitem [{\citenamefont {Cong}\ \emph {et~al.}(2019)\citenamefont {Cong}, \citenamefont {Choi},\ and\ \citenamefont {Lukin}}]{cong2019quantum}%
  \BibitemOpen
  \bibfield  {author} {\bibinfo {author} {\bibfnamefont {I.}~\bibnamefont {Cong}}, \bibinfo {author} {\bibfnamefont {S.}~\bibnamefont {Choi}}, \ and\ \bibinfo {author} {\bibfnamefont {M.~D.}\ \bibnamefont {Lukin}},\ }\href@noop {} {\bibfield  {journal} {\bibinfo  {journal} {Nature Physics}\ }\textbf {\bibinfo {volume} {15}},\ \bibinfo {pages} {1273} (\bibinfo {year} {2019})}\BibitemShut {NoStop}%
\bibitem [{\citenamefont {Kauzmann}(1970)}]{kauzmann1970quantum}%
  \BibitemOpen
  \bibfield  {author} {\bibinfo {author} {\bibfnamefont {W.}~\bibnamefont {Kauzmann}},\ }\href {https://books.google.com.br/books?id=HRZZAAAAYAAJ} {\emph {\bibinfo {title} {Quantum Chemistry: An Introduction}}}\ (\bibinfo  {publisher} {Academic Press},\ \bibinfo {year} {1970})\BibitemShut {NoStop}%
\bibitem [{\citenamefont {Levine}(2014)}]{levine2014quantum}%
  \BibitemOpen
  \bibfield  {author} {\bibinfo {author} {\bibfnamefont {I.}~\bibnamefont {Levine}},\ }\href {https://books.google.com.br/books?id=dvyJMQEACAAJ} {\emph {\bibinfo {title} {Quantum Chemistry}}},\ Pearson advanced chemistry series\ (\bibinfo  {publisher} {Pearson},\ \bibinfo {year} {2014})\BibitemShut {NoStop}%
\bibitem [{\citenamefont {Costain}(1958)}]{costain1958determination}%
  \BibitemOpen
  \bibfield  {author} {\bibinfo {author} {\bibfnamefont {C.}~\bibnamefont {Costain}},\ }\href@noop {} {\bibfield  {journal} {\bibinfo  {journal} {The Journal of Chemical Physics}\ }\textbf {\bibinfo {volume} {29}},\ \bibinfo {pages} {864} (\bibinfo {year} {1958})}\BibitemShut {NoStop}%
\bibitem [{\citenamefont {Parr}(1989)}]{Parr1989}%
  \BibitemOpen
  \bibfield  {author} {\bibinfo {author} {\bibfnamefont {R.~G.}\ \bibnamefont {Parr}},\ }\href {https://lens.org/187-490-165-841-69X} {\emph {\bibinfo {title} {Density-functional theory of atoms and molecules}}}\ (\bibinfo {year} {1989})\BibitemShut {NoStop}%
\bibitem [{\citenamefont {Kitaev}(1995)}]{kitaev1995quantummeasurementsabelianstabilizer}%
  \BibitemOpen
  \bibfield  {author} {\bibinfo {author} {\bibfnamefont {A.~Y.}\ \bibnamefont {Kitaev}},\ }\href {https://arxiv.org/abs/quant-ph/9511026} {\enquote {\bibinfo {title} {Quantum measurements and the abelian stabilizer problem},}\ } (\bibinfo {year} {1995})\BibitemShut {NoStop}%
\bibitem [{\citenamefont {van Dam}\ \emph {et~al.}(2007)\citenamefont {van Dam}, \citenamefont {D'Ariano}, \citenamefont {Ekert}, \citenamefont {Macchiavello},\ and\ \citenamefont {Mosca}}]{PhysRevLett.98.090501}%
  \BibitemOpen
  \bibfield  {author} {\bibinfo {author} {\bibfnamefont {W.}~\bibnamefont {van Dam}}, \bibinfo {author} {\bibfnamefont {G.~M.}\ \bibnamefont {D'Ariano}}, \bibinfo {author} {\bibfnamefont {A.}~\bibnamefont {Ekert}}, \bibinfo {author} {\bibfnamefont {C.}~\bibnamefont {Macchiavello}}, \ and\ \bibinfo {author} {\bibfnamefont {M.}~\bibnamefont {Mosca}},\ }\href {\doibase 10.1103/PhysRevLett.98.090501} {\bibfield  {journal} {\bibinfo  {journal} {Phys. Rev. Lett.}\ }\textbf {\bibinfo {volume} {98}},\ \bibinfo {pages} {090501} (\bibinfo {year} {2007})}\BibitemShut {NoStop}%
\bibitem [{\citenamefont {Abrams}\ and\ \citenamefont {Lloyd}(1997)}]{PhysRevLett.79.2586}%
  \BibitemOpen
  \bibfield  {author} {\bibinfo {author} {\bibfnamefont {D.~S.}\ \bibnamefont {Abrams}}\ and\ \bibinfo {author} {\bibfnamefont {S.}~\bibnamefont {Lloyd}},\ }\href {\doibase 10.1103/PhysRevLett.79.2586} {\bibfield  {journal} {\bibinfo  {journal} {Phys. Rev. Lett.}\ }\textbf {\bibinfo {volume} {79}},\ \bibinfo {pages} {2586} (\bibinfo {year} {1997})}\BibitemShut {NoStop}%
\bibitem [{\citenamefont {Abrams}\ and\ \citenamefont {Lloyd}(1999)}]{PhysRevLett.83.5162}%
  \BibitemOpen
  \bibfield  {author} {\bibinfo {author} {\bibfnamefont {D.~S.}\ \bibnamefont {Abrams}}\ and\ \bibinfo {author} {\bibfnamefont {S.}~\bibnamefont {Lloyd}},\ }\href {\doibase 10.1103/PhysRevLett.83.5162} {\bibfield  {journal} {\bibinfo  {journal} {Phys. Rev. Lett.}\ }\textbf {\bibinfo {volume} {83}},\ \bibinfo {pages} {5162} (\bibinfo {year} {1999})}\BibitemShut {NoStop}%
\bibitem [{\citenamefont {Aspuru-Guzik}\ \emph {et~al.}(2005)\citenamefont {Aspuru-Guzik}, \citenamefont {Dutoi}, \citenamefont {Love},\ and\ \citenamefont {Head-Gordon}}]{doi:10.1126/science.1113479}%
  \BibitemOpen
  \bibfield  {author} {\bibinfo {author} {\bibfnamefont {A.}~\bibnamefont {Aspuru-Guzik}}, \bibinfo {author} {\bibfnamefont {A.~D.}\ \bibnamefont {Dutoi}}, \bibinfo {author} {\bibfnamefont {P.~J.}\ \bibnamefont {Love}}, \ and\ \bibinfo {author} {\bibfnamefont {M.}~\bibnamefont {Head-Gordon}},\ }\href {\doibase 10.1126/science.1113479} {\bibfield  {journal} {\bibinfo  {journal} {Science}\ }\textbf {\bibinfo {volume} {309}},\ \bibinfo {pages} {1704} (\bibinfo {year} {2005})}\BibitemShut {NoStop}%
\bibitem [{\citenamefont {McClean}\ \emph {et~al.}(2016)\citenamefont {McClean}, \citenamefont {Romero}, \citenamefont {Babbush},\ and\ \citenamefont {Aspuru-Guzik}}]{mcclean2016theory}%
  \BibitemOpen
  \bibfield  {author} {\bibinfo {author} {\bibfnamefont {J.~R.}\ \bibnamefont {McClean}}, \bibinfo {author} {\bibfnamefont {J.}~\bibnamefont {Romero}}, \bibinfo {author} {\bibfnamefont {R.}~\bibnamefont {Babbush}}, \ and\ \bibinfo {author} {\bibfnamefont {A.}~\bibnamefont {Aspuru-Guzik}},\ }\href@noop {} {\bibfield  {journal} {\bibinfo  {journal} {New Journal of Physics}\ }\textbf {\bibinfo {volume} {18}},\ \bibinfo {pages} {023023} (\bibinfo {year} {2016})}\BibitemShut {NoStop}%
\bibitem [{\citenamefont {Choy}\ and\ \citenamefont {Wales}(2023)}]{doi:10.1021/acs.jctc.2c01057}%
  \BibitemOpen
  \bibfield  {author} {\bibinfo {author} {\bibfnamefont {B.}~\bibnamefont {Choy}}\ and\ \bibinfo {author} {\bibfnamefont {D.~J.}\ \bibnamefont {Wales}},\ }\href {\doibase 10.1021/acs.jctc.2c01057} {\bibfield  {journal} {\bibinfo  {journal} {Journal of Chemical Theory and Computation}\ }\textbf {\bibinfo {volume} {19}},\ \bibinfo {pages} {1197} (\bibinfo {year} {2023})},\ \bibinfo {note} {pMID: 36749922}\BibitemShut {NoStop}%
\bibitem [{\citenamefont {Sennane}\ \emph {et~al.}(2023)\citenamefont {Sennane}, \citenamefont {Piquemal},\ and\ \citenamefont {Ran{\v{c}}i{\'c}}}]{sennane2023calculating}%
  \BibitemOpen
  \bibfield  {author} {\bibinfo {author} {\bibfnamefont {W.}~\bibnamefont {Sennane}}, \bibinfo {author} {\bibfnamefont {J.-P.}\ \bibnamefont {Piquemal}}, \ and\ \bibinfo {author} {\bibfnamefont {M.~J.}\ \bibnamefont {Ran{\v{c}}i{\'c}}},\ }\href@noop {} {\bibfield  {journal} {\bibinfo  {journal} {Physical Review A}\ }\textbf {\bibinfo {volume} {107}},\ \bibinfo {pages} {012416} (\bibinfo {year} {2023})}\BibitemShut {NoStop}%
\bibitem [{\citenamefont {Bartlett}\ and\ \citenamefont {Musia\l{}}(2007)}]{RevModPhys.79.291}%
  \BibitemOpen
  \bibfield  {author} {\bibinfo {author} {\bibfnamefont {R.~J.}\ \bibnamefont {Bartlett}}\ and\ \bibinfo {author} {\bibfnamefont {M.}~\bibnamefont {Musia\l{}}},\ }\href {\doibase 10.1103/RevModPhys.79.291} {\bibfield  {journal} {\bibinfo  {journal} {Rev. Mod. Phys.}\ }\textbf {\bibinfo {volume} {79}},\ \bibinfo {pages} {291} (\bibinfo {year} {2007})}\BibitemShut {NoStop}%
\bibitem [{\citenamefont {Ravi}\ \emph {et~al.}(2023)\citenamefont {Ravi}, \citenamefont {Perera}, \citenamefont {Park},\ and\ \citenamefont {Bartlett}}]{10.1063/5.0161368}%
  \BibitemOpen
  \bibfield  {author} {\bibinfo {author} {\bibfnamefont {M.}~\bibnamefont {Ravi}}, \bibinfo {author} {\bibfnamefont {A.}~\bibnamefont {Perera}}, \bibinfo {author} {\bibfnamefont {Y.~C.}\ \bibnamefont {Park}}, \ and\ \bibinfo {author} {\bibfnamefont {R.~J.}\ \bibnamefont {Bartlett}},\ }\href {\doibase 10.1063/5.0161368} {\bibfield  {journal} {\bibinfo  {journal} {The Journal of Chemical Physics}\ }\textbf {\bibinfo {volume} {159}},\ \bibinfo {pages} {094101} (\bibinfo {year} {2023})}\BibitemShut {NoStop}%
\bibitem [{\citenamefont {Romero}\ \emph {et~al.}(2018)\citenamefont {Romero}, \citenamefont {Babbush}, \citenamefont {McClean}, \citenamefont {Hempel}, \citenamefont {Love},\ and\ \citenamefont {Aspuru-Guzik}}]{Romero_2019}%
  \BibitemOpen
  \bibfield  {author} {\bibinfo {author} {\bibfnamefont {J.}~\bibnamefont {Romero}}, \bibinfo {author} {\bibfnamefont {R.}~\bibnamefont {Babbush}}, \bibinfo {author} {\bibfnamefont {J.~R.}\ \bibnamefont {McClean}}, \bibinfo {author} {\bibfnamefont {C.}~\bibnamefont {Hempel}}, \bibinfo {author} {\bibfnamefont {P.~J.}\ \bibnamefont {Love}}, \ and\ \bibinfo {author} {\bibfnamefont {A.}~\bibnamefont {Aspuru-Guzik}},\ }\href {\doibase 10.1088/2058-9565/aad3e4} {\bibfield  {journal} {\bibinfo  {journal} {Quantum Science and Technology}\ }\textbf {\bibinfo {volume} {4}},\ \bibinfo {pages} {014008} (\bibinfo {year} {2018})}\BibitemShut {NoStop}%
\bibitem [{\citenamefont {Lee}\ \emph {et~al.}(2019)\citenamefont {Lee}, \citenamefont {Huggins}, \citenamefont {Head-Gordon},\ and\ \citenamefont {Whaley}}]{doi:10.1021/acs.jctc.8b01004}%
  \BibitemOpen
  \bibfield  {author} {\bibinfo {author} {\bibfnamefont {J.}~\bibnamefont {Lee}}, \bibinfo {author} {\bibfnamefont {W.~J.}\ \bibnamefont {Huggins}}, \bibinfo {author} {\bibfnamefont {M.}~\bibnamefont {Head-Gordon}}, \ and\ \bibinfo {author} {\bibfnamefont {K.~B.}\ \bibnamefont {Whaley}},\ }\href {\doibase 10.1021/acs.jctc.8b01004} {\bibfield  {journal} {\bibinfo  {journal} {Journal of Chemical Theory and Computation}\ }\textbf {\bibinfo {volume} {15}},\ \bibinfo {pages} {311} (\bibinfo {year} {2019})}\BibitemShut {NoStop}%
\bibitem [{\citenamefont {McLachlan}(1964)}]{McLachlan01011964}%
  \BibitemOpen
  \bibfield  {author} {\bibinfo {author} {\bibfnamefont {A.}~\bibnamefont {McLachlan}},\ }\href {\doibase 10.1080/00268976400100041} {\bibfield  {journal} {\bibinfo  {journal} {Molecular Physics}\ }\textbf {\bibinfo {volume} {8}},\ \bibinfo {pages} {39} (\bibinfo {year} {1964})}\BibitemShut {NoStop}%
\bibitem [{\citenamefont {McArdle}\ \emph {et~al.}(2019)\citenamefont {McArdle}, \citenamefont {Jones}, \citenamefont {Endo}, \citenamefont {Li}, \citenamefont {Benjamin},\ and\ \citenamefont {Yuan}}]{mcardle2019variational}%
  \BibitemOpen
  \bibfield  {author} {\bibinfo {author} {\bibfnamefont {S.}~\bibnamefont {McArdle}}, \bibinfo {author} {\bibfnamefont {T.}~\bibnamefont {Jones}}, \bibinfo {author} {\bibfnamefont {S.}~\bibnamefont {Endo}}, \bibinfo {author} {\bibfnamefont {Y.}~\bibnamefont {Li}}, \bibinfo {author} {\bibfnamefont {S.~C.}\ \bibnamefont {Benjamin}}, \ and\ \bibinfo {author} {\bibfnamefont {X.}~\bibnamefont {Yuan}},\ }\href@noop {} {\bibfield  {journal} {\bibinfo  {journal} {npj Quantum Information}\ }\textbf {\bibinfo {volume} {5}},\ \bibinfo {pages} {75} (\bibinfo {year} {2019})}\BibitemShut {NoStop}%
\bibitem [{\citenamefont {Nishi}\ \emph {et~al.}(2021)\citenamefont {Nishi}, \citenamefont {Kosugi},\ and\ \citenamefont {Matsushita}}]{nishi2021implementation}%
  \BibitemOpen
  \bibfield  {author} {\bibinfo {author} {\bibfnamefont {H.}~\bibnamefont {Nishi}}, \bibinfo {author} {\bibfnamefont {T.}~\bibnamefont {Kosugi}}, \ and\ \bibinfo {author} {\bibfnamefont {Y.-i.}\ \bibnamefont {Matsushita}},\ }\href@noop {} {\bibfield  {journal} {\bibinfo  {journal} {npj Quantum Information}\ }\textbf {\bibinfo {volume} {7}},\ \bibinfo {pages} {85} (\bibinfo {year} {2021})}\BibitemShut {NoStop}%
\bibitem [{\citenamefont {Kamakari}\ \emph {et~al.}(2022{\natexlab{a}})\citenamefont {Kamakari}, \citenamefont {Sun}, \citenamefont {Motta},\ and\ \citenamefont {Minnich}}]{kamakari2022digital}%
  \BibitemOpen
  \bibfield  {author} {\bibinfo {author} {\bibfnamefont {H.}~\bibnamefont {Kamakari}}, \bibinfo {author} {\bibfnamefont {S.-N.}\ \bibnamefont {Sun}}, \bibinfo {author} {\bibfnamefont {M.}~\bibnamefont {Motta}}, \ and\ \bibinfo {author} {\bibfnamefont {A.~J.}\ \bibnamefont {Minnich}},\ }\href@noop {} {\bibfield  {journal} {\bibinfo  {journal} {PRX quantum}\ }\textbf {\bibinfo {volume} {3}},\ \bibinfo {pages} {010320} (\bibinfo {year} {2022}{\natexlab{a}})}\BibitemShut {NoStop}%
\bibitem [{\citenamefont {Knill}\ \emph {et~al.}(2007)\citenamefont {Knill}, \citenamefont {Ortiz},\ and\ \citenamefont {Somma}}]{PhysRevA.75.012328}%
  \BibitemOpen
  \bibfield  {author} {\bibinfo {author} {\bibfnamefont {E.}~\bibnamefont {Knill}}, \bibinfo {author} {\bibfnamefont {G.}~\bibnamefont {Ortiz}}, \ and\ \bibinfo {author} {\bibfnamefont {R.~D.}\ \bibnamefont {Somma}},\ }\href {\doibase 10.1103/PhysRevA.75.012328} {\bibfield  {journal} {\bibinfo  {journal} {Phys. Rev. A}\ }\textbf {\bibinfo {volume} {75}},\ \bibinfo {pages} {012328} (\bibinfo {year} {2007})}\BibitemShut {NoStop}%
\bibitem [{\citenamefont {Hodecker}\ and\ \citenamefont {Dreuw}(2020)}]{10.1063/5.0019055}%
  \BibitemOpen
  \bibfield  {author} {\bibinfo {author} {\bibfnamefont {M.}~\bibnamefont {Hodecker}}\ and\ \bibinfo {author} {\bibfnamefont {A.}~\bibnamefont {Dreuw}},\ }\href {\doibase 10.1063/5.0019055} {\bibfield  {journal} {\bibinfo  {journal} {The Journal of Chemical Physics}\ }\textbf {\bibinfo {volume} {153}},\ \bibinfo {pages} {084112} (\bibinfo {year} {2020})}\BibitemShut {NoStop}%
\bibitem [{\citenamefont {Shen}\ \emph {et~al.}(2017)\citenamefont {Shen}, \citenamefont {Zhang}, \citenamefont {Zhang}, \citenamefont {Zhang}, \citenamefont {Yung},\ and\ \citenamefont {Kim}}]{PhysRevA.95.020501}%
  \BibitemOpen
  \bibfield  {author} {\bibinfo {author} {\bibfnamefont {Y.}~\bibnamefont {Shen}}, \bibinfo {author} {\bibfnamefont {X.}~\bibnamefont {Zhang}}, \bibinfo {author} {\bibfnamefont {S.}~\bibnamefont {Zhang}}, \bibinfo {author} {\bibfnamefont {J.-N.}\ \bibnamefont {Zhang}}, \bibinfo {author} {\bibfnamefont {M.-H.}\ \bibnamefont {Yung}}, \ and\ \bibinfo {author} {\bibfnamefont {K.}~\bibnamefont {Kim}},\ }\href {\doibase 10.1103/PhysRevA.95.020501} {\bibfield  {journal} {\bibinfo  {journal} {Phys. Rev. A}\ }\textbf {\bibinfo {volume} {95}},\ \bibinfo {pages} {020501} (\bibinfo {year} {2017})}\BibitemShut {NoStop}%
\bibitem [{\citenamefont {O'Malley}\ \emph {et~al.}(2016)\citenamefont {O'Malley}, \citenamefont {Babbush}, \citenamefont {Kivlichan}, \citenamefont {Romero}, \citenamefont {McClean}, \citenamefont {Barends}, \citenamefont {Kelly}, \citenamefont {Roushan}, \citenamefont {Tranter}, \citenamefont {Ding}, \citenamefont {Campbell}, \citenamefont {Chen}, \citenamefont {Chen}, \citenamefont {Chiaro}, \citenamefont {Dunsworth}, \citenamefont {Fowler}, \citenamefont {Jeffrey}, \citenamefont {Lucero}, \citenamefont {Megrant}, \citenamefont {Mutus}, \citenamefont {Neeley}, \citenamefont {Neill}, \citenamefont {Quintana}, \citenamefont {Sank}, \citenamefont {Vainsencher}, \citenamefont {Wenner}, \citenamefont {White}, \citenamefont {Coveney}, \citenamefont {Love}, \citenamefont {Neven}, \citenamefont {Aspuru-Guzik},\ and\ \citenamefont {Martinis}}]{PhysRevX.6.031007}%
  \BibitemOpen
  \bibfield  {author} {\bibinfo {author} {\bibfnamefont {P.~J.~J.}\ \bibnamefont {O'Malley}}, \bibinfo {author} {\bibfnamefont {R.}~\bibnamefont {Babbush}}, \bibinfo {author} {\bibfnamefont {I.~D.}\ \bibnamefont {Kivlichan}}, \bibinfo {author} {\bibfnamefont {J.}~\bibnamefont {Romero}}, \bibinfo {author} {\bibfnamefont {J.~R.}\ \bibnamefont {McClean}}, \bibinfo {author} {\bibfnamefont {R.}~\bibnamefont {Barends}}, \bibinfo {author} {\bibfnamefont {J.}~\bibnamefont {Kelly}}, \bibinfo {author} {\bibfnamefont {P.}~\bibnamefont {Roushan}}, \bibinfo {author} {\bibfnamefont {A.}~\bibnamefont {Tranter}}, \bibinfo {author} {\bibfnamefont {N.}~\bibnamefont {Ding}}, \bibinfo {author} {\bibfnamefont {B.}~\bibnamefont {Campbell}}, \bibinfo {author} {\bibfnamefont {Y.}~\bibnamefont {Chen}}, \bibinfo {author} {\bibfnamefont {Z.}~\bibnamefont {Chen}}, \bibinfo {author} {\bibfnamefont {B.}~\bibnamefont {Chiaro}}, \bibinfo {author} {\bibfnamefont {A.}~\bibnamefont {Dunsworth}}, \bibinfo {author} {\bibfnamefont {A.~G.}\
  \bibnamefont {Fowler}}, \bibinfo {author} {\bibfnamefont {E.}~\bibnamefont {Jeffrey}}, \bibinfo {author} {\bibfnamefont {E.}~\bibnamefont {Lucero}}, \bibinfo {author} {\bibfnamefont {A.}~\bibnamefont {Megrant}}, \bibinfo {author} {\bibfnamefont {J.~Y.}\ \bibnamefont {Mutus}}, \bibinfo {author} {\bibfnamefont {M.}~\bibnamefont {Neeley}}, \bibinfo {author} {\bibfnamefont {C.}~\bibnamefont {Neill}}, \bibinfo {author} {\bibfnamefont {C.}~\bibnamefont {Quintana}}, \bibinfo {author} {\bibfnamefont {D.}~\bibnamefont {Sank}}, \bibinfo {author} {\bibfnamefont {A.}~\bibnamefont {Vainsencher}}, \bibinfo {author} {\bibfnamefont {J.}~\bibnamefont {Wenner}}, \bibinfo {author} {\bibfnamefont {T.~C.}\ \bibnamefont {White}}, \bibinfo {author} {\bibfnamefont {P.~V.}\ \bibnamefont {Coveney}}, \bibinfo {author} {\bibfnamefont {P.~J.}\ \bibnamefont {Love}}, \bibinfo {author} {\bibfnamefont {H.}~\bibnamefont {Neven}}, \bibinfo {author} {\bibfnamefont {A.}~\bibnamefont {Aspuru-Guzik}}, \ and\ \bibinfo {author} {\bibfnamefont
  {J.~M.}\ \bibnamefont {Martinis}},\ }\href {\doibase 10.1103/PhysRevX.6.031007} {\bibfield  {journal} {\bibinfo  {journal} {Phys. Rev. X}\ }\textbf {\bibinfo {volume} {6}},\ \bibinfo {pages} {031007} (\bibinfo {year} {2016})}\BibitemShut {NoStop}%
\bibitem [{\citenamefont {Mao}\ \emph {et~al.}(2024)\citenamefont {Mao}, \citenamefont {Tian},\ and\ \citenamefont {Sun}}]{Mao2024}%
  \BibitemOpen
  \bibfield  {author} {\bibinfo {author} {\bibfnamefont {R.}~\bibnamefont {Mao}}, \bibinfo {author} {\bibfnamefont {G.}~\bibnamefont {Tian}}, \ and\ \bibinfo {author} {\bibfnamefont {X.}~\bibnamefont {Sun}},\ }\href {\doibase 10.1038/s42005-024-01798-0} {\bibfield  {journal} {\bibinfo  {journal} {Communications Physics}\ }\textbf {\bibinfo {volume} {7}},\ \bibinfo {pages} {342} (\bibinfo {year} {2024})}\BibitemShut {NoStop}%
\bibitem [{\citenamefont {Breuer}\ and\ \citenamefont {Petruccione}(2002)}]{breuer2002theory}%
  \BibitemOpen
  \bibfield  {author} {\bibinfo {author} {\bibfnamefont {H.-P.}\ \bibnamefont {Breuer}}\ and\ \bibinfo {author} {\bibfnamefont {F.}~\bibnamefont {Petruccione}},\ }\href@noop {} {\emph {\bibinfo {title} {The theory of open quantum systems}}}\ (\bibinfo  {publisher} {OUP Oxford},\ \bibinfo {year} {2002})\BibitemShut {NoStop}%
\bibitem [{\citenamefont {Dormand}\ and\ \citenamefont {Prince}(1980)}]{DORMAND198019}%
  \BibitemOpen
  \bibfield  {author} {\bibinfo {author} {\bibfnamefont {J.}~\bibnamefont {Dormand}}\ and\ \bibinfo {author} {\bibfnamefont {P.}~\bibnamefont {Prince}},\ }\href {\doibase https://doi.org/10.1016/0771-050X(80)90013-3} {\bibfield  {journal} {\bibinfo  {journal} {Journal of Computational and Applied Mathematics}\ }\textbf {\bibinfo {volume} {6}},\ \bibinfo {pages} {19} (\bibinfo {year} {1980})}\BibitemShut {NoStop}%
\bibitem [{\citenamefont {Wu}\ \emph {et~al.}(2007)\citenamefont {Wu}, \citenamefont {Pechen}, \citenamefont {Brif},\ and\ \citenamefont {Rabitz}}]{Wu_2007}%
  \BibitemOpen
  \bibfield  {author} {\bibinfo {author} {\bibfnamefont {R.}~\bibnamefont {Wu}}, \bibinfo {author} {\bibfnamefont {A.}~\bibnamefont {Pechen}}, \bibinfo {author} {\bibfnamefont {C.}~\bibnamefont {Brif}}, \ and\ \bibinfo {author} {\bibfnamefont {H.}~\bibnamefont {Rabitz}},\ }\href {\doibase 10.1088/1751-8113/40/21/015} {\bibfield  {journal} {\bibinfo  {journal} {Journal of Physics A: Mathematical and Theoretical}\ }\textbf {\bibinfo {volume} {40}},\ \bibinfo {pages} {5681} (\bibinfo {year} {2007})}\BibitemShut {NoStop}%
\bibitem [{\citenamefont {{\v{S}}telmachovi{\v{c}}}\ and\ \citenamefont {Bu{\v{z}}ek}(2001)}]{vstelmachovivc2001dynamics}%
  \BibitemOpen
  \bibfield  {author} {\bibinfo {author} {\bibfnamefont {P.}~\bibnamefont {{\v{S}}telmachovi{\v{c}}}}\ and\ \bibinfo {author} {\bibfnamefont {V.}~\bibnamefont {Bu{\v{z}}ek}},\ }\href {\doibase 10.1103/PhysRevA.64.062106} {\bibfield  {journal} {\bibinfo  {journal} {Physical Review A}\ }\textbf {\bibinfo {volume} {64}},\ \bibinfo {pages} {062106} (\bibinfo {year} {2001})}\BibitemShut {NoStop}%
\bibitem [{\citenamefont {Manzano}(2020)}]{manzano2020short}%
  \BibitemOpen
  \bibfield  {author} {\bibinfo {author} {\bibfnamefont {D.}~\bibnamefont {Manzano}},\ }\href@noop {} {\bibfield  {journal} {\bibinfo  {journal} {Aip advances}\ }\textbf {\bibinfo {volume} {10}} (\bibinfo {year} {2020})}\BibitemShut {NoStop}%
\bibitem [{\citenamefont {Su}\ and\ \citenamefont {Li}(2020{\natexlab{a}})}]{PhysRevA.101.012328}%
  \BibitemOpen
  \bibfield  {author} {\bibinfo {author} {\bibfnamefont {H.-Y.}\ \bibnamefont {Su}}\ and\ \bibinfo {author} {\bibfnamefont {Y.}~\bibnamefont {Li}},\ }\href {\doibase 10.1103/PhysRevA.101.012328} {\bibfield  {journal} {\bibinfo  {journal} {Phys. Rev. A}\ }\textbf {\bibinfo {volume} {101}},\ \bibinfo {pages} {012328} (\bibinfo {year} {2020}{\natexlab{a}})}\BibitemShut {NoStop}%
\bibitem [{\citenamefont {Barison}\ \emph {et~al.}(2021)\citenamefont {Barison}, \citenamefont {Vicentini},\ and\ \citenamefont {Dalmonte}}]{barison2021efficient}%
  \BibitemOpen
  \bibfield  {author} {\bibinfo {author} {\bibfnamefont {S.}~\bibnamefont {Barison}}, \bibinfo {author} {\bibfnamefont {F.}~\bibnamefont {Vicentini}}, \ and\ \bibinfo {author} {\bibfnamefont {M.}~\bibnamefont {Dalmonte}},\ }\href@noop {} {\bibfield  {journal} {\bibinfo  {journal} {Quantum}\ }\textbf {\bibinfo {volume} {5}},\ \bibinfo {pages} {512} (\bibinfo {year} {2021})}\BibitemShut {NoStop}%
\bibitem [{\citenamefont {Kamakari}\ \emph {et~al.}(2022{\natexlab{b}})\citenamefont {Kamakari}, \citenamefont {Sun}, \citenamefont {Motta},\ and\ \citenamefont {Minnich}}]{PRXQuantum.3.010320}%
  \BibitemOpen
  \bibfield  {author} {\bibinfo {author} {\bibfnamefont {H.}~\bibnamefont {Kamakari}}, \bibinfo {author} {\bibfnamefont {S.-N.}\ \bibnamefont {Sun}}, \bibinfo {author} {\bibfnamefont {M.}~\bibnamefont {Motta}}, \ and\ \bibinfo {author} {\bibfnamefont {A.~J.}\ \bibnamefont {Minnich}},\ }\href {\doibase 10.1103/PRXQuantum.3.010320} {\bibfield  {journal} {\bibinfo  {journal} {PRX Quantum}\ }\textbf {\bibinfo {volume} {3}},\ \bibinfo {pages} {010320} (\bibinfo {year} {2022}{\natexlab{b}})}\BibitemShut {NoStop}%
\bibitem [{\citenamefont {Su}\ and\ \citenamefont {Li}(2020{\natexlab{b}})}]{su2020quantum}%
  \BibitemOpen
  \bibfield  {author} {\bibinfo {author} {\bibfnamefont {H.-Y.}\ \bibnamefont {Su}}\ and\ \bibinfo {author} {\bibfnamefont {Y.}~\bibnamefont {Li}},\ }\href@noop {} {\bibfield  {journal} {\bibinfo  {journal} {Physical Review A}\ }\textbf {\bibinfo {volume} {101}},\ \bibinfo {pages} {012328} (\bibinfo {year} {2020}{\natexlab{b}})}\BibitemShut {NoStop}%
\bibitem [{\citenamefont {Chen}\ \emph {et~al.}(2024)\citenamefont {Chen}, \citenamefont {Gomes}, \citenamefont {Niu},\ and\ \citenamefont {Jong}}]{Chen2024adaptivevariational}%
  \BibitemOpen
  \bibfield  {author} {\bibinfo {author} {\bibfnamefont {H.}~\bibnamefont {Chen}}, \bibinfo {author} {\bibfnamefont {N.}~\bibnamefont {Gomes}}, \bibinfo {author} {\bibfnamefont {S.}~\bibnamefont {Niu}}, \ and\ \bibinfo {author} {\bibfnamefont {W.~A.~d.}\ \bibnamefont {Jong}},\ }\href {\doibase 10.22331/q-2024-02-13-1252} {\bibfield  {journal} {\bibinfo  {journal} {{Quantum}}\ }\textbf {\bibinfo {volume} {8}},\ \bibinfo {pages} {1252} (\bibinfo {year} {2024})}\BibitemShut {NoStop}%
\bibitem [{\citenamefont {Kokail}\ \emph {et~al.}(2019)\citenamefont {Kokail}, \citenamefont {Maier}, \citenamefont {van Bijnen}, \citenamefont {Brydges}, \citenamefont {Joshi}, \citenamefont {Jurcevic}, \citenamefont {Muschik}, \citenamefont {Silvi}, \citenamefont {Blatt}, \citenamefont {Roos},\ and\ \citenamefont {Zoller}}]{Kokail2019}%
  \BibitemOpen
  \bibfield  {author} {\bibinfo {author} {\bibfnamefont {C.}~\bibnamefont {Kokail}}, \bibinfo {author} {\bibfnamefont {C.}~\bibnamefont {Maier}}, \bibinfo {author} {\bibfnamefont {R.}~\bibnamefont {van Bijnen}}, \bibinfo {author} {\bibfnamefont {T.}~\bibnamefont {Brydges}}, \bibinfo {author} {\bibfnamefont {M.~K.}\ \bibnamefont {Joshi}}, \bibinfo {author} {\bibfnamefont {P.}~\bibnamefont {Jurcevic}}, \bibinfo {author} {\bibfnamefont {C.~A.}\ \bibnamefont {Muschik}}, \bibinfo {author} {\bibfnamefont {P.}~\bibnamefont {Silvi}}, \bibinfo {author} {\bibfnamefont {R.}~\bibnamefont {Blatt}}, \bibinfo {author} {\bibfnamefont {C.~F.}\ \bibnamefont {Roos}}, \ and\ \bibinfo {author} {\bibfnamefont {P.}~\bibnamefont {Zoller}},\ }\href {\doibase 10.1038/s41586-019-1177-4} {\bibfield  {journal} {\bibinfo  {journal} {Nature}\ }\textbf {\bibinfo {volume} {569}},\ \bibinfo {pages} {355} (\bibinfo {year} {2019})}\BibitemShut {NoStop}%
\bibitem [{\citenamefont {Yoshioka}\ \emph {et~al.}(2022)\citenamefont {Yoshioka}, \citenamefont {Sato}, \citenamefont {Nakagawa}, \citenamefont {Ohnishi},\ and\ \citenamefont {Mizukami}}]{PhysRevResearch.4.013052}%
  \BibitemOpen
  \bibfield  {author} {\bibinfo {author} {\bibfnamefont {N.}~\bibnamefont {Yoshioka}}, \bibinfo {author} {\bibfnamefont {T.}~\bibnamefont {Sato}}, \bibinfo {author} {\bibfnamefont {Y.~O.}\ \bibnamefont {Nakagawa}}, \bibinfo {author} {\bibfnamefont {Y.-y.}\ \bibnamefont {Ohnishi}}, \ and\ \bibinfo {author} {\bibfnamefont {W.}~\bibnamefont {Mizukami}},\ }\href {\doibase 10.1103/PhysRevResearch.4.013052} {\bibfield  {journal} {\bibinfo  {journal} {Phys. Rev. Res.}\ }\textbf {\bibinfo {volume} {4}},\ \bibinfo {pages} {013052} (\bibinfo {year} {2022})}\BibitemShut {NoStop}%
\bibitem [{\citenamefont {Galvão}\ \emph {et~al.}(2025)\citenamefont {Galvão}, \citenamefont {Cruz}, \citenamefont {do~Prado Rosa~Junior},\ and\ \citenamefont {Moret}}]{galvão2025variationalquantumsimulationnonadditive}%
  \BibitemOpen
  \bibfield  {author} {\bibinfo {author} {\bibfnamefont {L.~Q.}\ \bibnamefont {Galvão}}, \bibinfo {author} {\bibfnamefont {C.}~\bibnamefont {Cruz}}, \bibinfo {author} {\bibfnamefont {A.~C.}\ \bibnamefont {do~Prado Rosa~Junior}}, \ and\ \bibinfo {author} {\bibfnamefont {M.~A.}\ \bibnamefont {Moret}},\ }\href {https://arxiv.org/abs/2505.12013} {\enquote {\bibinfo {title} {Variational quantum simulation of a nonadditive relaxation dynamics in a qubit coupled to a finite-temperature bath},}\ } (\bibinfo {year} {2025}),\ \Eprint {http://arxiv.org/abs/2505.12013} {arXiv:2505.12013 [quant-ph]} \BibitemShut {NoStop}%
\bibitem [{\citenamefont {Chen}\ \emph {et~al.}(2023)\citenamefont {Chen}, \citenamefont {Kastoryano}, \citenamefont {Brandão},\ and\ \citenamefont {Gilyén}}]{chen2023quantumthermalstatepreparation}%
  \BibitemOpen
  \bibfield  {author} {\bibinfo {author} {\bibfnamefont {C.-F.}\ \bibnamefont {Chen}}, \bibinfo {author} {\bibfnamefont {M.~J.}\ \bibnamefont {Kastoryano}}, \bibinfo {author} {\bibfnamefont {F.~G. S.~L.}\ \bibnamefont {Brandão}}, \ and\ \bibinfo {author} {\bibfnamefont {A.}~\bibnamefont {Gilyén}},\ }\href {https://arxiv.org/abs/2303.18224} {\enquote {\bibinfo {title} {Quantum thermal state preparation},}\ } (\bibinfo {year} {2023})\BibitemShut {NoStop}%
\bibitem [{\citenamefont {Sagastizabal}\ \emph {et~al.}(2021)\citenamefont {Sagastizabal}, \citenamefont {Premaratne}, \citenamefont {Klaver}, \citenamefont {Rol}, \citenamefont {Neg{\^\i}rneac}, \citenamefont {Moreira}, \citenamefont {Zou}, \citenamefont {Johri}, \citenamefont {Muthusubramanian}, \citenamefont {Beekman} \emph {et~al.}}]{sagastizabal2021variational}%
  \BibitemOpen
  \bibfield  {author} {\bibinfo {author} {\bibfnamefont {R.}~\bibnamefont {Sagastizabal}}, \bibinfo {author} {\bibfnamefont {S.}~\bibnamefont {Premaratne}}, \bibinfo {author} {\bibfnamefont {B.}~\bibnamefont {Klaver}}, \bibinfo {author} {\bibfnamefont {M.}~\bibnamefont {Rol}}, \bibinfo {author} {\bibfnamefont {V.}~\bibnamefont {Neg{\^\i}rneac}}, \bibinfo {author} {\bibfnamefont {M.}~\bibnamefont {Moreira}}, \bibinfo {author} {\bibfnamefont {X.}~\bibnamefont {Zou}}, \bibinfo {author} {\bibfnamefont {S.}~\bibnamefont {Johri}}, \bibinfo {author} {\bibfnamefont {N.}~\bibnamefont {Muthusubramanian}}, \bibinfo {author} {\bibfnamefont {M.}~\bibnamefont {Beekman}},  \emph {et~al.},\ }\href@noop {} {\bibfield  {journal} {\bibinfo  {journal} {npj Quantum Information}\ }\textbf {\bibinfo {volume} {7}},\ \bibinfo {pages} {130} (\bibinfo {year} {2021})}\BibitemShut {NoStop}%
\bibitem [{\citenamefont {Consiglio}\ \emph {et~al.}(2024)\citenamefont {Consiglio}, \citenamefont {Settino}, \citenamefont {Giordano}, \citenamefont {Mastroianni}, \citenamefont {Plastina}, \citenamefont {Lorenzo}, \citenamefont {Maniscalco}, \citenamefont {Goold},\ and\ \citenamefont {Apollaro}}]{PhysRevA.110.012445}%
  \BibitemOpen
  \bibfield  {author} {\bibinfo {author} {\bibfnamefont {M.}~\bibnamefont {Consiglio}}, \bibinfo {author} {\bibfnamefont {J.}~\bibnamefont {Settino}}, \bibinfo {author} {\bibfnamefont {A.}~\bibnamefont {Giordano}}, \bibinfo {author} {\bibfnamefont {C.}~\bibnamefont {Mastroianni}}, \bibinfo {author} {\bibfnamefont {F.}~\bibnamefont {Plastina}}, \bibinfo {author} {\bibfnamefont {S.}~\bibnamefont {Lorenzo}}, \bibinfo {author} {\bibfnamefont {S.}~\bibnamefont {Maniscalco}}, \bibinfo {author} {\bibfnamefont {J.}~\bibnamefont {Goold}}, \ and\ \bibinfo {author} {\bibfnamefont {T.~J.~G.}\ \bibnamefont {Apollaro}},\ }\href {\doibase 10.1103/PhysRevA.110.012445} {\bibfield  {journal} {\bibinfo  {journal} {Phys. Rev. A}\ }\textbf {\bibinfo {volume} {110}},\ \bibinfo {pages} {012445} (\bibinfo {year} {2024})}\BibitemShut {NoStop}%
\bibitem [{\citenamefont {das Neves~Silva}\ \emph {et~al.}(2024)\citenamefont {das Neves~Silva}, \citenamefont {Queiroz~Galvão},\ and\ \citenamefont {Cruz}}]{Silva_2024}%
  \BibitemOpen
  \bibfield  {author} {\bibinfo {author} {\bibfnamefont {A.~C.}\ \bibnamefont {das Neves~Silva}}, \bibinfo {author} {\bibfnamefont {L.}~\bibnamefont {Queiroz~Galvão}}, \ and\ \bibinfo {author} {\bibfnamefont {C.}~\bibnamefont {Cruz}},\ }\href {\doibase 10.1088/1402-4896/ad6ec3} {\bibfield  {journal} {\bibinfo  {journal} {Physica Scripta}\ }\textbf {\bibinfo {volume} {99}},\ \bibinfo {pages} {095131} (\bibinfo {year} {2024})}\BibitemShut {NoStop}%
\bibitem [{\citenamefont {Galv\~ao}\ \emph {et~al.}(2025)\citenamefont {Galv\~ao}, \citenamefont {das Neves}, \citenamefont {Anka},\ and\ \citenamefont {Cruz}}]{6bm48ckl}%
  \BibitemOpen
  \bibfield  {author} {\bibinfo {author} {\bibfnamefont {L.~Q.}\ \bibnamefont {Galv\~ao}}, \bibinfo {author} {\bibfnamefont {A.~C.}\ \bibnamefont {das Neves}}, \bibinfo {author} {\bibfnamefont {M.~F.}\ \bibnamefont {Anka}}, \ and\ \bibinfo {author} {\bibfnamefont {C.}~\bibnamefont {Cruz}},\ }\href {\doibase 10.1103/6bm4-8ckl} {\bibfield  {journal} {\bibinfo  {journal} {Phys. Rev. E}\ }\textbf {\bibinfo {volume} {111}},\ \bibinfo {pages} {064119} (\bibinfo {year} {2025})}\BibitemShut {NoStop}%
\bibitem [{\citenamefont {Hoang}\ \emph {et~al.}(2024)\citenamefont {Hoang}, \citenamefont {Metz}, \citenamefont {Thomasen}, \citenamefont {Anh-Tai}, \citenamefont {Busch},\ and\ \citenamefont {Fogarty}}]{PhysRevResearch.6.013038}%
  \BibitemOpen
  \bibfield  {author} {\bibinfo {author} {\bibfnamefont {D.~T.}\ \bibnamefont {Hoang}}, \bibinfo {author} {\bibfnamefont {F.}~\bibnamefont {Metz}}, \bibinfo {author} {\bibfnamefont {A.}~\bibnamefont {Thomasen}}, \bibinfo {author} {\bibfnamefont {T.~D.}\ \bibnamefont {Anh-Tai}}, \bibinfo {author} {\bibfnamefont {T.}~\bibnamefont {Busch}}, \ and\ \bibinfo {author} {\bibfnamefont {T.}~\bibnamefont {Fogarty}},\ }\href {\doibase 10.1103/PhysRevResearch.6.013038} {\bibfield  {journal} {\bibinfo  {journal} {Phys. Rev. Res.}\ }\textbf {\bibinfo {volume} {6}},\ \bibinfo {pages} {013038} (\bibinfo {year} {2024})}\BibitemShut {NoStop}%
\bibitem [{\citenamefont {Medina}\ \emph {et~al.}(2024)\citenamefont {Medina}, \citenamefont {Drinko}, \citenamefont {Correr}, \citenamefont {Azado},\ and\ \citenamefont {Soares-Pinto}}]{PhysRevA.110.012443}%
  \BibitemOpen
  \bibfield  {author} {\bibinfo {author} {\bibfnamefont {I.}~\bibnamefont {Medina}}, \bibinfo {author} {\bibfnamefont {A.}~\bibnamefont {Drinko}}, \bibinfo {author} {\bibfnamefont {G.~I.}\ \bibnamefont {Correr}}, \bibinfo {author} {\bibfnamefont {P.~C.}\ \bibnamefont {Azado}}, \ and\ \bibinfo {author} {\bibfnamefont {D.~O.}\ \bibnamefont {Soares-Pinto}},\ }\href {\doibase 10.1103/PhysRevA.110.012443} {\bibfield  {journal} {\bibinfo  {journal} {Phys. Rev. A}\ }\textbf {\bibinfo {volume} {110}},\ \bibinfo {pages} {012443} (\bibinfo {year} {2024})}\BibitemShut {NoStop}%
\bibitem [{\citenamefont {Schuld}\ \emph {et~al.}(2014)\citenamefont {Schuld}, \citenamefont {Sinayskiy},\ and\ \citenamefont {Petruccione}}]{schuld2014quest}%
  \BibitemOpen
  \bibfield  {author} {\bibinfo {author} {\bibfnamefont {M.}~\bibnamefont {Schuld}}, \bibinfo {author} {\bibfnamefont {I.}~\bibnamefont {Sinayskiy}}, \ and\ \bibinfo {author} {\bibfnamefont {F.}~\bibnamefont {Petruccione}},\ }\href@noop {} {\bibfield  {journal} {\bibinfo  {journal} {Quantum Information Processing}\ }\textbf {\bibinfo {volume} {13}},\ \bibinfo {pages} {2567} (\bibinfo {year} {2014})}\BibitemShut {NoStop}%
\bibitem [{\citenamefont {Jeswal}\ and\ \citenamefont {Chakraverty}(2019)}]{jeswal2019recent}%
  \BibitemOpen
  \bibfield  {author} {\bibinfo {author} {\bibfnamefont {S.}~\bibnamefont {Jeswal}}\ and\ \bibinfo {author} {\bibfnamefont {S.}~\bibnamefont {Chakraverty}},\ }\href@noop {} {\bibfield  {journal} {\bibinfo  {journal} {Archives of Computational Methods in Engineering}\ }\textbf {\bibinfo {volume} {26}},\ \bibinfo {pages} {793} (\bibinfo {year} {2019})}\BibitemShut {NoStop}%
\bibitem [{\citenamefont {Sajjan}\ \emph {et~al.}(2022)\citenamefont {Sajjan}, \citenamefont {Li}, \citenamefont {Selvarajan}, \citenamefont {Sureshbabu}, \citenamefont {Kale}, \citenamefont {Gupta}, \citenamefont {Singh},\ and\ \citenamefont {Kais}}]{sajjan2022quantum}%
  \BibitemOpen
  \bibfield  {author} {\bibinfo {author} {\bibfnamefont {M.}~\bibnamefont {Sajjan}}, \bibinfo {author} {\bibfnamefont {J.}~\bibnamefont {Li}}, \bibinfo {author} {\bibfnamefont {R.}~\bibnamefont {Selvarajan}}, \bibinfo {author} {\bibfnamefont {S.~H.}\ \bibnamefont {Sureshbabu}}, \bibinfo {author} {\bibfnamefont {S.~S.}\ \bibnamefont {Kale}}, \bibinfo {author} {\bibfnamefont {R.}~\bibnamefont {Gupta}}, \bibinfo {author} {\bibfnamefont {V.}~\bibnamefont {Singh}}, \ and\ \bibinfo {author} {\bibfnamefont {S.}~\bibnamefont {Kais}},\ }\href {https://pubs.rsc.org/en/content/articlehtml/2022/cs/d2cs00203e} {\bibfield  {journal} {\bibinfo  {journal} {Chemical Society Reviews}\ }\textbf {\bibinfo {volume} {51}},\ \bibinfo {pages} {6475} (\bibinfo {year} {2022})}\BibitemShut {NoStop}%
\bibitem [{\citenamefont {Goodfellow}\ \emph {et~al.}(2014)\citenamefont {Goodfellow}, \citenamefont {Pouget-Abadie}, \citenamefont {Mirza}, \citenamefont {Xu}, \citenamefont {Warde-Farley}, \citenamefont {Ozair}, \citenamefont {Courville},\ and\ \citenamefont {Bengio}}]{goodfellow2014generative}%
  \BibitemOpen
  \bibfield  {author} {\bibinfo {author} {\bibfnamefont {I.~J.}\ \bibnamefont {Goodfellow}}, \bibinfo {author} {\bibfnamefont {J.}~\bibnamefont {Pouget-Abadie}}, \bibinfo {author} {\bibfnamefont {M.}~\bibnamefont {Mirza}}, \bibinfo {author} {\bibfnamefont {B.}~\bibnamefont {Xu}}, \bibinfo {author} {\bibfnamefont {D.}~\bibnamefont {Warde-Farley}}, \bibinfo {author} {\bibfnamefont {S.}~\bibnamefont {Ozair}}, \bibinfo {author} {\bibfnamefont {A.}~\bibnamefont {Courville}}, \ and\ \bibinfo {author} {\bibfnamefont {Y.}~\bibnamefont {Bengio}},\ }\href@noop {} {\bibfield  {journal} {\bibinfo  {journal} {Advances in neural information processing systems}\ }\textbf {\bibinfo {volume} {27}} (\bibinfo {year} {2014})}\BibitemShut {NoStop}%
\bibitem [{\citenamefont {Romero}\ and\ \citenamefont {Aspuru-Guzik}(2021)}]{romero2021variational}%
  \BibitemOpen
  \bibfield  {author} {\bibinfo {author} {\bibfnamefont {J.}~\bibnamefont {Romero}}\ and\ \bibinfo {author} {\bibfnamefont {A.}~\bibnamefont {Aspuru-Guzik}},\ }\href@noop {} {\bibfield  {journal} {\bibinfo  {journal} {Advanced Quantum Technologies}\ }\textbf {\bibinfo {volume} {4}},\ \bibinfo {pages} {2000003} (\bibinfo {year} {2021})}\BibitemShut {NoStop}%
\bibitem [{\citenamefont {Krizhevsky}\ \emph {et~al.}(2012)\citenamefont {Krizhevsky}, \citenamefont {Sutskever},\ and\ \citenamefont {Hinton}}]{NIPS2012_c399862d}%
  \BibitemOpen
  \bibfield  {author} {\bibinfo {author} {\bibfnamefont {A.}~\bibnamefont {Krizhevsky}}, \bibinfo {author} {\bibfnamefont {I.}~\bibnamefont {Sutskever}}, \ and\ \bibinfo {author} {\bibfnamefont {G.~E.}\ \bibnamefont {Hinton}},\ }in\ \href {https://proceedings.neurips.cc/paper_files/paper/2012/file/c399862d3b9d6b76c8436e924a68c45b-Paper.pdf} {\emph {\bibinfo {booktitle} {Advances in Neural Information Processing Systems}}},\ Vol.~\bibinfo {volume} {25},\ \bibinfo {editor} {edited by\ \bibinfo {editor} {\bibfnamefont {F.}~\bibnamefont {Pereira}}, \bibinfo {editor} {\bibfnamefont {C.}~\bibnamefont {Burges}}, \bibinfo {editor} {\bibfnamefont {L.}~\bibnamefont {Bottou}}, \ and\ \bibinfo {editor} {\bibfnamefont {K.}~\bibnamefont {Weinberger}}}\ (\bibinfo  {publisher} {Curran Associates, Inc.},\ \bibinfo {year} {2012})\BibitemShut {NoStop}%
\end{thebibliography}%

\end{document}